%% file: hopac.tex
\documentclass[preprint,authoryear]{elsarticle}
 
 \biboptions{longnamesfirst,semicolon}

\usepackage{graphicx}
\usepackage{epstopdf}
\usepackage{amsfonts}
\usepackage{amsmath}
\usepackage{subcaption}
\usepackage{bm}
\usepackage{bbm}
\usepackage{tabularx}
\usepackage{algorithm}
\usepackage{algorithmic}

\usepackage{color}
\definecolor{myRed}{RGB}{255,0,0}
\definecolor{myGreen}{RGB}{0,205,0}
\definecolor{myBlue}{RGB}{0,0,255}
\definecolor{myCyan}{RGB}{0,205,205}
\definecolor{myMagenta}{RGB}{205,0,205}

\newtheorem{theorem}{Theorem}
\newtheorem{proposition}{Proposition}
\newproof{proof}{Proof}

\newcommand{\argmax}{\operatornamewithlimits{argmax}}
\newcommand{\argmin}{\operatornamewithlimits{argmin}}
\newcommand{\avg}{\operatornamewithlimits{avg}}
\newcommand*{\x}{\bm{x}}

\newcommand*{\VV}{\mathcal{V}}
\newcommand*{\EE}{\mathcal{E}}

\newcommand*{\HAC}{C_{(\VV,\EE,\Psi)}}

\newcommand*{\LS}{\mathcal{LS}}
\newcommand*{\LSi}{\LS^{-1}}

\newcommand*{\VaR}{\operatorname{VaR}}

\begin{document}

\title{Outer power transformations of hierarchical Archimedean copulas: Construction, sampling and estimation}

\author[1]{Jan G\'{o}recki\corref{cor1}}
\ead{gorecki@opf.slu.cz}
\author[2]{Marius Hofert}
\ead{marius.hofert@uwaterloo.ca}
\author[3]{Ostap Okhrin}
\ead{ostap.okhrin@tu-dresden.de}

\cortext[cor1]{Corresponding author}
%\fntext[fn1]{This is the first author footnote.}
%\fntext[fn2]{Another author footnote, this is a very long
%	footnote and it should be a really long footnote. But this
%	footnote is not yet sufficiently long enough to make two
%	lines of footnote text.}
%\fntext[fn3]{Yet another author footnote.}
\address[1]{ Silesian University in Opava, Univerzitn\'{i} n\'{a}m\v{e}st\'{i} 1934/3, Karvin\'{a}, Czech Republic}
\address[2]{University of Waterloo, 200 University Avenue West,
	Waterloo, Canada}
\address[3]{Technische Universit\"at Dresden, W\"urzburger Stra{\ss}e 35, Dresden, Germany}

\begin{abstract}
A large number of commonly used parametric Archimedean copula (AC)
families are restricted to a
single parameter, connected to a concordance measure such as Kendall's
tau. This often leads to poor statistical fits, particularly in the
joint tails, and can
sometimes even limit the ability to model concordance or tail
dependence mathematically.
This work suggests outer power (OP) transformations
of Archimedean generators to overcome these limitations. The copulas
generated by OP-transformed generators can, for example, allow one to
capture both a given concordance measure and a tail dependence
coefficient simultaneously. For exchangeable OP-transformed ACs, a
formula for computing tail
dependence coefficients is obtained, as well as two feasible OP AC
estimators are proposed and their properties studied by simulation.
For hierarchical extensions of OP-transformed ACs, a new construction
principle, efficient sampling
and parameter estimation are addressed. By simulation, convergence rate and standard errors of
the proposed
estimator are studied. Excellent tail fitting capabilities of OP-transformed
hierarchical AC models are demonstrated in a risk management application.
The results show that the OP transformation is able to 
improve the statistical fit
of exchangeable ACs, particularly of those that cannot capture upper tail
dependence or strong concordance, as well as the statistical fit of hierarchical
ACs, especially in terms of tail dependence and higher dimensions.
Given how comparably simple it is to include OP transformations into
existing exchangeable and hierarchical AC models, this transformation
provides an attractive trade-off between computational effort and
statistical improvement.	
\end{abstract}

%maximum 85 characters, including spaces, per bullet point
%\begin{highlights}
%	\item Outer power transformed Archimedean copulas excel in tail dependence modeling
%	\item Exchangeable transformed models bring exceptional improvement/complexity trade-off
%	\item Non-exchangeable transformed models adding extra flexibility are developed
%	\item Complementary sampling and estimation methods are proposed
%	\item Transformed models keep analytically expressible relationship to dependence measures
%\end{highlights}


\maketitle

\input{body}

\section*{Acknowledgements}
The authors would like to thank Maximilian Coblenz for providing them early versions of the MATVines toolbox.
Also, the authors are grateful to the project No. CZ.02.2.69/0.0/0.0/16\_027/0008521 ``Support of International Mobility of Researchers at SU'' which supports international cooperation.

% There are no strict requirements on reference formatting at submission. References can be in any style or format as long as the style is consistent.

\bibliographystyle{elsarticle-harv}
\bibliography{lit} 

%\clearpage
\section*{Appendix}
\label{sec:appendix}
%Algorithm~\ref{alg:structure_estim} from \citet[Algorithm~1]{gorecki2017structure} for structure estimation from the sample version of a Kendall correlation matrix.
\begin{algorithm}[!h]
	\floatname{algorithm}{Algorithm}
	\caption{HACs structure estimation \citep[Algorithm~1]{gorecki2017structure}}
	\label{alg:structure_estim}
	
	\begin{algorithmic}
		\renewcommand{\algorithmicrequire}{\textbf{Input:}}
		\renewcommand{\algorithmicensure}{\textbf{The estimation:}}
		\REQUIRE 
		\STATE 1) $(\tau^n_{ij})$ -- the sample version of a Kendall correlation matrix
		%\STATE 2) $g$ ... an $\I$-aggregation function
		
		\STATE ~
		\ENSURE
		\STATE 1. $\hat{\VV} := \{1, ..., 2d-1\}, ~\hat{\EE} := \emptyset$, $\mathcal{I} := \{1, ..., d\}$ 
		\STATE ~~~ recall that $\downarrow\!(i) = \{i\}$ for $i \in \{1, ...,d\}$
		\FOR{$k = 1, ..., d - 1$}
		\STATE 2. find two nodes from $\mathcal{I}$ to join, i.e, 
		\STATE ~~~ $(i, j) := \argmax\limits_{\tilde{i} < \tilde{j},~ \tilde{i} \in \mathcal{I}, ~\tilde{j} \in \mathcal{I}} \avg((\tau^n_{\tilde{\tilde{i}}\tilde{\tilde{j}}})_{(\tilde{\tilde{i}},\tilde{\tilde{j}}) \in \downarrow(\tilde{i}) \times \downarrow(\tilde{j})})$
		\STATE 3. set the children of the fork $d+k$ to $\{i, j\}$, i.e., 
		\STATE ~~~ $\hat{\EE} := \hat{\EE} \cup \{\{i, d+k\}, \{j, d+k\}\}$,
		\STATE ~~~ which implies that $\wedge(d+k) = \{i, j\}$ and $\downarrow\!(d+k) = \downarrow\!(i) \cup \downarrow\!(j)$
		\STATE 4. remove the nodes $i$ and $j$ from the clustering process (as they have been
		\STATE ~~~ already joined) and add the fork $d+k$ to be considered for joining in 
		\STATE ~~~ the following steps, i.e., 
		\STATE ~~~ $\mathcal{I} := \mathcal{I} \cup \{d + k\} \backslash \{i, j\}$
		\STATE 5. estimate the Kendall's tau corresponding to the fork $d+k$, i.e., 
		\STATE ~~~ $\hat{\tau}_{d+k} := \avg((\tau^n_{\tilde{i}\tilde{j}})_{(\tilde{i},\tilde{j}) \in \downarrow(i) \times \downarrow(j)})$
		\ENDFOR
		\STATE ~
		\renewcommand{\algorithmicensure}{\textbf{Output:}}
		\ENSURE
		\STATE $(\hat{\VV}, \hat{\EE}), (\hat{\tau}_k)_{d+1}^{2d-1})$
		
	\end{algorithmic}
\end{algorithm}

\end{document}

%% file: body.tex
\section{Introduction}
\label{sec:introduction}
Archimedean copulas (ACs) are dependence models frequently used in finance, insurance and risk management, e.g., for stress testing. 
In contrast to elliptical copulas such as the prominent Gaussian and $t$ copulas, ACs allow for asymmetry in the joint tails, which is of particular interest, e.g., in risk management \citep[Chapter~5]{mcneil2015quantitative} or hydrology \citep{Gen07,liu2018hydrological}. ACs are also appreciated for their simple and explicit construction, for efficient sampling techniques and for likelihood-based inference; see \cite{Hofert13}.  

However, as follows from the construction, all multivariate margins of an AC are the same, which limits ACs' applicability particularly in high dimensions.
Also, a vast majority of known ACs are one-parametric, which, on the one hand, enables the mentioned advantages, but on the other hand causes the following limitation. The single parameter completely determines all properties of the AC, as well as for many families it is related in a one-to-one relationship  to the strength of the dependence, e.g., expressed by Kendall's tau or Spearman's rho, see Table 1 in \cite{Gen93}. It is thus natural that this parameter is frequently estimated in the method of moments fashion such that the model's dependency measure is close to its empirical counterpart. However, this often results in a model that fits well in its body, but not that well in its tails. 

To alleviate these limitations, several approaches have been introduced in the literature. This work particularly focuses on the following two: 
\begin{enumerate}
	\item Given a one-parameter family of ACs, a way to construct its two-parameter extension called the \emph{outer}\footnote{Note that the referred book uses the name \emph{exterior} instead of \emph{outer}.} \emph{power AC (OPAC) family} is proposed in \citet[Theorem 4.5.1]{Nel06}. 
	With an additional parameter, one can fix, e.g., the model's Kendall's tau ($\tau$) to a desired value (to keep a good fit in the body), and then fine-tune both parameters to get a good fit in one of the tails. Such a property is crucial, e.g., in risk management applications \citep{Hofert13,mcneil2015quantitative};
	\item \citet[pp.~87]{Joe97} proposed a way how to construct \emph{hierarchical} (or \emph{nested}) \emph{ACs} (\emph{HACs}) by nesting several ACs into each other. This allows certain multivariate margins to differ (which lead to asymmetric models) and extends the one-parameter model to allow up to $(d-1)$-parameters. 
	%Competitiveness of such constructions with even more flexible copula models like \emph{vine} or \emph{factor} copulas has been empirically studied, e.g., in \cite{okhrin2017copulae} or  \cite{bacigal2019state}.
	However, to this date, all contributions in the literature addressing HACs' estimation have been restricted to the case where all ACs nested in a HAC are one-parametric; see \cite{Okh13,goreckihofertholena2016approachjiis,gorecki2017structure} to mention a few.
\end{enumerate}

\begin{figure}[t]
	\centering
	\includegraphics[width=1\textwidth]{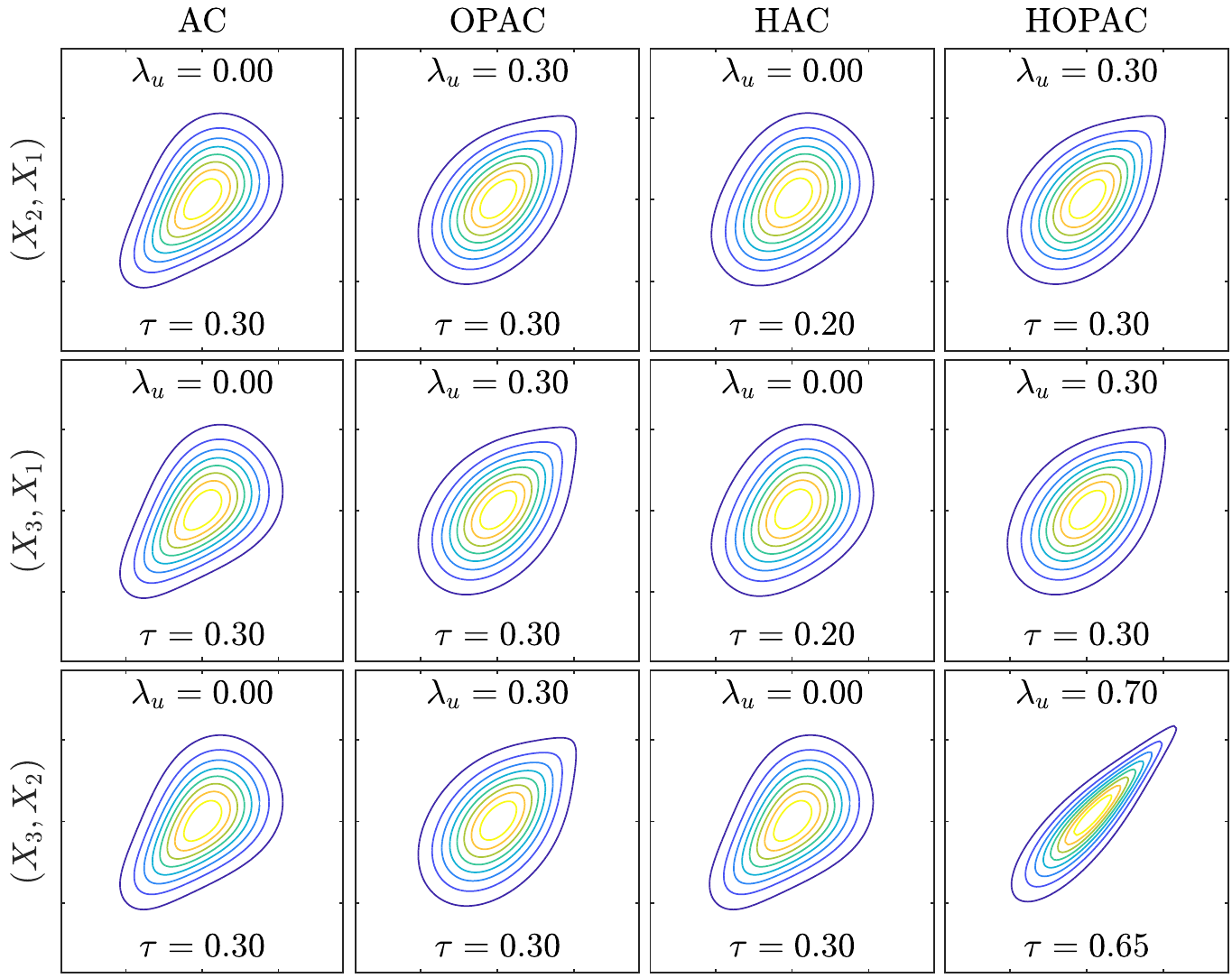}
	\caption{Contour plots of the bivariate marginal densities of a random vector
		$(X_1,X_2,X_3)$ with $\textrm{N}(0, 1)$ margins and varying (across columns) Ali-Mikhail-Haq copulas. Note that $\tau$ and $\lambda_u$ denote the corresponding pairwise Kendall's tau and upper-tail dependence coefficient, respectively.}
	\label{fig:hopacAcontour}
	% MATLAB outerpowerCSDA2transposed.m
\end{figure}

This work merges these two approaches, resulting in  \emph{hierarchical outer power ACs} (\emph{HOPACs}), which are tail-asymmetric copulas that allow for different multivariate margins with extra flexibility added by the outer power (OP) transformation. 
To illustrate the flexibility, Figure~\ref{fig:hopacAcontour} presents trivariate examples of HOPACs, ACs, OPACs and HACs based on Ali-Mikhail-Haq copulas that are combined with $\textrm{N}(0, 1)$ margins and displayed via contour plots of bivariate marginal densities.

In the first column, the parameter of the trivariate Ali-Mikhail-Haq copula is chosen such that Kendall's tau of all bivariate margins is 0.3. Such a choice provides a good fit in the model's body if we \emph{observe} Kendall's tau close to this value in data of interest.
Following the nature of Ali-Mikhail-Haq copulas, the dependence is clearly tail-asymmetric, i.e., for small joint values (found in the bottom-left corners of each density plot) it is stronger than for large values (top-right corners). However, if we observe stronger dependence for large values in our data, such a model does not allow to adjust to such a situation and require one to discard the Ali-Mikhail-Haq family for modeling purposes. Involving the OP transformation, this adjustment is possible, as illustrated in the second column, where the extra parameter allows us to fine-tune the upper-tail dependence coefficient, e.g., to 0.3, while still \emph{keeping} the same value of Kendall's tau. This allows one to build dependence models that fit well \emph{both} in the body and tail, and, as we demonstrate in Section~\ref{sec:applic}, just this transition from ACs to OPACs already yields substantial improvement in tail-dependence modeling for risk management applications.

In contrast to ACs and OPACs, which are exchangeable and thus all their multivariate margins are the same, HACs and HOPACs provide asymmetry in the bivariate margins. It can be seen from the last two columns that the margin corresponding to $(X_2, X_3)$ differs (in $\tau$) from the remaining two. Moreover, as Ali-Mikhail-Haq ACs are limited in $\tau$ to  $[0, 1/3)$, the OP transformation allows to attain any $\tau \in [0, 1)$, which makes (H)OPACs based on this family more suitable for modeling purposes. 
An Ali-Mikhail-Haq OPAC with $\tau \geq 1/3$ is, e.g., the bivariate margin corresponding to $(X_2, X_3)$ in the fourth column. 
Similarly, as  Ali-Mikhail-Haq ACs are tail independent, i.e., $\lambda_u = 0$ for all parameter values, their OP transformation allows to attain $\lambda_u$ with any value from $[0, 1)$, which is illustrated by the same example. This extended flexibility in tail dependence modeling enabled the Ali-Mikhail-Haq HOPACs to gain the best results in several scenarios out of all copula constructions/families considered in a risk management application provided in Section~\ref{sec:applic}.

Also note that in \cite{Hof11CDO}, a sub-class of HOPAC models based on Clayton copulas shows the best performance among all the copula models considered. 
The limitation of the mentioned sub-class lied in the fact that those HOPACs were restricted to one certain copula structure with two nesting levels. Moreover, they were constructed in a way that one of the parameters of each two-parameter OPAC nested in a HOPAC was arbitrarily fixed to a certain value that was the same for all of them, i.e., these nested two-parameter OPACs were in fact turned to one-parameter ACs. We relax both such limitations.

Finally, note that the OP transformation establishes a connection between several one-parameter AC families from the list in \citet[pp. 116--119]{Nel06}. For example, the three families denoted 4.2.1 (Clayton), 4.2.12 and 4.2.14 are special cases of the OP Clayton family. 
To have these families in a single HAC, it was necessary to develop sampling and estimation strategies allowing to nest \emph{different} families, which was done in \cite{Hof11} and \cite{gorecki2017structure}, respectively. 
With regard to OP transformations, sampling and estimation of HACs involving these different one-parameter families can be addressed by restricting to hierarchical models involving just \emph{one} but \emph{OP-transformed} AC family, i.e., Clayton HOPACs.

The paper is organized as follows. Section~\ref{sec:prelim} recalls basic concepts concerning ACs, the OP transformation and their relationship to three measures of association. For the tail dependence coefficients a new result simplifying their computation is proposed. This section also recalls an efficient sampling strategy for OPACs and presents a simulation study on the viability of two OPAC estimators.
Efficient sampling and estimation strategies for HOPACs are then developed in 	Section~\ref{sec:nestedcase}, and their excellent abilities in tail dependence modeling are demonstrated in an application from risk management reported in Section~\ref{sec:applic}. Section~\ref{sec:conclu} provides concluding remarks. 

\section{The exchangeable case}
\label{sec:prelim}

\subsection{Archimedean copulas}
An \emph{Archimedean generator}, or simply \emph{generator}, is a continuous, decreasing function $\psi : [0,\infty) \rightarrow [0, 1]$ that is strictly decreasing on $[0, \inf\{t:\psi(t) = 0\}]$ and satisfies $\psi(0) = 1$ and $\lim_{t \rightarrow \infty} \psi(t) = 0$. If $(-1)^k \psi^{(k)}(t)$ $\geq 0$ for all $k \in \mathbb{N}, ~t \in [0,\infty)$, then $\psi$ is called \emph{completely monotone} (shortly, \emph{c.m.}); see \cite{kimberling1974} or \citet[p.~54]{Hof10book}.
As follows from \cite{McNei09}, given a c.m.~generator $\psi$, the function $C_{\psi}:[0,1]^d \rightarrow [0, 1]$ defined by
\begin{equation}
\label{eq:AC}
C_{\psi}(u_1, ..., u_d) = \psi \{\psi^{-1}(u_1)+ ... + \psi^{-1}(u_d)\}
\end{equation}
is a $d$-dimensional \emph{Archimedean} copula ($d$-AC) for any $d \geq 2$,
where  $\psi^{-1}$ is the generalized inverse of $\psi$ given by $\psi^{-1}(s) = \inf \{t \in [0, \infty]~|~\psi(t) = s\}, s \in [0, 1]$. In what follows, we assume all appearing generators to be c.m.

Table \ref{tab:geners} shows the popular Archimedean generators of Ali-Mikhail-Haq (A), Clayton (C), Frank (F), Gumbel (G) and Joe (J), which will serve as working examples throughout the paper. Also note that $\psi_{(a, \theta)}$ will denote the generator of a family $a$ with a real parameter $\theta \in \Theta_a \subseteq [0, \infty)$.
\begin{table*}[tb]
		\caption{Five popular families of c.m.~one-parameter generators. For each family, the table shows its family label $a$, parameter range $\Theta_a  \subseteq [0, \infty)$,  form of $\psi_{(a,\theta)}$, and the lower- and upper-tail dependence coefficients $\lambda_l := \lim_{t \downarrow 0} C_{\psi}(t, t)/t$ and  $\lambda_u := \lim_{t \downarrow 0} \{1-2t +C_{\psi}(t,t)\}/(1-t)$, where $\beta \in [1, \infty)$ is the OP transform parameter.}		
	\label{tab:geners}
\begin{tabularx}{\linewidth}{@{}llllX@{}}
			\hline
			~~~~$a$~~~~ & $\Theta_a$~~~~~~ &$\psi_{(a,\theta)}(t)$ & $\lambda_l$ & $\lambda_u$ \\
			\hline
			Ali-Mikhail-Haq (A) & [0, 1)         & $(1-\theta)/(\mathrm{e}^t - \theta)$               & 0                         & $2- 2^{1/\beta}$ \\
			Clayton (C)         & $(0, \infty)$  & $(1+t)^{-1/\theta}$                                & $2^{-1/(\theta \beta)}$   & $2- 2^{1/\beta}$ \\
			Frank (F)           & $(0, \infty)$  & $\frac{-\log\{1 -(1 - \mathrm{e}^{-\theta})\mathrm{e}^{-t}\}}{\theta}$ & 0                         & $2- 2^{1/\beta}$ \\
			Gumbel (G)          & $[1, \infty)$  & $\mathrm{e}^{-t^{1/\theta}}$                              & 0                         & $2- 2^{1/(\theta\beta)}$\\
			Joe (J)             & $[1, \infty) $ & $1 - (1 - \mathrm{e}^{-t})^{1/\theta} $                     & 0                         & $2- 2^{1/(\theta\beta)}$\\
			%12 & $[1, \infty)$ & $(1 + t^{1/\theta})^{-1}$ & $\theta_1 \leq \theta_2$ & $2^{-1/\theta}$ & $2 - 2^{1/\theta}$\\
			%14 & $[1, \infty)$ & $(1 + t^{1/\theta})^{-\theta}$ & unknown & $1/2$ & $2- 2^{1/\theta}$\\
			%19 & $(0, \infty)$ & $\theta / {\ln\left(t + \mathrm{e}^{\theta}\right)}$ & $\theta_1 \leq \theta_2$ & 1 & 0\\
			%20 & $(0, \infty)$ & $\ln ^{-1/\theta}(t + \mathrm{e})$ & $\theta_1 \leq \theta_2$  & 1 & 0\\
			\hline
\end{tabularx}
\end{table*}

\subsection{Outer power transformation} 

Theorem 4.5.1 in \cite{Nel06} implies that for any $\beta \in [1, \infty)$ and any generator $\psi$ of a 2-AC
\begin{equation}
\label{eq:OP_trans}
\psi_{\beta}(t) = \psi(t^{1/\beta})
\end{equation}
generates again a proper 2-AC. Parametric families generated in this way are referred to as \emph{outer power families}, where the unintuitive use of ``outer'' relates to the fact that they were named with reference to generator inverses. Given a one-parameter generator $\psi_{(a, \theta)}$, its OP transformed version with parameter $\beta  \in [1, \infty)$ is denoted by $\psi_{(a, \theta, \beta)}$. A well-known example of an OPAC family is the \emph{generalized Clayton copula} (also denoted BB1, see \cite{joe2014dependence}), % given by
%\begin{equation*}
%\label{eq:opC}
%C_{\psi_{(\textrm{C}, \theta, \beta})}(u_1, u_2) = [\{(u_1^{-\theta}-1)^{\beta} + (u_2^{-\theta}-1)^{\beta}\}^{1/\beta} + 1]^{-1/\theta}, 
%\end{equation*}
which, as mentioned before, encompasses the three one-parameter families from \citet[pp. 116--119]{Nel06} denoted 4.2.1 (Clayton), 4.2.12 and 4.2.14 as special cases. 

Note that the Gumbel family has generator $\psi_{(\textrm{G}, \theta)} = \textrm{e}^{-t^{1/\theta}}$, so OP Gumbel copulas are simply Gumbel copulas with parameter $\theta\beta$ instead of $\theta$ since 
\begin{equation*}
\psi_{(\textrm{G}, \theta, \beta)}(t) = \psi_{(\textrm{G}, \theta)}(t^{1/\beta}) = \textrm{e}^{-(t^{1/\beta})^{1/\theta}} = \textrm{e}^{-t^{1/(\theta\beta)}} = \psi_{(\textrm{G}, \theta\beta)},
\end{equation*}
where $\theta \in [1, \infty)$, is again a one-parameter Gumbel with the parameter $\theta\beta$. For this reason, this family is not further considered.

\subsection{Outer power transformed dependence measures}

As can be observed from the examples in Figure~\ref{fig:hopacAcontour}, the OP transformation can have an impact on measures of association such as Kendall's tau (e.g., its values reached beyond $\frac{1}{3}$ reached for the Ali-Mikhail-Haq family) or the tail dependence coefficients  (the upper-tail independent Ali-Mikhail-Haq family reached $\lambda_u > 0$). We now consider these three measures of association in more detail for OPACs.

Given a one-parameter 2-AC $C_{\psi_{(a, \theta)}}$, there exists a functional relationship between the parameter $\theta$ and Kendall's tau that can sometimes be expressed in a closed form, e.g., $\tau_{(\text{C})}(\theta) = \theta/(\theta+2)$ for the Clayton family.
This relationship can easily be extended to OPACs. As follows from Proposition 3.7 in \cite{Hof11}, given a 2-OPAC $C_{\psi_{(a, \theta, \beta)}}$, its corresponding Kendall's tau $\tau_{(a)}(\theta, \beta)$ is
\begin{equation}
  \label{eq:OPTAU}
  \tau_{(a)}(\theta, \beta) = 1 - \{1 - \tau_{(a)}(\theta)\}/\beta.
\end{equation}
We thus see how Kendall's tau of Ali-Mikhail-Haq copulas can cover the whole $[0,1)$, while  $\tau_{(\textrm{A})}(\theta)$ only covers $\left[0,\frac{1}{3}\right)$. A similar result can be derived for the coefficients of tail dependence under additional assumptions on $\psi_{(a, \theta)}$ (or $\psi_{(a, \theta)}'$) such as regular variation. Note that $\lambda_l(C)$ ($\lambda_u(C)$) denotes the lower (upper) tail dependence coefficient of a 2-AC $C$.\\
\begin{proposition}
	Let $\psi$ be a generator of a 2-AC $C_{\psi}$ and $\psi_\beta(t) =  \psi(t^{\frac{1}{\beta}})$ for all $t \in [0, \infty)$ and $\beta \in [1, \infty)$. Then:
	\begin{enumerate}
		\item If $\psi$ is regularly varying at infinity with index $\alpha \in \mathbb{R}$, i.e., 
		$\lim_{t \rightarrow \infty} \frac{\psi(ct)}{\psi(t)} = c^{\alpha}$ for all $c \in (0, \infty)$,
		then
		$\lambda_l(C_{\psi_\beta}) =2^{\frac{\alpha}{\beta}}$.
		\item If $\psi'$ is regularly varying at zero with index $\alpha_0 \in \mathbb{R}$, i.e., 
		$\lim_{t \downarrow 0} \frac{\psi'(ct)}{\psi'(t)} = c^{\alpha_0}$ for all $c \in (0, \infty)$
		then $\lambda_u(C_{\psi_\beta}) = 2 -  2^{\frac{\alpha_0+1}{\beta}}$.
	\end{enumerate}
	\label{thm:lambdas}
\end{proposition}

\begin{proof}
	\begin{enumerate}
		\item Using (2.11) from \cite{Hof10book}, 
		$\lambda_l(C_{\psi_\beta}) = 
		\lim\limits_{t\rightarrow\infty} \frac{\psi_\beta(2t)}{\psi_\beta(t)} =$ $ \lim\limits_{t\rightarrow\infty}$ $\frac{\psi(2^{1/\beta}t^{1/\beta})}{\psi(t^{1/\beta})}= $ $
		\lim\limits_{s\rightarrow\infty}\frac{\psi(2^{1/\beta}s)}{\psi(s)}$, where $s = t^{1/\beta}$.
		If $\psi$ is regularly varying at infinity with index $\alpha$, $c = 2^{1/\beta}$ establishes the proof.
		\item Using (2.12) from \cite{Hof10book}, 
		$\lambda_u(C_{\psi_\beta}) = 
		2 - 2\lim\limits_{t\downarrow 0} \frac{1 - \psi_\beta(2t)}{1 -\psi_\beta(t)} =$ 
		$ 2 - \lim\limits_{t\downarrow 0}\frac{1 -\psi(2^{1/\beta}t^{1/\beta})}{\psi(t^{1/\beta})}= $ $
		2 - \lim\limits_{s\downarrow 0}\frac{1 -\psi(2^{1/\beta}s)}{1-\psi(s)}$, where $s = t^{1/\beta}$. Applying l'H\^{o}pital's rule,  $2 - \lim\limits_{s\downarrow 0}\frac{1 -\psi(2^{1/\beta}s)}{1-\psi(s)} =
		 2 - 2^{1/\beta}\lim\limits_{s\downarrow 0}\frac{\psi'(2^{1/\beta}s)}{\psi'(s)}$. 
		If $\psi'$ is regularly varying at zero with index $\alpha_0$, using $c = 2^{1/\beta}$ implies that 
		$2 - 2^{1/\beta}\lim\limits_{s\downarrow 0}\frac{\psi'(2^{1/\beta}s)}{\psi'(s)} =
		2 - 2^{1/\beta}(2^{1/\beta})^{\alpha_0} = 2 -  2^{\frac{\alpha_0+1}{\beta}}$.  \qed
	\end{enumerate}
\end{proof} 

\begin{figure}[t]
	\centering
	\includegraphics[width=1\textwidth]{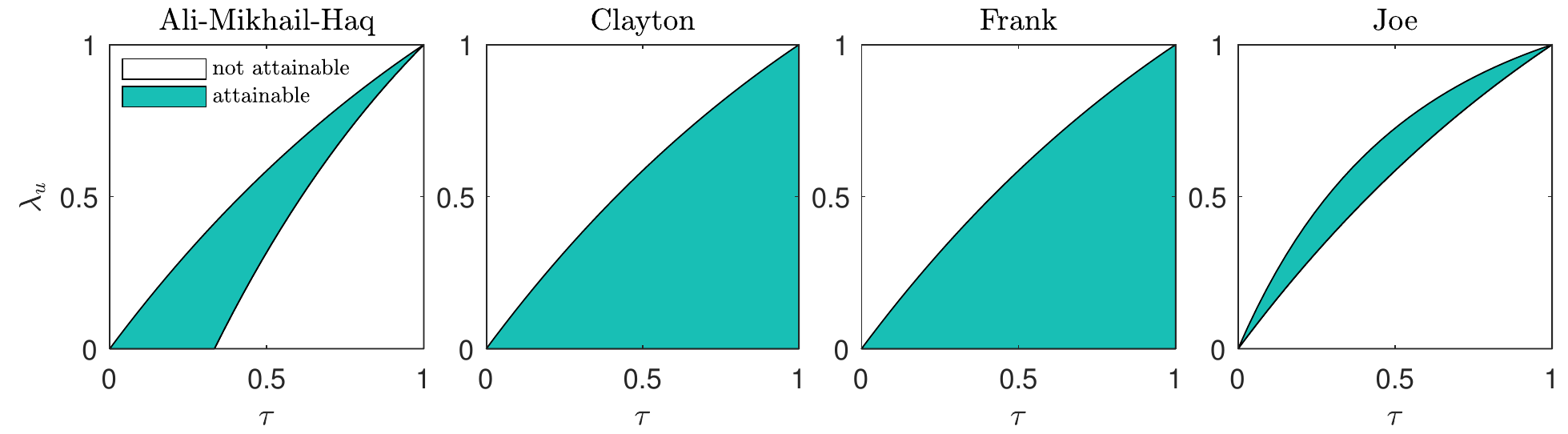}
	\caption{Attainable pairs of $\tau$ and $\lambda_u$ for four OPAC families.}
	\label{fig:attainables}
	% MATLAB outerpowerCSDA3.m
\end{figure}

Having the index $\alpha$ ($\alpha_0$) for a one-parameter family, the proposition provides $\lambda_l$ ($\lambda_u$) for its OP family. 
It can easily be verified that $\alpha = -\theta^{-1}$ for $a =$ C, whereas $\psi_{(a, \theta)}$ is not regularly varying at $\infty$ for $a \in $ \{A, F, J\}, and that $\alpha_0 = 1$ for $a \in $ \{A, C, F\}, whereas $\alpha_0 = \theta^{-1} - 1$ for $a =$ J; see also the last two columns of Table \ref{tab:geners}.
These two columns also reveal that $\beta$ influences $\lambda_u$ for all listed families, whereas $\lambda_l$ only for the Clayton family. From this point of view, $\beta$ is playing an important role particularly for upper-tail dependence modeling. Given a bivariate OPAC $C_{\psi_{(a,\theta,\beta)}}$ and also considering its Kendall's tau via \eqref{eq:OPTAU}, an interesting question is if, given $(\tau,\lambda_u) \in [0, 1]^2$, there exist values of $\theta$ and $\beta$ such that $\tau_{(a)}(\theta,\beta) = \tau$ and $\lambda_u(C_{\psi_{(a,\theta,\beta)}}) = \lambda_u$. This question is answered in Figure~\ref{fig:attainables}, which highlights the pairs of $\tau$ and $\lambda_u$ for which such $\theta$ and $\beta$ exist (so which $\tau$ and $\lambda_u$ are \emph{attainable}).
We observe a larger region of attainable pairs for the upper-tail independent AC families of Ali-Mikhail-Haq, Clayton and Frank; for the Joe family, $\lambda_u > 0$ even with $\beta=1$. 
Hence, fixing $\tau$ to a desired value in order to obtain a good fit in an OPAC's body, the most flexibility in the upper tail is provided by, somewhat unexpected, the upper-tail \emph{independent} AC family.

\begin{figure}[tbh!]
	\centering
	\begin{subfigure}[t]{0.49\textwidth}
		\centering
		\includegraphics[width=1\textwidth]{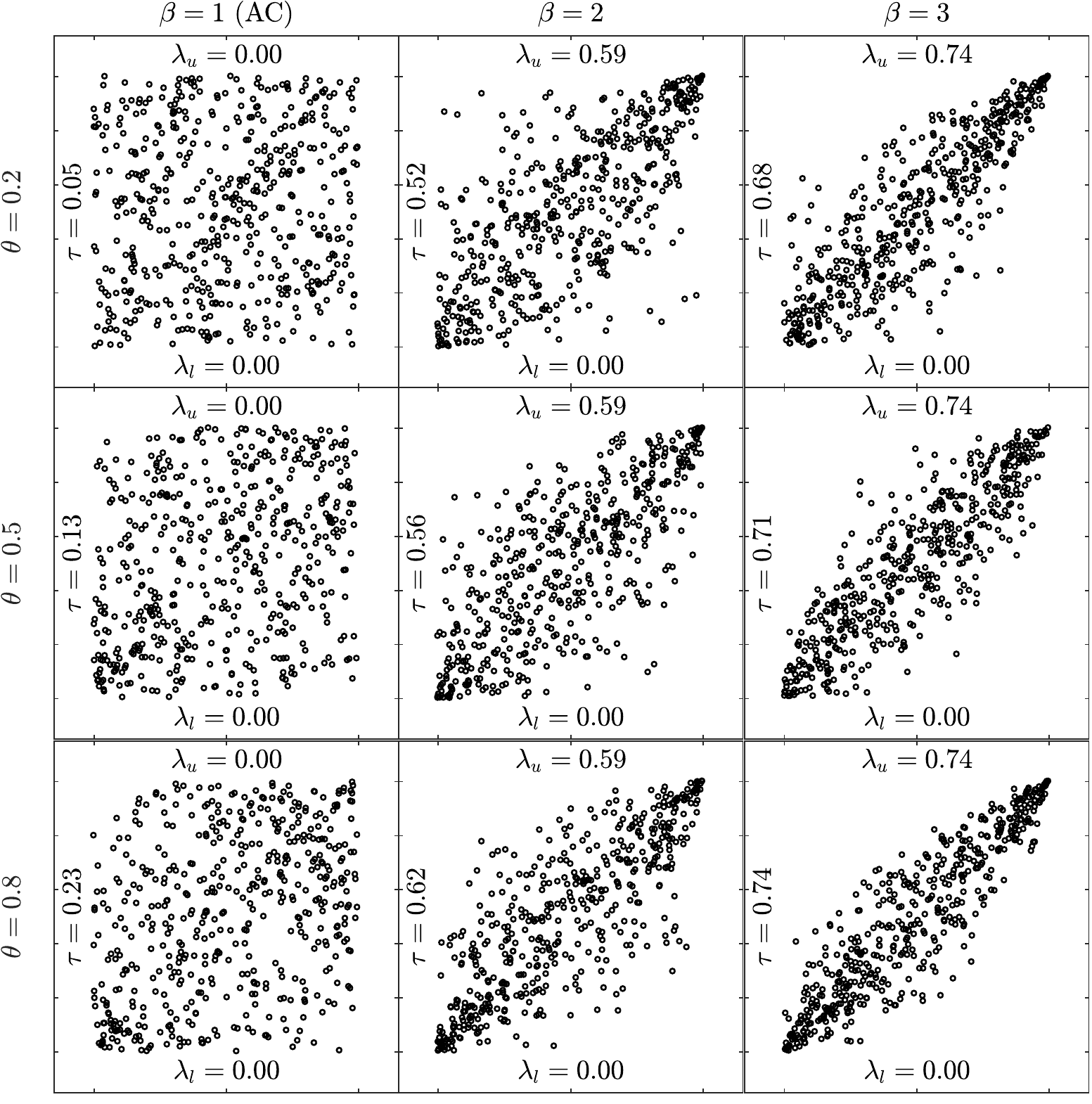}
		\caption{Ali-Mikhail-Haq}
		\label{fig:gridA}
	\end{subfigure}
	\hfill
	\begin{subfigure}[t]{0.49\textwidth}
		\centering
		\includegraphics[width=1\textwidth]{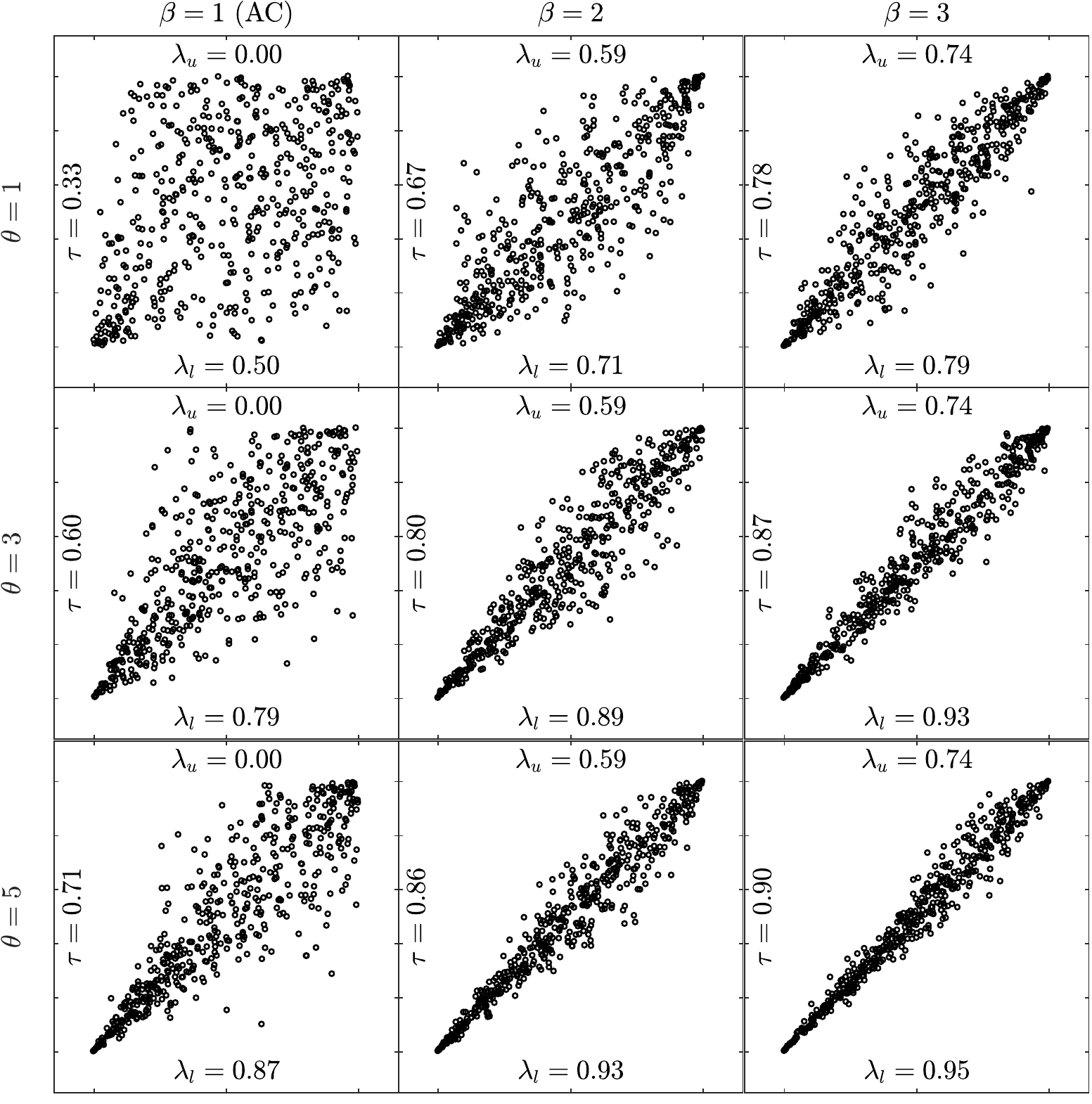}
		\caption{Clayton}
		\label{fig:gridC}
	\end{subfigure}
	\begin{subfigure}[t]{0.49\textwidth}
	\centering
	\includegraphics[width=1\textwidth]{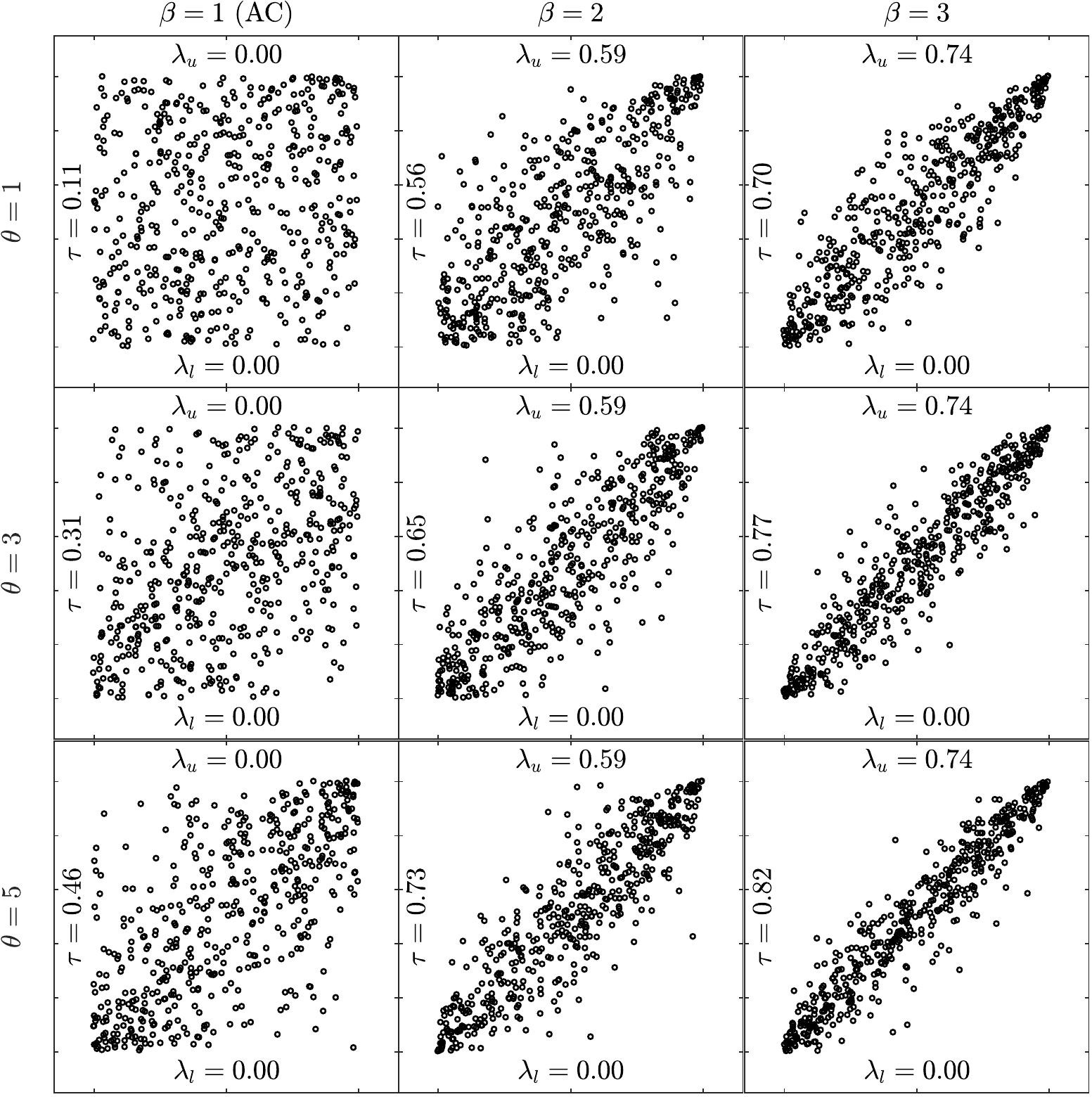}
	\caption{Frank}
	\label{fig:gridF}
\end{subfigure}
\hfill
\begin{subfigure}[t]{0.49\textwidth}
	\centering
	\includegraphics[width=1\textwidth]{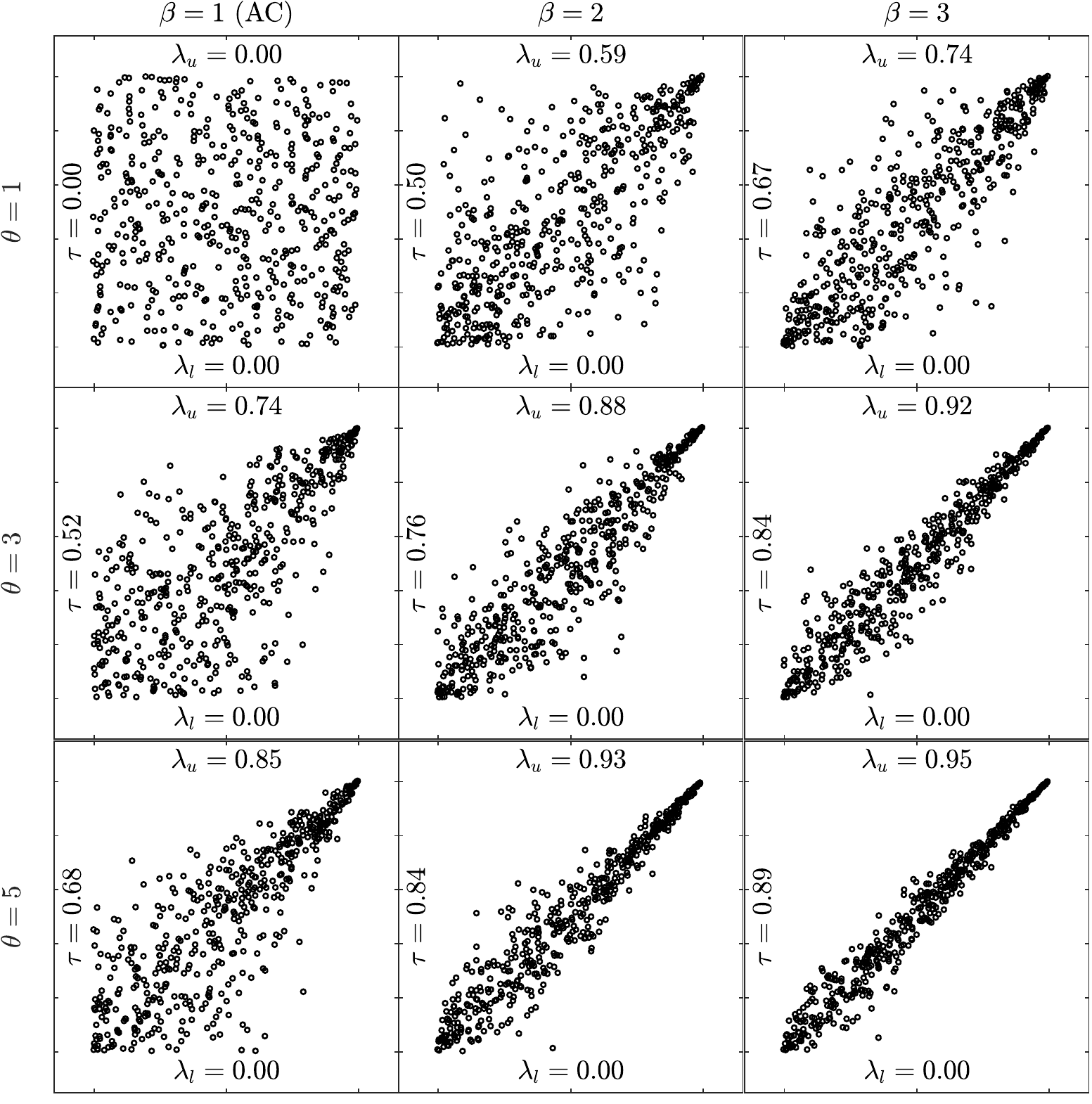}
	\caption{Joe}
	\label{fig:gridJ}
\end{subfigure}
	\caption{Samples from OPACs $C_{\psi_{(a, \theta, \beta)}}$ with $n = 500$.}
	\label{fig:OPACgrids}
	% MATLAB % MATLAB OPACsgrid.m
\end{figure}

Finally note that all three considered measures of association are monotone with respect to $\beta$, which, for $\lambda_l$ and $\lambda_u$, follows from Proposition~3.7 in \cite{Hof11}. This is also clearly visible in Figure \ref{fig:OPACgrids}, where samples of size $n = 500$ from a bivariate OPAC with different values of $\theta$ and $\beta$ are shown for each working family.

\subsection{Sampling from OPACs}
\label{sec:OPACsampling}
The samples in Figure~\ref{fig:OPACgrids} were obtained using the well-known sampling algorithm of \cite{marshall1988families} together with Theorem 3.6 from \cite{Hof11}; for the sake of completeness, we recall the latter below as it is needed later in Section \ref{sec:constandsamp}. Note that $\mathcal{LS}^{-1}[f]$ denotes the inverse Laplace–Stieltjes transform of a function $f$.
\begin{theorem}[\cite{Hof11}]
	Let $\beta \in [1, \infty)$, $\psi$ be a c.m.~generator and $\psi_\beta$ be its OP transformation given by \eqref{eq:OP_trans}. Then
	\begin{equation*}
	\tilde{V} := SV^{\beta} \sim \tilde{F} := \mathcal{LS}^{-1}[\psi_{\beta}],
	\end{equation*}
	where $V \sim \mathcal{LS}^{-1}[\psi]$ and $S \sim \textrm{S}\{1/\beta, 1, \cos^{\beta}(\pi /2\beta), \mathbbm{1}_{\{\beta = 1\}}; 1\}$ $($$1/\beta$-stable distribution{\normalfont)}.
\label{thm:samplingOPAC}
\end{theorem}
As sampling from $S$ is a standard routine, sampling from an OPAC only requires a sampling strategy for $\mathcal{LS}^{-1}[\psi]$, which is known for many one-parameter families; see \citet[Table 2.1]{Hof10book}.

\subsection{Estimating OPACs}
\label{sec:estimatingOPAC}
For the one-parameter case, two AC estimators are particularly popular, 1) the~maximum likelihood (ML) estimator and 2) the Kendall's tau inverse estimator; see \cite{Gen93}. The latter can be viewed as a generalized method of moments estimator and is statistically not as efficient as ML. In contrast, the ML estimator naturally extends to any parameter dimension and is also feasible for estimating OPACs. To complement this estimator, we also consider a distance-based estimator (based on a goodness-of-fit test statistic) in what follows.

Given a copula family $a$, the ML estimator for the parameters $\theta$ and $\beta$ of a $d$-OPAC $C_{\psi_{(a,\theta,\beta)}}$ is defined by
\begin{equation}
(\hat{\theta}_{\textrm{ML}}, \hat{\beta}_{\textrm{ML}}
) = \argmax_{(\theta, \beta)}\sum_{i=1}^{n} \log c_{\psi_{(a,\theta,\beta)}}(\bm{u}_i), 
\label{eq:mleopac}
\end{equation}
and a distance-based estimator, denoted $S_n$, by
\begin{equation}
(\hat{\theta}_{S_n}, \hat{\beta}_{S_n}
) = \argmin_{(\theta, \beta)}\sum_{i=1}^{n}\{C_{\psi_{(a,\theta,\beta)}}(\bm{u}_i) - C_n(\bm{u}_i)\}^2,
\label{eq:snopac}
\end{equation}
where $c_{\psi_{(a,\theta,\beta)}}$ is the density of $C_{\psi_{(a,\theta,\beta)}}$ and $C_n (\bm{u}) = \frac{1}{n} \sum_{i=1}^n{\mathbbm{1}_{\{\bm{u}_i \leq \bm{u}\}}}$ is the empirical copula of a sample of (pseudo-)observations $\bm{u}_i = (u_{i1}, ..., u_{id}) \in [0, 1]^d, ~ i \in \{1, \ldots, n\}$ and $n \in \mathbb{N}$ for all $\bm{u} = (u_{1}, ..., u_{d}) \in [0, 1]^d$. 

\begin{figure}[tb!]
	\centering
	\begin{subfigure}{1\textwidth}
		\includegraphics[width=1\textwidth]{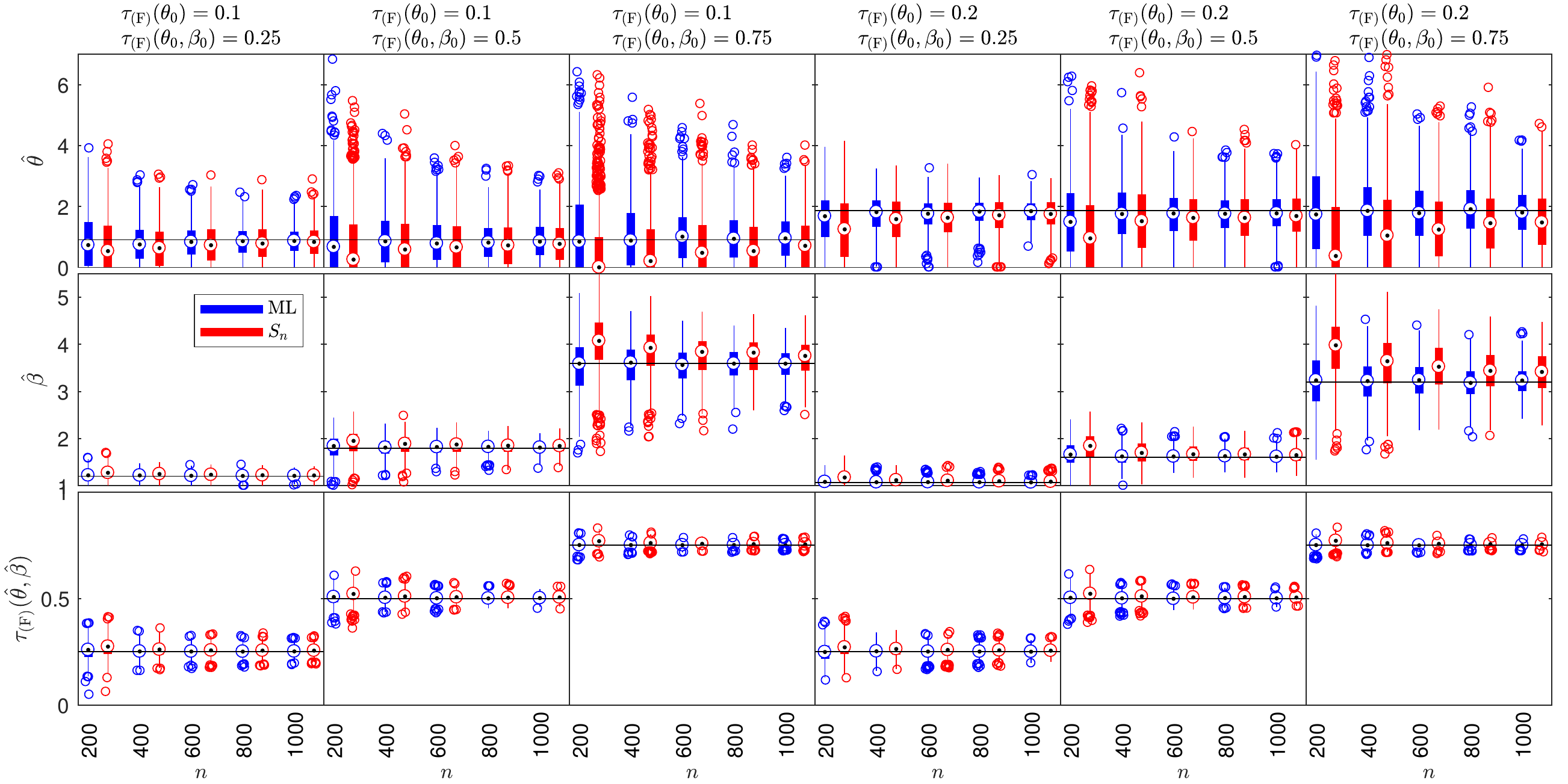}
		\caption{Frank}
	\end{subfigure}
	\vspace*{2mm}
	
	\begin{subfigure}{1\textwidth}
		\includegraphics[width=1\textwidth]{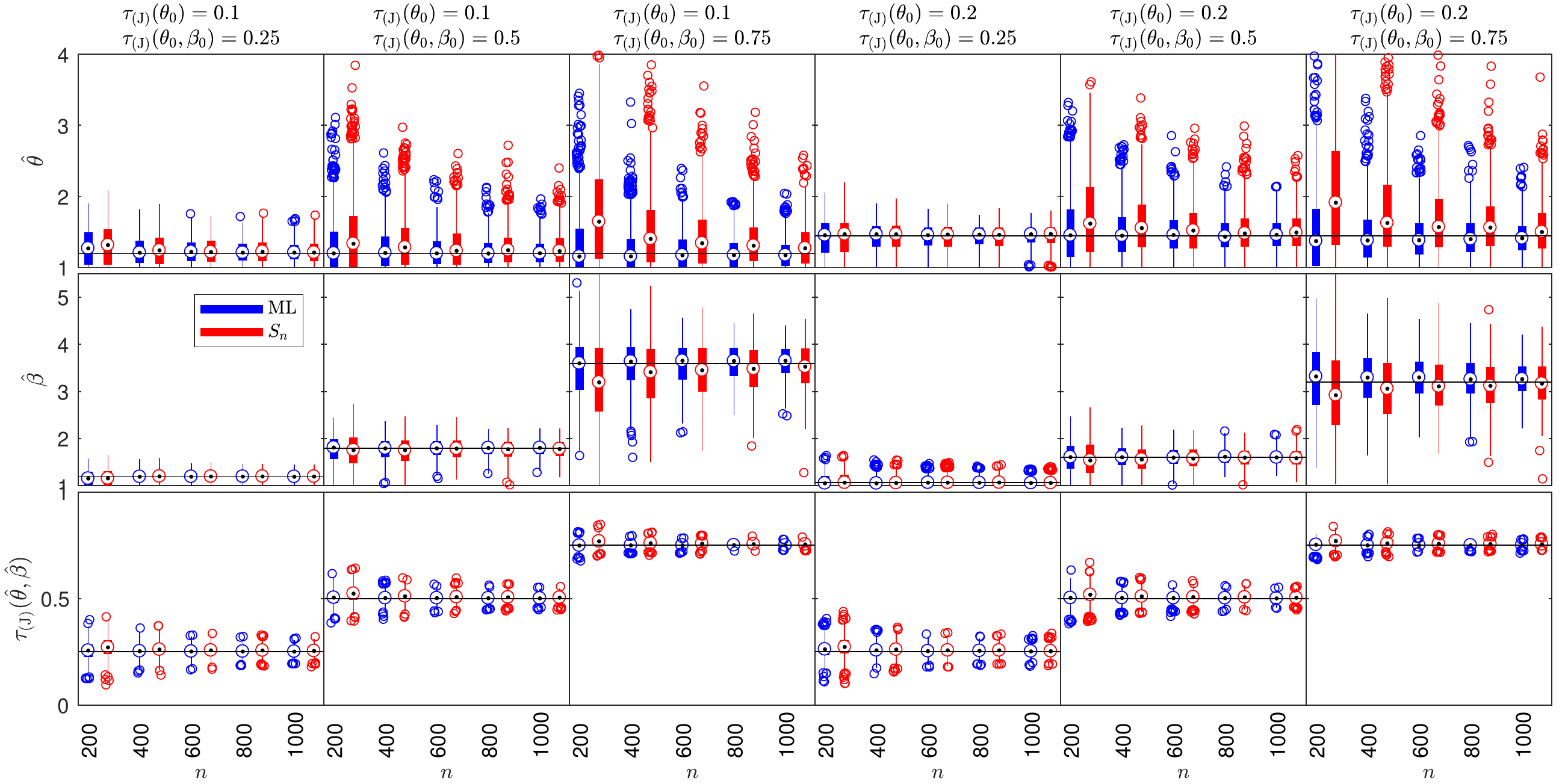}
		\caption{Joe}
	\end{subfigure}
	\caption{Results of the simulation study for the estimators {\color{myBlue}ML} and {\color{myRed}$S_n$} of the Archimedean families A and C. The black line in each plot shows $\theta_0$, $\beta_0$ or $\tau_{a}(\theta_0, \beta_0)$, respectively.}
	\label{fig:mlegof2}
	% MATLAB OPAC_mle_gofplot3.m
\end{figure}

These two estimators are compared in the following simulation study, where all estimates are replicated $N = 1000$ times for sample sizes $n = 200, 400, ..., 1000$, and 6 OPAC models $C_{\psi_{(a,\theta_0,\beta_0)}}$ with
\begin{itemize}
	\item $\theta_0$ chose such that $\tau_{(a)}(\theta_0) \in \{0.1, 0.2\}$, and
	\item given the $\theta_0$ from the previous step, $\beta_0$ is set such that $\tau_{(a)}(\theta_0, \beta_0) \in \{0.25, 0.5,$ $0.75\}$.
\end{itemize}
Based on the results for $a \in$ \{A, C, F, J\} shown in Figures~\ref{fig:mlegof1} and~\ref{fig:mlegof2}, we conclude that: 
\begin{figure}[tb!]
	\centering
	\begin{subfigure}{1\textwidth}
	\includegraphics[width=1\textwidth]{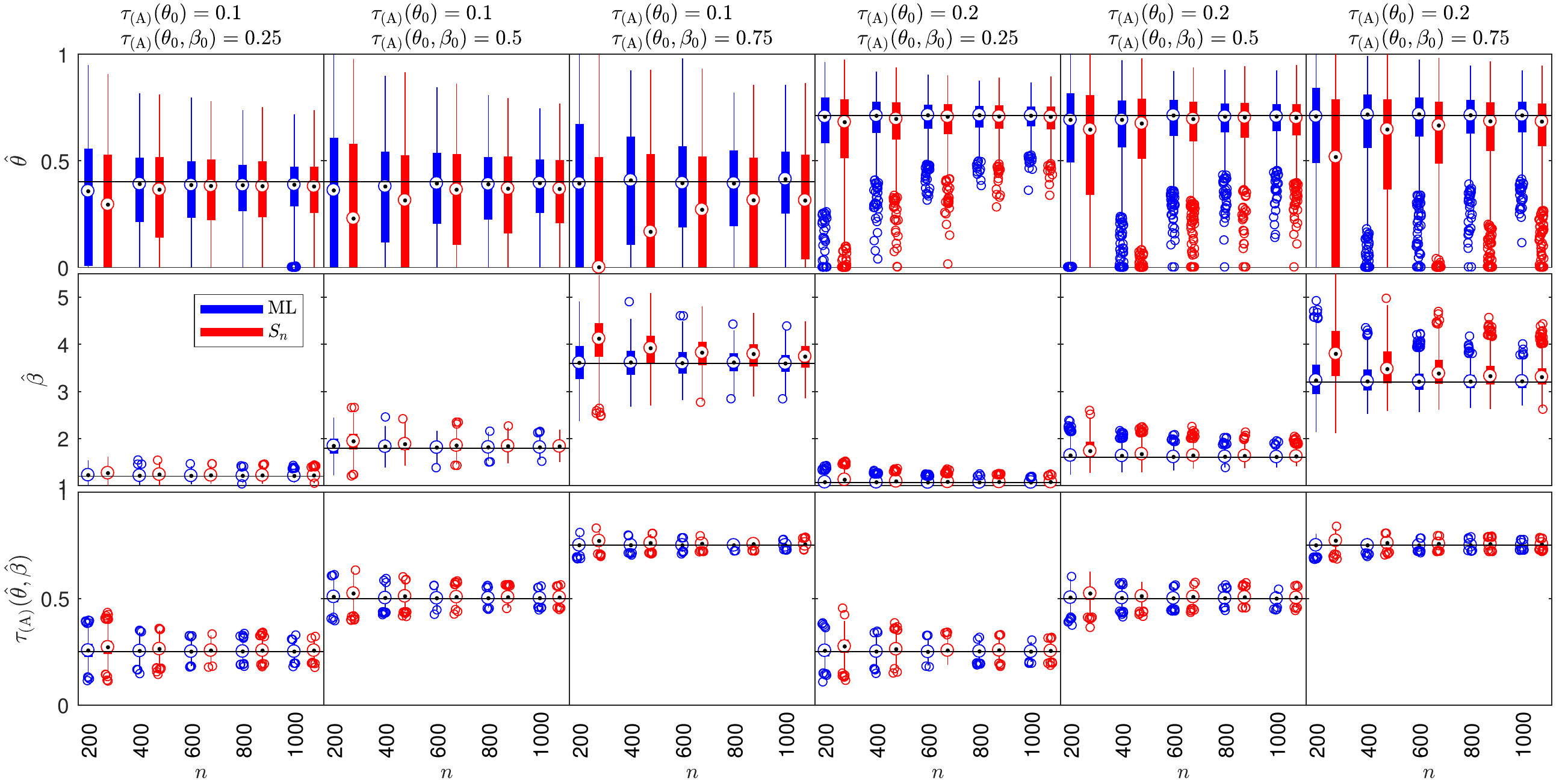}
	\caption{Ali-Mikhail-Haq}
	\end{subfigure}
	\vspace*{2mm}

	\begin{subfigure}{1\textwidth}
	\includegraphics[width=1\textwidth]{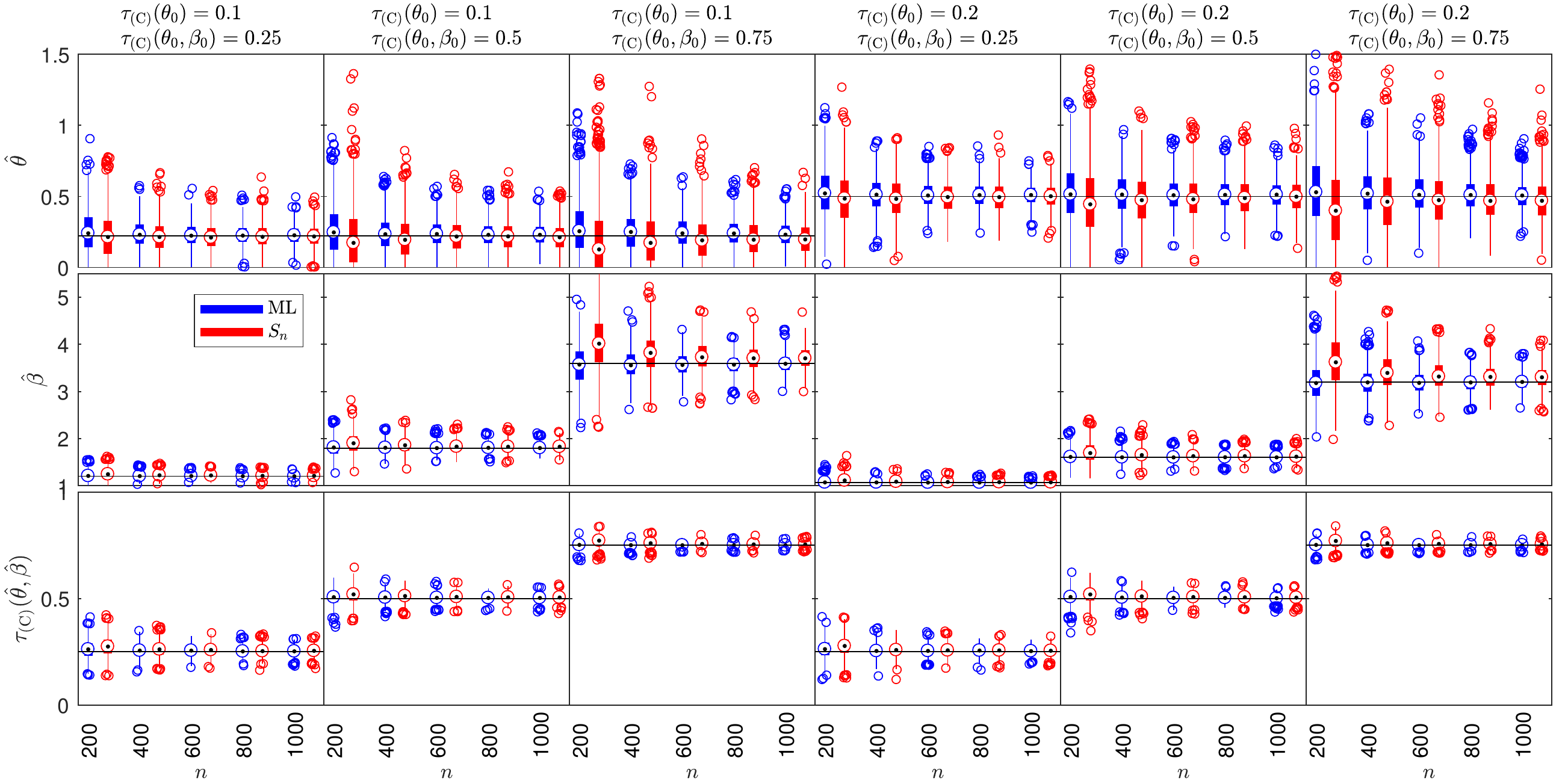}
	\caption{Clayton}
\end{subfigure}
	\caption{Results of the simulation study for the estimators {\color{myBlue}ML} and {\color{myRed}$S_n$} of the Archimedean families A and C. The black line in each plot shows $\theta_0$, $\beta_0$ or $\tau_{a}(\theta_0, \beta_0)$, respectively.}
	\label{fig:mlegof1}
	% MATLAB OPAC_mle_gofplot3.m
\end{figure}
\begin{enumerate}
	\item Both estimators converge to the true values ($\theta_0$ and $\beta_0$) with increasing $n$;
	\item The ML estimator is unbiased and more efficient than $S_n$, which is expected from classical statistical estimation theory;
	\item The standard errors of $\hat{\theta}$ and $\hat{\beta}$ increase with $\tau_{(a)}(\theta_0, \beta_0)$, whereas for $\tau_{(a)}(\hat{\theta}, \hat{\beta})$ they decrease; and
	%\item $\beta$ is monotone with $\tau$.
	\item Conclusions 1.-3.~are independent of the family $a$ considered.
\end{enumerate}
To summarize, both estimators are viable for OPAC estimation. We thus use these estimators also for HOPAC estimation considered in Sections~\ref{sec:estim} and~\ref{sec:sim_study}.

\section{The nested case}
\label{sec:nestedcase}

\subsection{Hierarchical Archimedean copulas}

\begin{figure*}
	\centering
		\begin{subfigure}[t]{0.32\textwidth}
        	\includegraphics[width=1\textwidth]{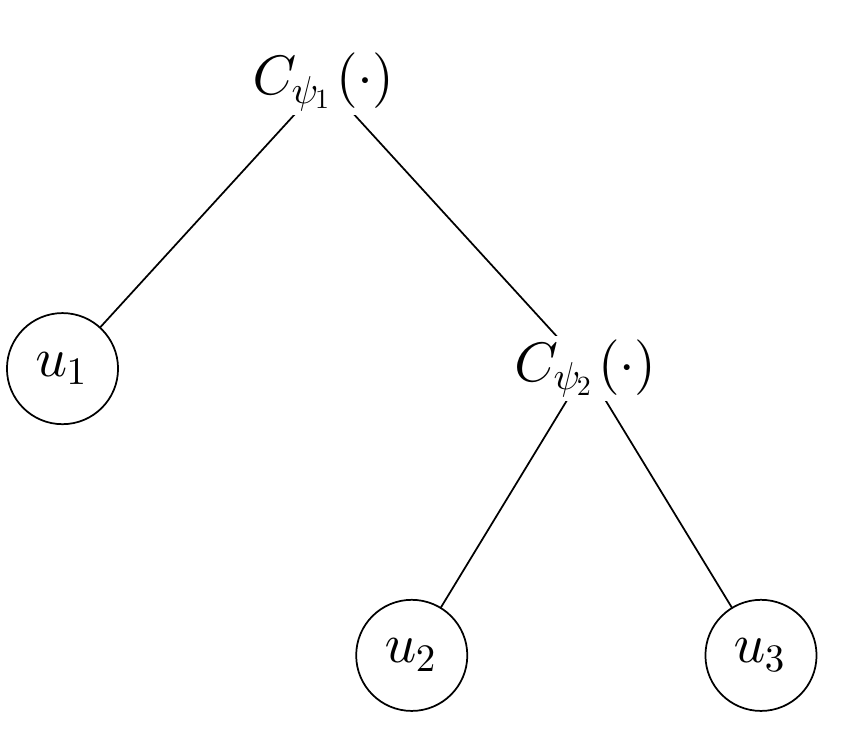}
            \caption{}
            \label{fig:cop3a}
		\end{subfigure}				
		\begin{subfigure}[t]{0.32\textwidth}
			\includegraphics[width=1\textwidth]{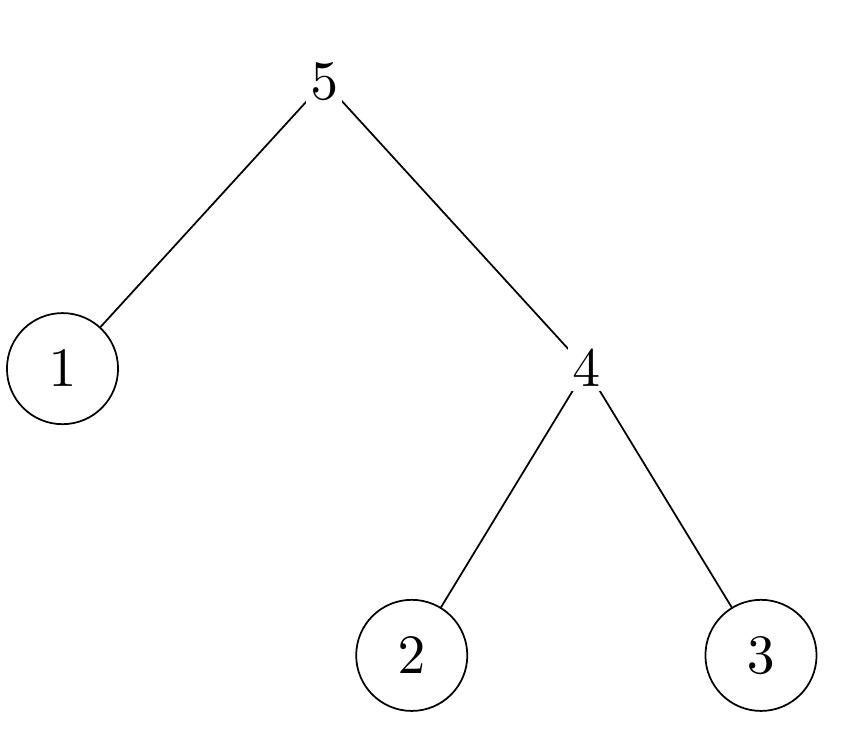}%
			\caption{}
 			\label{fig:cop3b}
		\end{subfigure}	
		\begin{subfigure}[t]{0.32\textwidth}
			\includegraphics[width=1\textwidth]{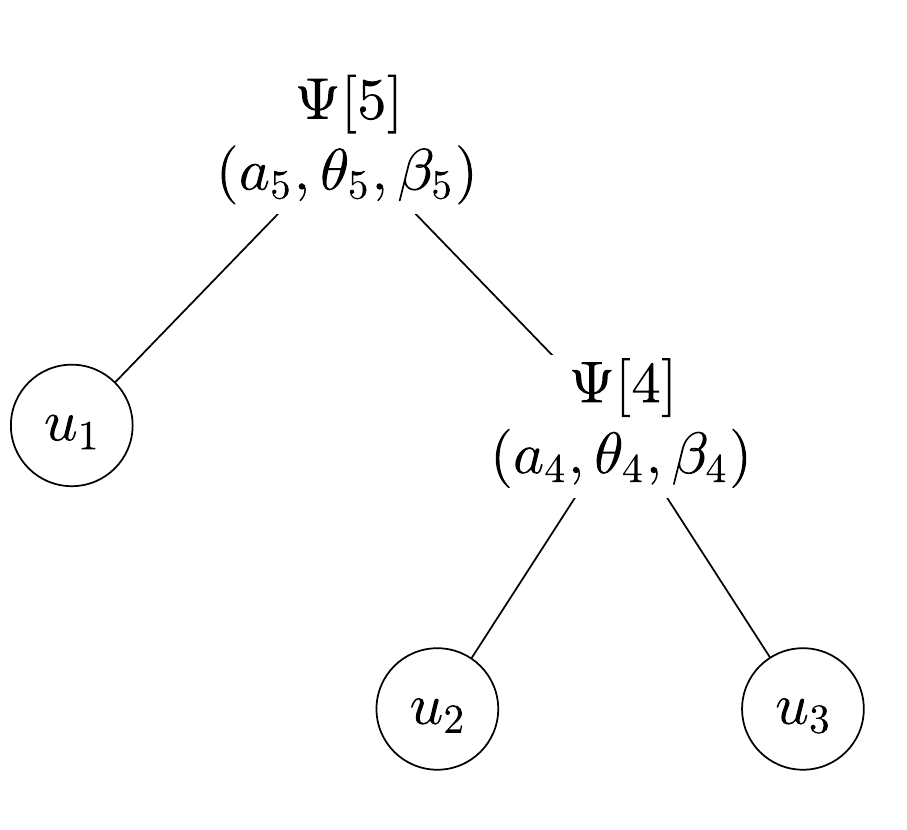}
			\caption{}
			\label{fig:cop3c}
		\end{subfigure}		
	\caption{(a) A tree-like representation of $C_{\psi_1, \psi_2}(u_1, u_2, u_3) = C_{\psi_1}\{u_1, C_{\psi_2}(u_2, u_3)\}$.	
	(b)~An~undirected tree $(\VV,\EE), ~  \VV = \{1, ..., 5\},~  \EE = \{ \{1, 5\}, \{2, 4\}, \{3, 4\}, \{4, 5\} \}$ derived for the tree in Figure~\ref{fig:cop3a}. (c) Our representation of $\HAC(u_1, u_2, u_3) = C_{\Psi[5]}\{u_1, C_{\Psi[4]}(u_2, u_3)\}$, where  $\Psi[4] = \psi_{(a_4, \theta_4, \beta_4)}$, and $\Psi[5] = \psi_{(a_5, \theta_5, \beta_5)}$ and $(\VV, \EE)$ is given by Figure \ref{fig:cop3b}.}
	\label{fig:cop3}
\end{figure*}

To construct a \emph{hierarchical Archimedean copula} (\emph{HAC}), one replaces some arguments of an AC by other (H)ACs, see \citet[pp.~87]{Joe97}. To obtain a proper copula, one also needs to verify certain nesting conditions. 
For example, given two 2-ACs $C_{\psi_1}$ and $C_{\psi_2}$, a 3-variate HAC, denoted $C_{\psi_1, \psi_2}$, can be constructed by 
\begin{equation}
\label{eq:3HAC}
C_{\psi_1, \psi_2}(u_1, u_2, u_3) = C_{\psi_1}\{u_1, C_{\psi_2}(u_2, u_3)\}.
\end{equation}
A tree representation of such a construction is given in Figure~\ref{fig:cop3a}.
Using the language of graph theory, an \emph{undirected tree} $(\VV,\EE)$ related to this representation can be derived by enumerating all of its nodes; here $\VV$ is a set of nodes $\{1, ..., m\}, ~m \in \mathbb{N}$, and $\EE \subset \VV \times \VV$. For example, in Figure~\ref{fig:cop3b}, $\VV = \{1, ..., 5\}$ and $\EE = \{ \{1, 5\}, \{2, 4\}, \{3, 4\}, \{4, 5\} \}$. The \emph{leaves} $\{1, 2, 3\}$ correspond to the HAC variables $u_1, u_2$ and  $u_3$, whereas the non-leaf nodes $\{4, 5\}$, called \emph{forks}, correspond to the ACs (uniquely determined by the corresponding generators) nested in $C_{\psi_1, \psi_2}$. 
When deriving a particular (undirected) tree for the tree representation in Figure~\ref{fig:cop3a}, we assume that
\begin{enumerate}
	\item the leaves 1, 2 and 3 in Figure~\ref{fig:cop3b} correspond to $u_1, u_2$ and $u_3$ in Figure~\ref{fig:cop3a}, respectively, and that
	\item the fork indices 4 and 5 are set arbitrarily (we could have also derives an  undirected tree where the fork indices 4 and 5 are switched). In order to enumerate the forks uniquely, we set them according to the corresponding Kendall's tau, meaning that the root, which has always the lowest Kendall's tau, is numbered by $m$, the node with the second lowest Kendall's tau by $m-1$, etc.  
\end{enumerate}
As each fork corresponds to a generator, we represent this relationship using a labeling denoted $\Psi$, which maps the forks to the corresponding generators. In our example, 
\begin{flalign}
\label{eq:lambda}
\Psi[4] = \psi_2 \text{ and } \Psi[5] = \psi_1. 
\end{flalign}
Using this notation,~\eqref{eq:3HAC} can be rewritten as 
\begin{flalign}
\label{eq:HAC3rewrit}
C_{\Psi[5]}\{u_1, C_{\Psi[4]}(u_2, u_3)\}. 
\end{flalign}
Observe that the indices of the arguments of the inner copula $C_{\Psi[4]}$ correspond to the set of children of fork 4, i.e., to \{2, 3\}, and the indices of the arguments of the copula $C_{\Psi[5]}$ correspond to the set of children of fork 5, i.e., to \{1, 4\} where the number 4 represents the copula $C_{\Psi[4]}(u_2, u_3)$. This implies that 
one can express $C_{\psi_1, \psi_2}(u_1, u_2, u_3)$ in terms of the triplet $(\VV, \EE, \Psi)$. Following this observation, we denote a HAC in an arbitrary dimension by $\HAC$; see also Definition 3.1 in \cite{gorecki2017structure}. 
Finally, let $\Psi[i] = \psi_{(a_i, \theta_i, \beta_i)},~i \in \{4,5\}$ be two OPAC generators. The graphical representation depicted in Figure \ref{fig:cop3c} fully determines the parametric HAC $\HAC = C_{\psi_1, \psi_2}$ given by \eqref{eq:3HAC} and \eqref{eq:lambda}, i.e., its structure, the families of its generators and its parameters.

To guarantee that a proper copula results from nesting ACs, we will use the \emph{sufficient nesting condition} (SNC) proposed by \citet[pp.~87]{Joe97} and \cite{McNeil08}. It states that if for all parent-child pairs of forks $(i,j)$ appearing in a nested construction $\HAC$ the first derivative of $\Psi[i]^{-1}\{\Psi[j](t)\}$ is completely monotone, then $\HAC$ is a copula. 
This SNC has three important practical advantages, which are that 
1) its expression in terms of the corresponding parameters is known, 
2) this expression does not depend on the copula dimension $d$ for all pairs for which it is known, and, most importantly, 
3) efficient sampling strategies based on a stochastic representation for HACs satisfying the SNC are known; see \cite{Hof2012stoch}. 
Note that there also exists a weaker sufficient condition, see \cite{rezapour2015construction}, which however lacks those three advantages and is thus of limited practical use.

\subsection{Constructing and sampling HOPACs}
\label{sec:constandsamp}
Starting with a simple trivariate nested copula structure, e.g., the one  depicted in Figure~\ref{fig:cop3}, we can sample from it according to the algorithm proposed by \cite{McNeil08}. Note that one can easily apply the same strategy also to the general $d$-variate HAC ($d$-HAC) case with $d \geq 3$. 
The sampling algorithm, extending the one of \cite{marshall1988families} for ACs, requires one to know how to sample the two following random variables:
\begin{enumerate}
	\item $V_1 \sim\mathcal{LS}^{-1}[\widetilde{\psi}_1]$ and
	\item $V_{12} \sim \mathcal{LS}^{-1}[\exp(-V_1\widetilde{\psi}_1^{-1} \circ \widetilde{\psi}_2)]$,
\end{enumerate}	
where the tildes emphasize that the generators are OP transformed, say $\widetilde{\psi}_1 = \psi_{(a_1, \theta_1, \beta_1)}$ and $\widetilde{\psi}_2 = \psi_{(a_2, \theta_2, \beta_2)}$, where $a_1$ and $a_2$ are labels of c.m.~families of one-parameter generators, 
$\theta_1 \in \Theta_{a_1}$, $\theta_2 \in \Theta_{a_2}$ and $\beta_1, \beta_2 \geq 1$.

A strategy for sampling from $V_1$ has been recalled in Section~\ref{sec:OPACsampling}. It is important to note that $V_1$ is non-negative and, in fact,
strictly positive with probability 1, as a result of Bernstein's
Theorem \citep{bernstein1929fonctions} and the fact that $\widetilde{\psi}_1$ is a c.m.~generator.

To sample from $V_{12}$, consider that for all $t \in [0, \infty)$, 
\begin{flalign}
\widetilde{\psi}_{12}(t;V_1)  :&=  \exp[-V_1\widetilde{\psi}_1^{-1} \{\widetilde{\psi}_2(t)\}]  \nonumber\\
 & = \exp[-V_1\psi_{(a_1, \theta_1, \beta_1)}^{-1} \{\psi_{(a_2, \theta_2, \beta_2)}(t)\}]  \nonumber\\ 
 &=\exp[-V_1(\psi_{(a_1, \theta_1)}^{-1} \{\psi_{(a_2, \theta_2)}(t^{\frac{1}{\beta_2}})\})^{\beta_1}]   \label{eq:psi12basic}\\ 
 &= \exp\big[-V_1\big\{-\log\big(\exp[-\psi_{(a_1, \theta_1)}^{-1} \{\psi_{(a_2, \theta_2)}(t^{\frac{1}{\beta_2}})\}]^{\beta_1}\big)\big\}\big] \nonumber\\
 &= \bar{\psi}_1[-\log\{\bar{\psi}_{12}(t^{\frac{1}{\beta_2}})\} ;V_1],
 \end{flalign}
where $\bar{\psi}_1(t;V_1) := \exp(-V_1t^{\beta_1})$ 
and
$\bar{\psi}_{12}(t) := \exp[-\psi_{(a_1, \theta_1)}^{-1} \{\psi_{(a_2, \theta_2)}(t)\}]$.
%Considering $\bar{\psi}_1$, it is a valid Gumbel generator only for $\beta_1 = 1$, as for $\beta_1 > 1$ its inverse Laplace-Stieltjes transform  $\text{S}(\beta_1, 1, (\cos(\beta_1\frac{\pi}{2})V_1)^{\frac{1}{\beta_1}},$ $V_1 \mathbbm{1}_{\{\beta_1 = 1\}}; 1)$ does not generate real values.
Note that $V_1$ acts as a parameter for $\widetilde{\psi}_{12}$. The following proposition provides an explicit way to sample from $V_{12}$.

\begin{proposition}
Let $\beta_1 = 1$ and $[\psi_{(a_1, \theta_1)}^{-1} \{\psi_{(a_2, 
		\theta_2)}\}]'$ be c.m. Then $\widetilde{\psi}_{12}(t;V_1)$ is c.m.~for all $t \in [0, \infty)$, $V_1 \in (0, \infty)$, and 
\begin{flalign*}
V_{12} :=  S\breve{V}_{12}^{\beta_2}  \sim \LSi[\widetilde{\psi}_{12}], 
\end{flalign*}	
where $S \sim \text{S}(1/\beta_2, 1, \cos^{\beta_2}(\frac{\pi}{2\beta_2}), \mathbbm{1}_{\{\beta_2 = 1\}}; 1)$ and $\breve{V}_{12} \sim \LSi[(\bar{\psi}_{12})^{V_1}]$.
\label{thm:tildepsi12iscm}
\end{proposition}
\begin{proof}
$\beta_1 = 1$ implies that $\bar{\psi}_1$ is c.m. 
As $-\log\{\bar{\psi}_{12}(t)\}$ $=$ $\psi_{(a_1, \theta_1)}^{-1} \{\psi_{(a_2, 
	\theta_2)}(t)\}$, the assumptions and Proposition 2.1.5 (5) from \cite{Hof10book} imply that  $(\bar{\psi}_{12})^{V_1}$ is c.m.~for all $V_1 > 0$ .
For the special case $V_1 = 1$, Theorem \ref{thm:samplingOPAC} with $\beta = \beta_2$ further implies that also $\bar{\psi}_{12}(t^{\frac{1}{\beta_2}})$ is c.m. With Proposition 2.1.5 (5) from \cite{Hof10book}, the first derivative of  $\psi_{(a_1, \theta_1)}^{-1} \{\psi_{(a_2, \theta_2)}(t^{\frac{1}{\beta_2}})\}$ is also c.m.
Finally, as  $\bar{\psi}_1$ and $[-\log\{\bar{\psi}_{12}(t^{\frac{1}{\beta_2}})\}]'$ are c.m., Proposition 2.1.5 (2) from \cite{Hof10book} implies that also $\widetilde{\psi}_{12}$ is c.m.

Further, starting from \eqref{eq:psi12basic} and with $\beta_1 = 1$, $\widetilde{\psi}_{12}(t;V_1)$ can be rewritten as
\begin{flalign}
\widetilde{\psi}_{12}(t;V_1) =\bar{\psi}_{12}(t^{\frac{1}{\beta_2}})^{V_1}.
\label{eq:psi12dist}
\end{flalign}
Denoting by $\breve{V}_{12}$ a random variable distributed according to  $\LSi[(\bar{\psi}_{12})^{V_1}]$ and applying Theorem \ref{thm:samplingOPAC} with $\psi(t) = \bar{\psi}_{12}(t)^{V_1}$ and $\beta = \beta_2$ based on \eqref{eq:psi12dist} establishes the proof.
\qed
\end{proof}	

Proposition \ref{thm:tildepsi12iscm} implies that:
\begin{enumerate}
	\item \emph{Any} OP family ($\beta_2 \geq 1$) can be nested into a non-OP family ($\beta_1 = 1$) just under the \emph{one-parameter} SNC, i.e., if  $[\psi_{(a_1, \theta_1)}^{-1} \{\psi_{(a_2, \theta_2)}\}]'$ is c.m., which simplifies to $\theta_1 \leq \theta_2$ for many families when $a_1 = a_2$; see the second column of  Table 2.3 in \cite{Hof10book}. To the best of our knowledge, this nesting case has never been considered in the literature. 
	\item Know-how of sampling $\mathcal{LS}^{-1}[\exp(-V_1\widetilde{\psi}_1^{-1} \circ \widetilde{\psi}_2)]$, under $\beta_1 = 1$, can be fully translated to know-how of sampling its non-OP transformed version, i.e., to $\LSi[(\bar{\psi}_{12})^{V_1}]$, which is known for many families; see the third column of Table 2.3 in \cite{Hof10book}. Also note that free implementations of sampling algorithms for $\LSi[(\bar{\psi}_{12})^{V_1}]$ are available, e.g., in the \textsf{R} package \textbf{copula} \citep{hofert2017copula} or in the \textbf{HACopula} toolbox for \textsf{MATLAB} and \textsf{Octave} \citep{gorecki2017hacopula}.
\end{enumerate}

For $\beta_1 > 1$, an OP family cannot be nested into a proper OP ($\beta_2 > 1$) family in the way mentioned above except for a special case already considered in \cite{Hof11}. Namely if $a_1 = a_2$ and $\theta_1 = \theta_2$, then 
\begin{flalign*}
\widetilde{\psi}_{12}(t;V_1)  &=\exp\big(-V_1[\psi_{(a_1, \theta_1)}^{-1} \{\psi_{(a_1, \theta_1)}(t^{\frac{1}{\beta_2}})\}]^{\beta_1}\big) = 
\exp(-V_1t^{\frac{\beta_1}{\beta_2}}),
\end{flalign*}
which is a proper Gumbel generator provided that $\beta_1 \leq \beta_2$, with  inverse Laplace-Stieltjes transform $\text{S}(\beta_1/\beta_2, 1, \{\cos(\frac{\beta_1}{\beta_2}\frac{\pi}{2})V_1\}^{\frac{\beta_2}{\beta_1}},$ $V_1 \mathbbm{1}_{\{\beta_1/\beta_2 = 1\}}; 1)$.

Hence, to construct a proper HOPAC of the form $C_{\widetilde{\psi}_1}\{u_1, C_{\widetilde{\psi}_2}(u_2, u_3)\}$ under the SNC, it must hold that either 
\begin{enumerate}
		\item $[\psi_{(a_1, \theta_1)}^{-1} \{\psi_{(a_2, \theta_2)}\}]'$ is c.m, if $\beta_1 = 1$,  or
		\item $\beta_1 \leq \beta_2$,  if $a_1 = a_2$ and $\theta_1 = \theta_2$.
\end{enumerate}
As mentioned above, these constraints can easily be translated to the general nesting case just by checking these constraints to hold for every parent-child pair of nodes in the nested copula structure.

\begin{figure}
	\centering
		\includegraphics[width=0.55\textwidth]{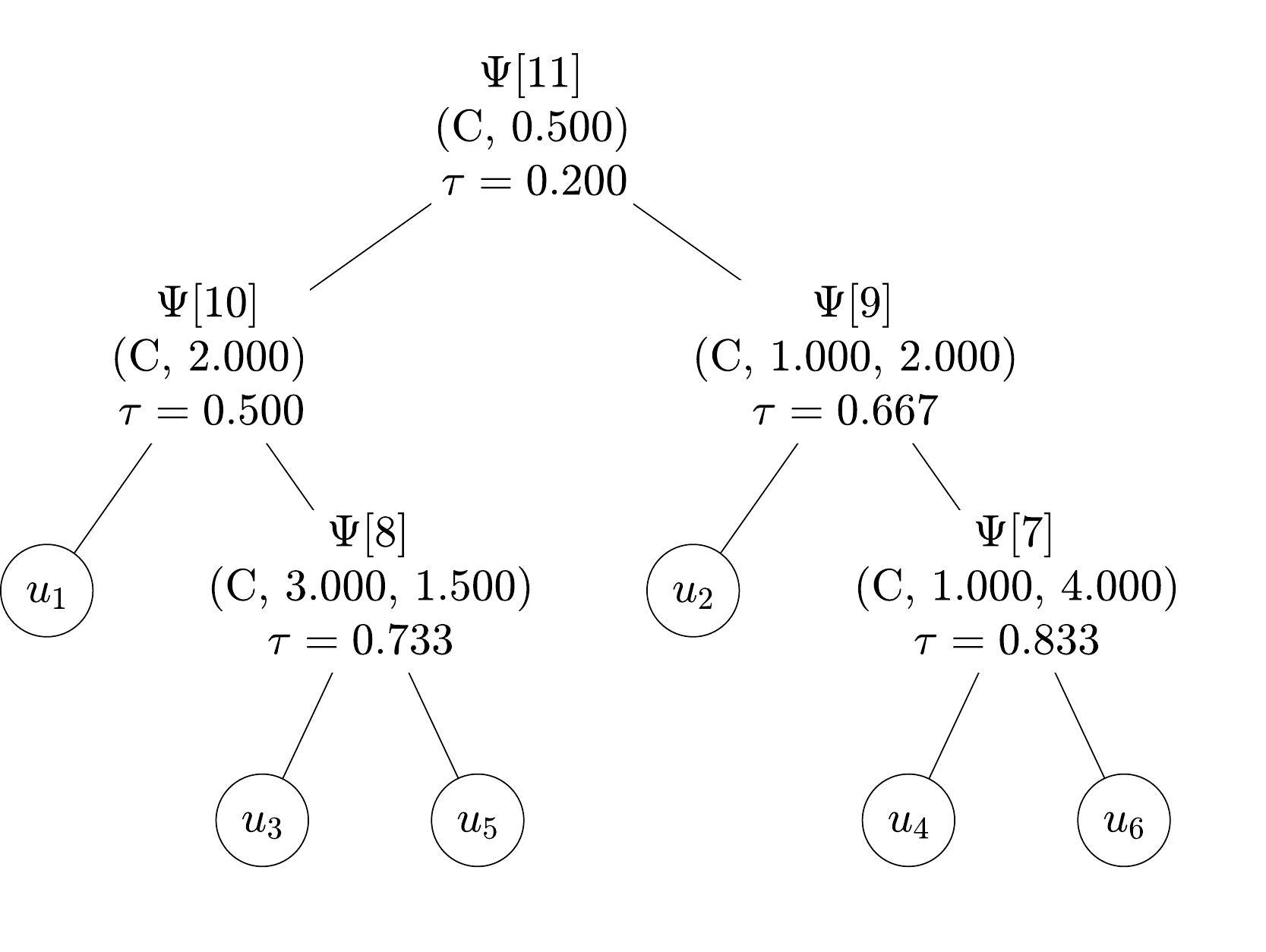}
		\includegraphics[width=0.44\textwidth]{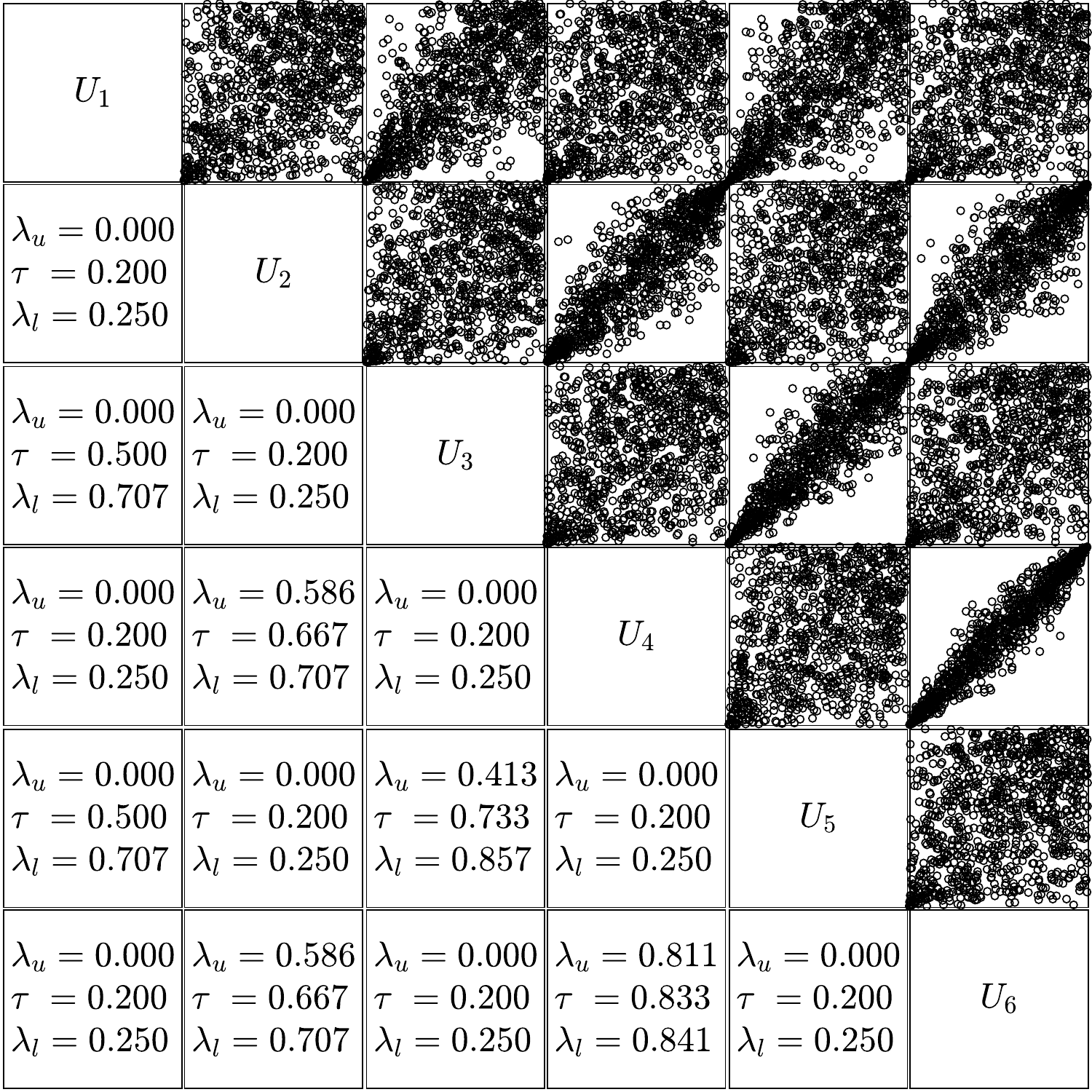}
	\caption{(left) A 6-variate Clayton HOPAC satisfying the SNC. If $\beta = 1$ then the value of $\beta$ is omitted (forks 10 and 11). (right) A pairwise plot of 1000 observations from this HOPAC, including three pairwise dependence measures $\lambda_u$, $\tau$ and $\lambda_l$ computed from the HOPAC.}
	\label{fig:opExsModels}
	% MATLAB outerpowerSTCO.m
\end{figure}

Let us now consider the case when $a_1 = a_2$ and the condition on $[\psi_{(a_1, \theta_1)}^{-1}\allowbreak\{\psi_{(a_2, \theta_2)}\}]'$ to be c.m.~is simplified to $\theta_1 \leq \theta_2$. Construction of HOPACs under the SNC is then similar to the one-parameter-generator HACs, where the parameters $\theta_i$ involved have to be increased as one goes further down in a branch of the copula structure. In the HOPAC structure, the parameters also have to increase but one can choose which one of them. Let us illustrate this with the model depicted on the left-hand side of Figure~\ref{fig:opExsModels}.
One can observe that $\theta$ or $\beta$ increases as one goes down in a branch of the copula structure. More precisely, if $\beta_{parent} = 1$, then $\theta_{child} \geq \theta_{parent}$ and of course $\beta_{child} \geq \beta_{parent}$, see, e.g., the parent-child pair of forks $(10,8)$ or $(11,9)$. This new flexible case is enabled by Proposition \ref{thm:tildepsi12iscm}. Or, if $\beta_{parent} > 1$ then $\theta_{child}$ is fixed to $\theta_{parent}$'s value and only $\beta$ increases as one goes down in a branch, as the pair of forks $(9,7)$ indicates. This case has already been used in \cite{Hof11CDO}. For completeness, let us mention that the pair of forks $(11,10)$ represents nesting of two one-parameter ACs.
In the following section, we develop HOPAC estimators under the SNC for HOPACs mentioned above in which the condition on $[\psi_{(a_1, \theta_1)}^{-1}\allowbreak\{\psi_{(a_2, \theta_2)}\}]'$ to be c.m.~can be simplified to $\theta_1 \leq \theta_2$ provided $a_1 = a_2$.

\subsection{Estimating HOPACs}
\label{sec:estim}
The literature provides a set of diverse HAC estimation methods; see  \cite{Okh13} or \cite{goreckihofertholena2016approachjiis,gorecki2017structure} for those concerning estimation of both structure and parameters.
However, as already mentioned in Section~\ref{sec:introduction}, all of them restrict to HACs involving just one-parameter ACs. On the one hand, these methods cannot be directly used for HOPAC estimation, on the other hand, they  serve as a natural starting point for development of such estimators.

In general, three ingredients are necessary to get a HAC estimated, 1)~its structure, 2)~the AC families and 3)~their parameters. 1) typically encompasses a kind of agglomerative clustering, where the structure finally results from clustering of variables under concern according to some measure of similarity between pairs of the variables, e.g., according to Kendall's tau. 
As the structure estimator in \cite{goreckihofertholena2016approachjiis,gorecki2017structure} does not, by contrast to \cite{Okh13}, require any assumptions on the family of the nested ACs, it is thus immediately feasible also for HOPAC estimation. Moreover, there exist theoretical justifications for such an estimator -- given a HAC, \cite{Okh13prop} show that its structure can be uniquely recovered from all its bivariate margins, and Theorem~2 in \cite{gorecki2017kendalls} show that it is possible just from all its pairwise Kendall's coefficients. 
Finally, as this estimator, formalized by Algorithm 1 in \cite{gorecki2017structure} (see also our Appendix), showed the best results in the ratio of successfully estimated true HAC structures on the basis of simulation studies, see \cite{goreckihofertholena2016approachjiis} or  \cite{uyttendaele2016estimation}, we adopt it to our HOPAC estimation approach.
 
To estimate the families, the mentioned references provide two different approaches. The one used in \cite{Okh13}  and \cite{goreckihofertholena2016approachjiis} arbitrarily assumes the same family for all nested ACs, whereas then one in \cite{gorecki2017structure} involves an extra model selection step, which chooses for each AC the best fitting family out of some predetermined pool of families, and thus allows the families in the estimated HAC to differ. 
%Although we shall be very careful with the goodness-of-fit procedure as the model selection method, see \cite{Gneiting07}. 
Whereas the first approach substantially simplifies the parametric constraints following from the SNC (to the condition $\theta_{parent} \leq \theta_{child}$ for most of the cases),	the latter requires an extensive analysis of the input family pool as not all families can be mixed together or since the parameter ranges of the families in the pool have to be adjusted before the estimation process. Hence, when the OP transformation comes to play, which makes the estimation process substantially more complex even under the assumption of the same family for all nested ACs, the approach allowing for different nested families becomes pretty challenging. 

In this work, we focus on the case of having the same family for all ACs nested in a HOPAC. Note that this case still covers estimation of HACs with one-parameter ACs from different families, for example, the families of Clayton, 12 and 14, as stated in Section \ref{sec:introduction}, where the latter two family numbers correspond to the notation used in \citet[pp. 116--119]{Nel06}.

\begin{figure}
	\centering
	\begin{subfigure}[t]{0.45\textwidth}
		\includegraphics[width=1\textwidth]{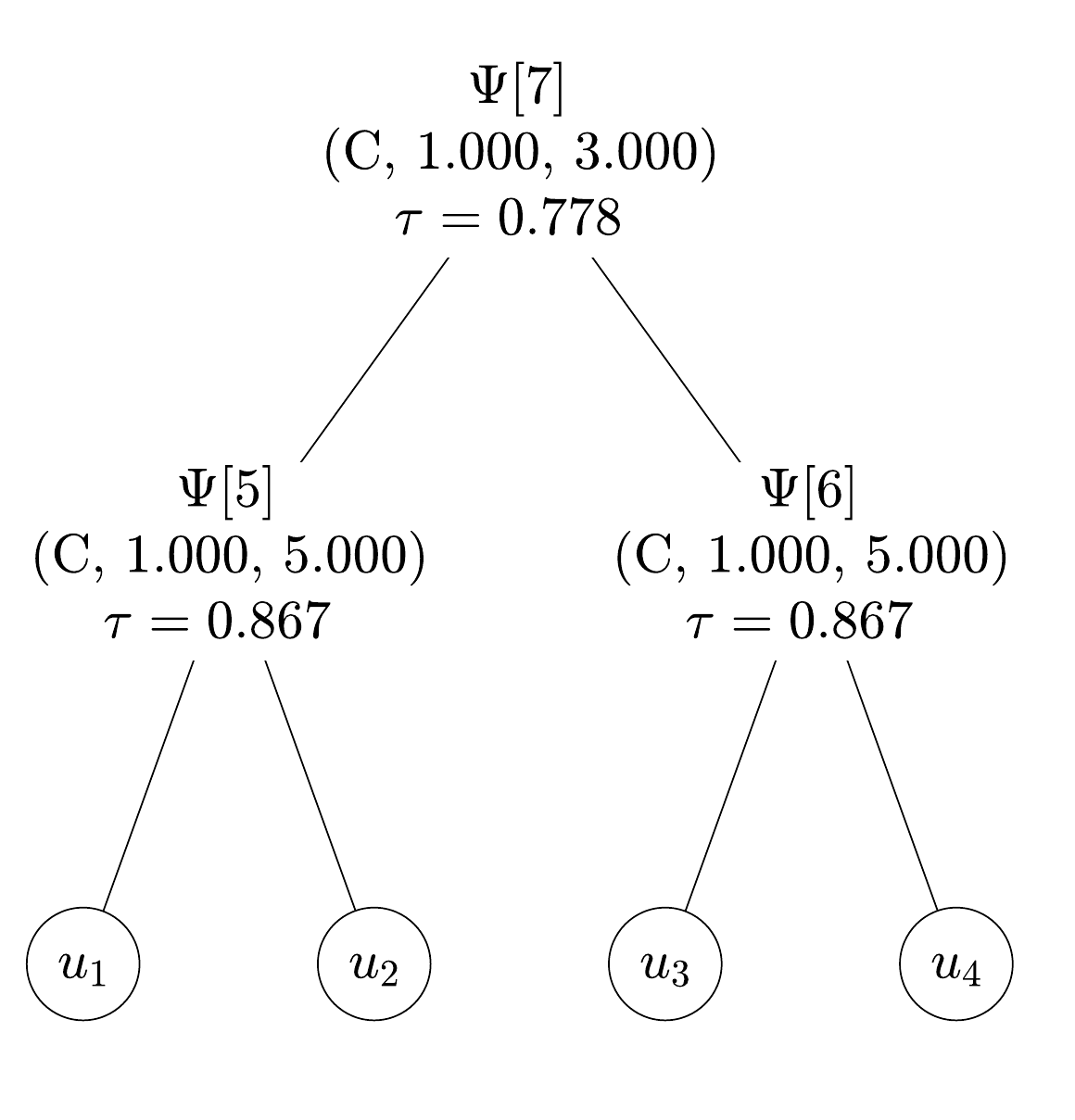}
		\caption{A HOPAC model built under the parametric SNC in the form  $\beta_{parent} \leq \beta_{child}$, if $\theta_{parent} = \theta_{child}$.}
		\label{fig:BUmodel}
	\end{subfigure}
	\begin{subfigure}[t]{0.45\textwidth}
		\includegraphics[width=1\textwidth]{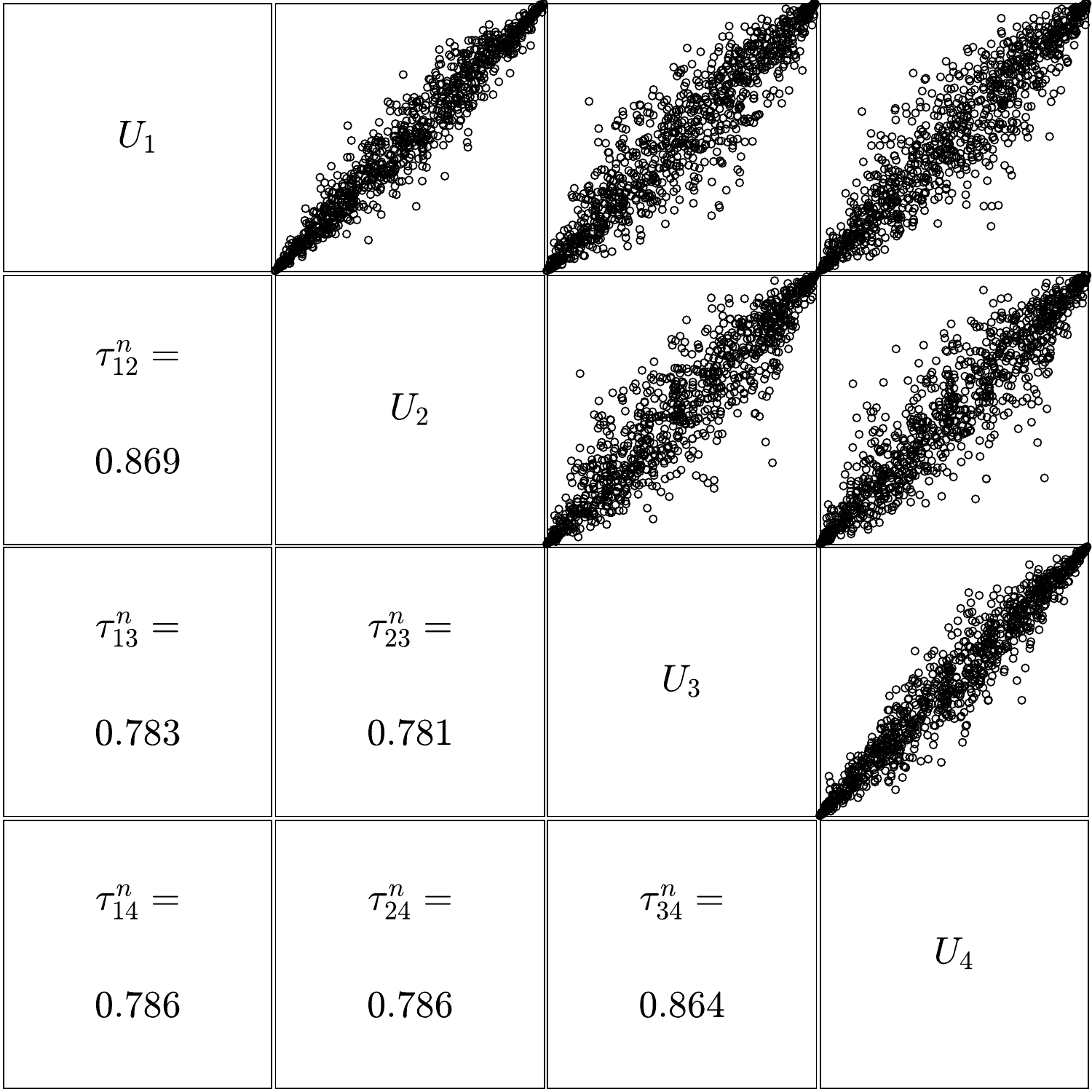}
		\caption{A sample of $n = 1000$ observations from the HOPAC at Figure~\ref{fig:BUmodel}. Note that $\tau_{ij}^n$ denotes the pairwise Kendall's tau for observations of $(U_i,U_j)$.}
		\label{fig:BUdata}
	\end{subfigure}
	\caption{An example for the Bottom-Up estimator.}
	\label{fig:BUmodeldata}
	% MATLAB: outerpowerSTCO
\end{figure}

For estimation of parameters, we use a mixture of existing step-wise procedures. This follows from the existence of a close relationship between bivariate margins of a HAC and the location of ACs nested in this HAC. 
To clarify, according to Proposition~3 in \cite{goreckihofertholena2016approachjiis}, given two random variables $U_i$ and $U_j$ from the vector $(U_1, ..., U_d)$ distributed according to a $d$-variate HAC, the copula of $(U_i, U_j)$ is the AC allocated in the HAC structure at the node where the two branches - 1) the one from the leaf $u_i$ to the root, and 2) the one from the leaf $u_j$ to the root -- meet for the first time. This AC (node) is called \emph{youngest common ancestor} of $u_i$ and $u_j$. In Figure~\ref{fig:BUmodel}, the youngest common ancestor of $u_1$ and $u_2$ is node 5 (AC $C_{\Psi[5]}$), whereas for $u_1$ and $u_3$ it is node 7 (AC $C_{\Psi[7]}$).
It follows that the parameters of this copula can be estimated from the observations of $U_i$ and $U_j$ only. To this end, we use the AC maximum-likelihood (ML) estimator as in \cite{Okh13}.
As there are often more of such pairs with the same youngest common ancestor, we use this fact in the way that we estimate the parameter(s) of the youngest common ancestor from the observations of \emph{all} pairs of random variables corresponding to this ancestor, and then aggregate these estimates by using the mean; the mean performed best in the simulation study involving the mean, maximum and minimum aggregation statistics implemented in Step 2 of Algorithm 3 in \cite{goreckihofertholena2016approachjiis}. 

The concept of our approach to the HOPAC estimation is summarized in Algorithm~\ref{alg:est_HOPAC_summary}, which requires two inputs: 1) a one-parameter Archimedean family $a$, e.g., from Table~\ref{tab:geners}, and 2) realizations $\bm{u} = (u_{ij}) \in [0, 1]^{n\times d}$ of \emph{pseudo-observations} $(U_{ij})_{i \in \{1, ..., n\}}^{j \in \{1, ..., d\}}$  given by 
\begin{flalign}
U_{ij}= \frac{n}{n+1}\hat{F}_{n, j}(X_{ij}) = \frac{R_{ij}}{n+1},
\label{eq:pseudo_observations}
\end{flalign}
\noindent where $\hat{F}_{n, j}$ denotes the \emph{empirical distribution function} corresponding to the $j$th margin, 
$R_{ij}$ denotes the \emph{rank} of $X_{ij}$ among $X_{1j}, ..., X_{nj}$,
and 
$(X_{i1}, ..., X_{id}),$ $i \in \{1, . . . , n\}$ are i.i.d.~random vectors distributed according to a joint distribution function
with continuous margins $F_j, ~j \in \{1, . . . , d\}$, and a HOPAC $C$. The algorithm returns a HOPAC from family $a$ with estimated structure and parameters.

\begin{algorithm}[t!]
	\floatname{algorithm}{Algorithm}
	\caption{The HOPAC estimation concept}
	\label{alg:est_HOPAC_summary}
	\begin{algorithmic}[1]
%		\renewcommand{\algorithmicensure}{\textbf{The estimation:}}
%		\ENSURE
		\STATE compute the matrix of pairwise Kendall's tau $(\tau_{ij})$ from $\bm{u}$
		\STATE estimate the structure using Algorithm~\ref{alg:structure_estim} with $(\tau_{ij})$
		\FOR{each fork $k$ in the structure}
			\FOR{each pair of leaves $(i_k, j_k)$ in the structure such that $k$ is the youngest common ancestor of $i_k$ and $j_k$}
			\STATE use ML to estimate OPAC's $\theta$ and $\beta$ for $(u_{m,i_k}, u_{m,j_k}), ~ m = 1, ..., n$
			\ENDFOR
			\STATE aggregate these estimated $\theta$s and $\beta$s by computing their means $\hat{\theta}_k$ and $\hat{\beta}_k$ 
			\STATE let $C_{\psi_{(a, \hat{\theta}_k, \hat{\beta}_k)}}$ be the OPAC estimate corresponding to the fork $k$ in the estimated HOPAC structure
		\ENDFOR 
	\end{algorithmic}
\end{algorithm} 

Now, let us discuss several properties of the concept using the following example.
Let $C$ be the Clayton HOPAC from Figure~\ref{fig:BUmodel}, with $d = 4$, and let  $(\bar{u}_{ij})_{i \in \{1, .., n\}}^{j \in \{1, ..., 4\}}$ be a sample from it shown in Figure~\ref{fig:BUdata}. Now assume $C$ to be unknown and let us estimate it based on the pseudo-observations $(u_{ij})_{i \in \{1, .., n\}}^{j \in \{1, ..., 4\}}$ of $(\bar{u}_{ij})_{i \in \{1, .., n\}}^{j \in \{1, ..., 4\}}$.

\begin{figure}[t!]
	\centering
	\begin{subfigure}[t]{0.49\textwidth}
		\includegraphics[width=1\textwidth]{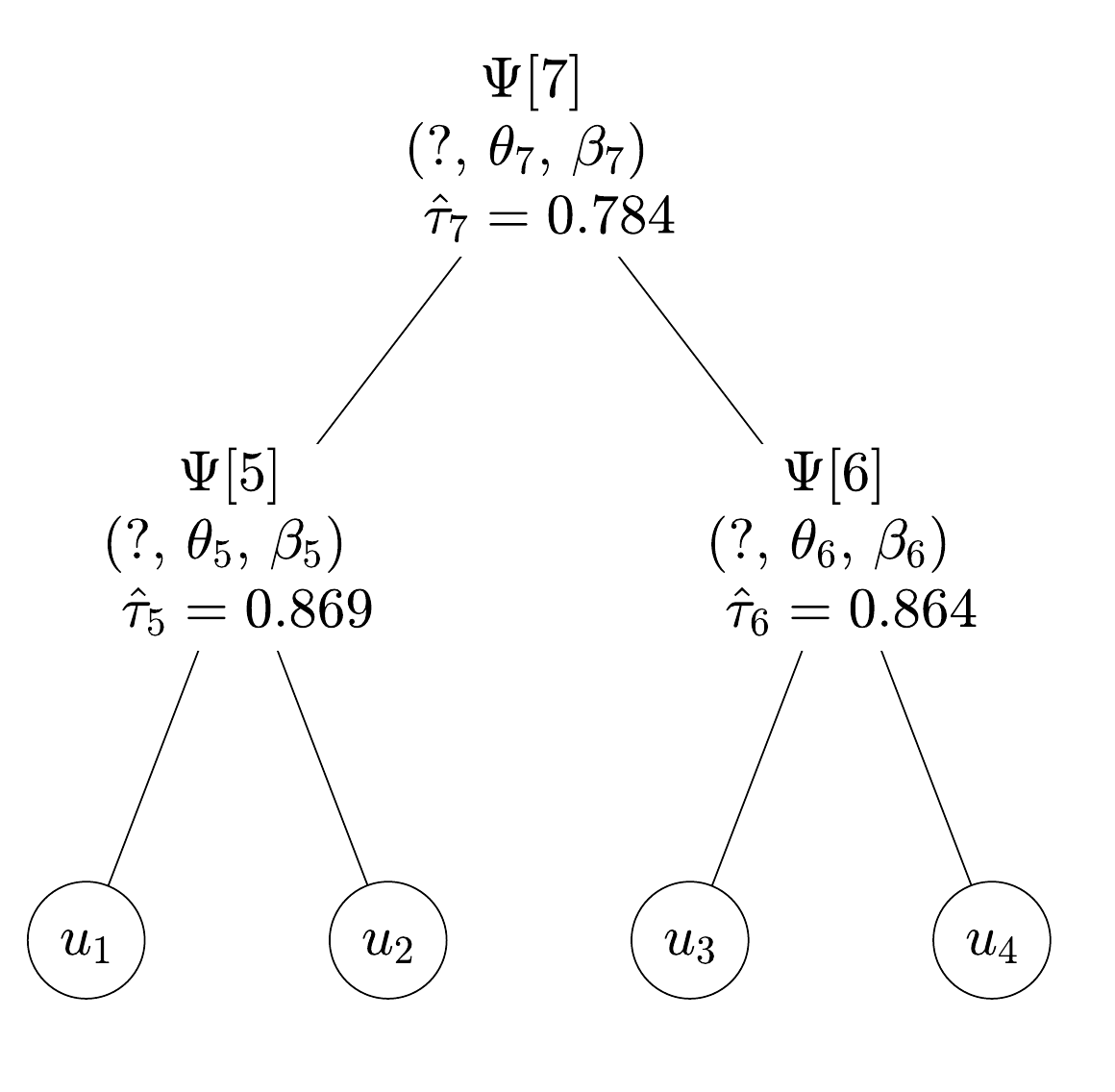}
		\caption{The binary tree $(\hat{\VV}, \hat{\EE})$ and estimates of Kendall's tau $(\hat{\tau}_5, \hat{\tau}_6, \hat{\tau}_7)$ for the sample from Figure~\ref{fig:BUdata}.}
		\label{fig:BUstruc}
	\end{subfigure}
	\begin{subfigure}[t]{0.49\textwidth}
		\includegraphics[width=1\textwidth]{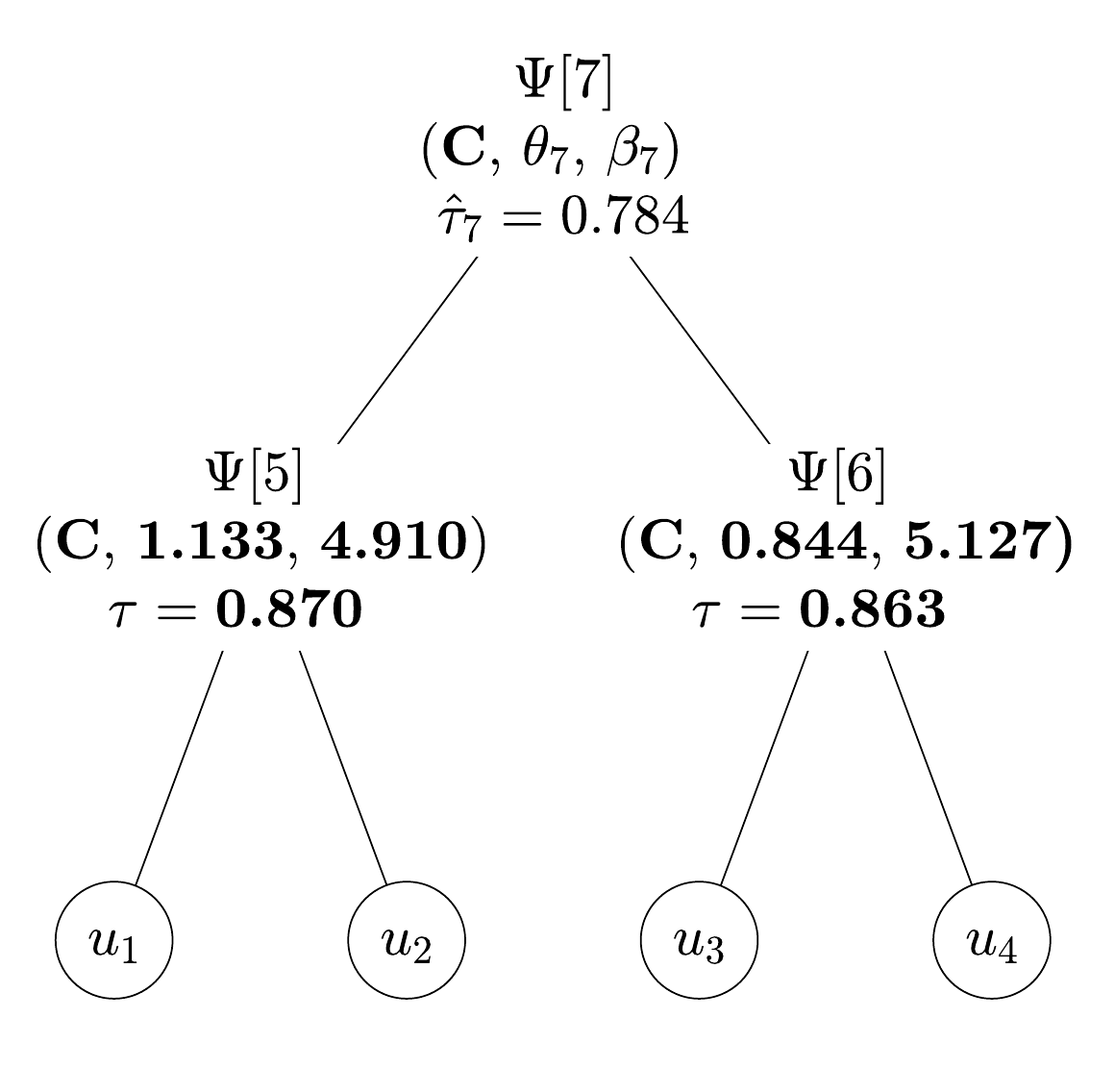}
		\caption{An update of the estimate from Figure~\ref{fig:BUstruc} after assuming the Clayton family and estimating the bottom-level using ML estimation.}
		\label{fig:BUest1}
	\end{subfigure}
	\begin{subfigure}[t]{0.49\textwidth}
		\includegraphics[width=1\textwidth]{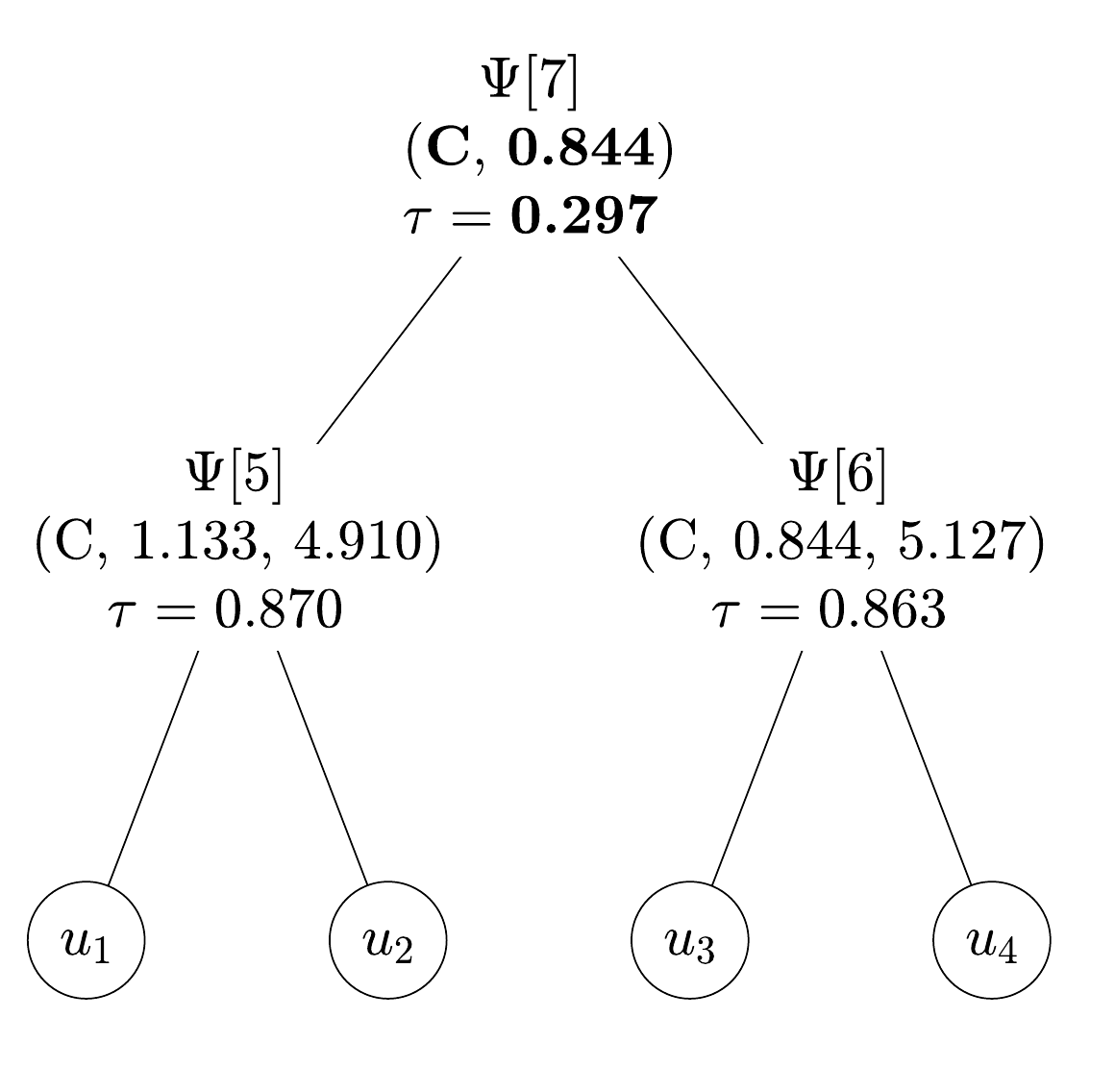}
		\caption{A final update of the estimate from Figure~\ref{fig:BUest1} after estimating the root under $\mathcal{R}_1$.}
		\label{fig:BUest2}
	\end{subfigure}
	\begin{subfigure}[t]{0.49\textwidth}
		\includegraphics[width=1\textwidth]{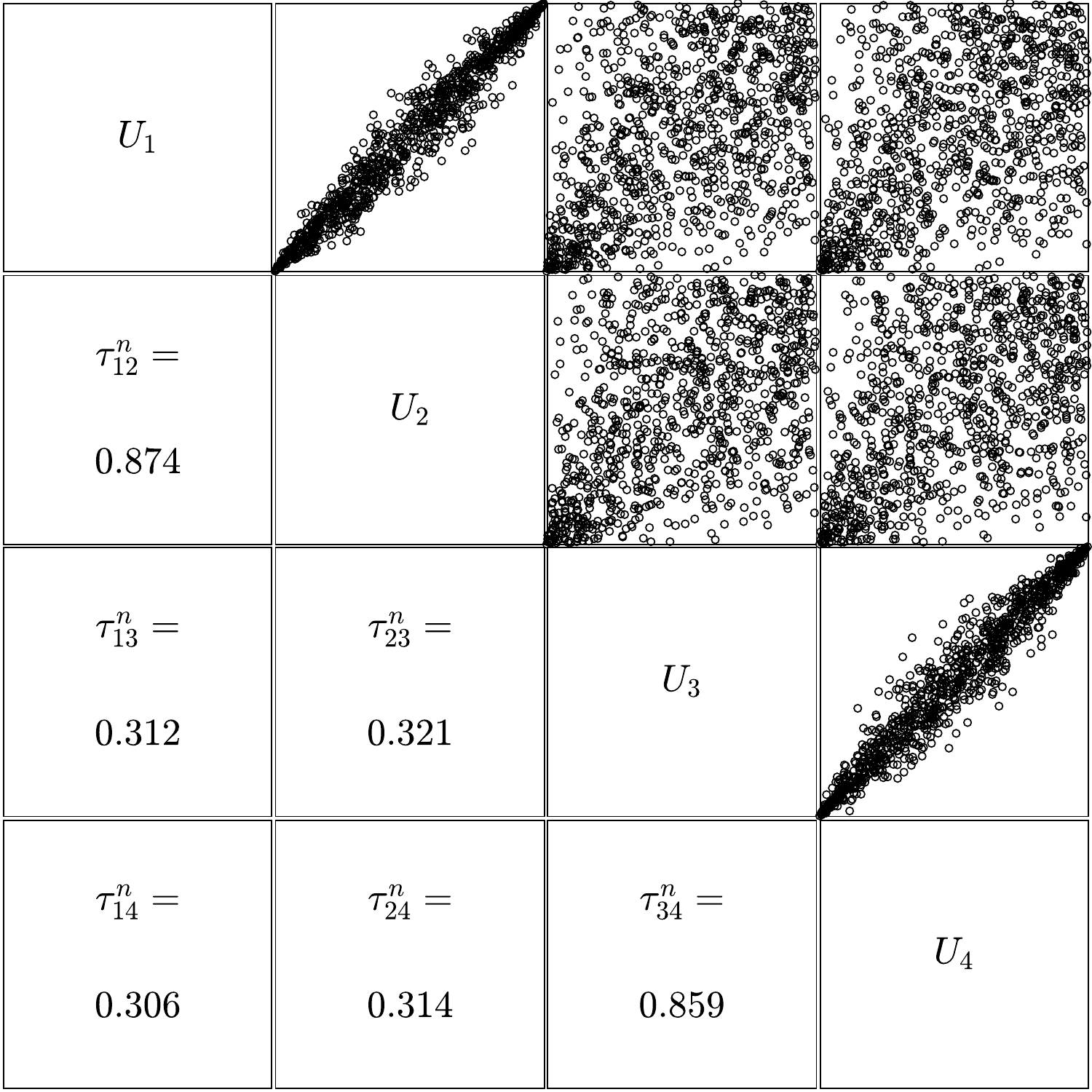}
		\caption{A sample of $n =1000$ observations from the HOPAC in Figure~\ref{fig:BUest2}.}
		\label{fig:BUest2sample}
	\end{subfigure}
	\caption{Evolution of a HOPAC model during Bottom-Up estimation. Note that the values in bold show what has been updated compared to a previous model.}
	\label{fig:BUstrucest11}
	% MATLAB: outerpowerSTCO	
\end{figure}

The algorithm estimates the structure in its first two steps using Algorithm~\ref{alg:structure_estim}, which returns a binary tree $(\hat{\VV}, \hat{\EE})$ and estimates of Kendall's tau $(\hat{\tau}_5, \hat{\tau}_6, \hat{\tau}_7)$ corresponding to each fork in that tree, all shown in Figure~\ref{fig:BUstruc}.

As described above, the families of nested ACs are assumed to be from some pool of available families, e.g., a pool implemented by the software toolbox we use. Such an assumption implies that the user, when deciding which family suits best for the considered data, should repeat the whole estimation process for each of the available families and then perform some extra evaluation of their fit. For simplicity, assume the Clayton family for all the nested ACs.
Recall that the HOPAC estimates will be built under the SNC in its simplified parametric form. This means that to satisfy the SNC, it must hold that for each parent-child pair of forks in the structure 
\begin{enumerate}
	\item[$\mathcal{R}_1$)] $\theta_{parent} \leq \theta_{child}$, if $\beta_{parent} = 1$,\label{r1}  or
	\item[$\mathcal{R}_2$)] $\beta_{parent} \leq \beta_{child}$, if $\theta_{parent} = \theta_{child}$.\label{r2}
\end{enumerate}

\noindent Let us consider two possible estimators under $\mathcal{R}_1$ and $\mathcal{R}_2$ following the concept in Algorithm~\ref{alg:est_HOPAC_summary}.

\subsubsection{Bottom-Up estimator}
\label{sec:bottom-up}
The idea of the \emph{Bottom-Up} estimator lies in traversing through the forks in the estimated structure in the way that one starts at the bottom of the structure and then continues up until the root is reached, similar to \cite{Okh13} and \cite{goreckihofertholena2016approachjiis,gorecki2017structure}. 
A way to guide the traversing process consistently is, e.g., according to Kendall's tau estimates returned by Algorithm~\ref{alg:structure_estim} -- starting with the fork with the highest value, then process the fork with the second highest value, until one gets to the fork with the lowest value, the root.

In our example, fork 5 corresponds to the bottom of the structure.
As it is the youngest common ancestor of leaves $u_1$ and $u_2$, we compute the ML estimator for $\theta_5$ and $\beta_5$ according to \eqref{eq:mleopac}:
\begin{equation}
(\hat{\theta}_5, \hat{\beta}_5
) = \argmax_{(\theta \in \Theta_{\textrm{C}}, \beta \in [1, \infty))}\sum_{i=1}^{n} \log c_{\psi_{(\textrm{C},\theta,\beta)}}(u_{i1}, u_{i2}). 
\label{eq:mleopac2}
\end{equation}
As node 6 is the youngest common ancestor of $u_3$ and $u_4$, the corresponding parameters $\theta_6$ and $\beta_6$ can be estimated according to \eqref{eq:mleopac2} with $u_{i1}$ and $u_{i2}$ being replaced by $u_{i3}$ and $u_{i4}$, respectively. The estimated parameter values are shown in Figure~\ref{fig:BUest1}.
Having the bottom level estimated, we continue to the upper levels until we reach to the root.  Here the SNC comes into play.

As $\hat{\theta}_5 \neq \hat{\theta}_6$, it clear that $\mathcal{R}_2$ is violated because it is impossible to satisfy both $\hat{\theta}_7 = \hat{\theta}_5$ and $\hat{\theta}_7 = \hat{\theta}_6$ for our example. The restriction $\mathcal{R}_2$ was, however, a constraint under which the model was built; see Figure~\ref{fig:BUmodel}. It follows that turning to the restriction $\mathcal{R}_1$ prevents a good fit for node 7, as $\mathcal{R}_1$ requires that $\hat{\beta}_7 = 1$ and that $\hat{\theta}_7$ has to be trimmed to the closest value allowed by $\mathcal{R}_1$, i.e., to 0.844; see Figure~\ref{fig:BUest2}. A sample of 1000 observations from this HOPAC estimate is shown in Figure~\ref{fig:BUest2sample}.
It is evident that choosing $\mathcal{R}_1$ instead of $\mathcal{R}_2$ substantially affects the fit -- compare the distributions of the pairs $(U_1, U_3)$, $(U_2, U_3)$, $(U_2, U_4)$ and $(U_1, U_4)$ shown in  Figure~\ref{fig:BUdata} (which correspond to the true HOPAC) with the corresponding ones of the estimate in Figure~\ref{fig:BUest2sample}. A way to cope with this problem could be to test if the parameters $\theta$ of the children are all equal to some aggregated value like their mean. With $m$ children of a given fork, this would however require additional ${m}\choose{2}$ tests (and there is the problem of multiple testing), making the computation substantially more involved. An efficient solution requiring at most one test for each fork is presented in the following section.

\subsubsection{Top-Down estimator}
A solution to those problems consists of reversing the way in which the structure is traversed during the estimation, meaning starting at the root of the copula structure and then using the depth-first approach to go through all forks. 
This way of estimation already appeared in connection to other hierarchical copula structures; see \cite{zhu2016structure} or recently \cite{COSSETTE201959}. 
Before considering our example with the \emph{Top-Down} estimator, we first present it in term of pseudo-code in Algorithm \ref{alg:topdown}.  

Let $\bm{u} = (u_{ij}) \in [0, 1]^{n\times d}$ be realizations of $(U_{ij})_{i \in \{1, ..., n\}}^{j \in \{1, ..., d\}}$ given by~\eqref{eq:pseudo_observations}. 
As for the Bottom-Up estimator, these are first used to estimate the copula structure $(\hat{\VV}, \hat{\EE})$ with Algorithm~\ref{alg:structure_estim} based on the matrix of pairwise sample versions of Kendall's tau.

Estimation of the parameters is then performed by calling the function 
TopDown$\{\bm{u}, a,$ $(\hat{\VV}, \hat{\EE}), 2d-1, \Theta_a, [1, \infty), \beta_{\mathcal{R}}\}$ presented in Algorithm~\ref{alg:topdown}, where $a$ is an Archimedean family, $2d-1$ denotes the root in the binary tree $(\hat{\VV}, \hat{\EE})$, $\Theta_a$ (see Table~\ref{tab:geners}) and  $[1, \infty)$ are ranges for the parameters $\theta$ and $\beta$ of the OPAC estimate $C_{\hat{\Psi}[2d-1]}$ corresponding to the root, and $\beta_{\mathcal{R}} \in [1, \infty)$ is an upper bound for accepting $\beta_{parent} = 1$ in $\mathcal{R}_1$, commented on more below. Recall that \emph{descendants} of a given node refer to the set of the children of that node, the children of these children, etc.

\begin{algorithm}[t!]
	\floatname{algorithm}{Algorithm}
	\caption{The Top-Down HOPAC estimator}
	\label{alg:topdown}
	\begin{algorithmic}
		\renewcommand{\algorithmicrequire}{\textbf{Inputs:}}
		\REQUIRE 
		\STATE $\bm{u} = (u_{ij}) \in [0, 1]^{n\times d}$ -- realizations of $(U_{ij})_{i \in \{1, ..., n\}}^{j \in \{1, ..., d\}}$ given by \eqref{eq:pseudo_observations}
		\STATE $a$  -- a one-parameter Archimedean family, e.g., from Table~\ref{tab:geners}
		\STATE $(\hat{\VV}, \hat{\EE})$ -- a binary tree (copula structure)
		\STATE $k \in \{d+1, ..., 2d-1\}$ -- a fork to estimate its parameters
		\STATE $r_{\theta} \subset \mathbb{R}$ -- a range for parameter $\theta$
		\STATE $r_{\beta} \subseteq [1, \infty)$ -- a range for parameter $\beta$
		\STATE $\beta_{\mathcal{R}} \in [1, \infty)$ -- an upper bound for accepting $\beta_{parent} = 1$ in $\mathcal{R}_1$
		\STATE
		\renewcommand{\algorithmicensure}{\textbf{Output:}}
		\ENSURE
		\STATE $(\hat{\VV}, \hat{\EE}, \hat{\Psi})$
		\STATE
	\end{algorithmic}
	\begin{algorithmic}[1]
		\renewcommand{\algorithmicensure}{\textbf{Function TopDown$\{\bm{u}, a,(\hat{\VV}, \hat{\EE}), k, r_{\theta}, r_{\beta}, \beta_{\mathcal{R}} \}$}}
		\ENSURE
		\IF{$k \in \{1, ..., d\}$} 
			\RETURN -- stop recursion if $k$ is a leaf \label{alg:return_if_leaf}
		\ENDIF
		\STATE $\{i, j\} \leftarrow$ the children of $k$ in $(\hat{\VV}, \hat{\EE})$ -- assume $i < j$ \label{alg:ij_children}
		\STATE $l_i \leftarrow$ the set of descendant leaves of $i$ if $i$ is a fork, otherwise $l_i \leftarrow \{i\}$ \label{alg:li}
		\STATE $l_j \leftarrow$ analogous to $l_i$ (by replacing $i$ by $j$) \label{alg:lj}
		\STATE $(\hat{\theta}, \hat{\beta}) \leftarrow  			
		\frac{1}{\#l_i \#l_j}\sum\limits_{\tilde{i} \in l_i} \sum\limits_{\tilde{j} \in l_j} \argmax\limits_{(\theta_{\tilde{i}\tilde{j}}, \beta_{\tilde{i}\tilde{j}}) \in \Theta_a \times [1, \infty)}\sum\limits_{m=1}^{n} \log c_{\psi_{(a,\theta_{\tilde{i}\tilde{j}},\beta_{\tilde{i}\tilde{j}})}}(u_{m\tilde{i}}, u_{m\tilde{j}})$ \label{alg:meanaggmle}
		\IF{$\hat{\beta} \in [1,\beta_{\mathcal{R}}]$} \label{alg:ifbeta1}
			\STATE $\tilde{r}_\theta \leftarrow r_{\theta} \cap [\hat{\theta}, \infty)$ and  $\tilde{r}_\beta \leftarrow r_{\beta}$ -- restriction $\mathcal{R}_1$
		\ELSE 
			\STATE $\tilde{r}_\theta \leftarrow [\hat{\theta}, \hat{\theta}]$ and $\tilde{r}_\beta \leftarrow [\hat{\beta}, \infty)$ -- restriction $\mathcal{R}_2$
		\ENDIF
		\STATE TopDown$\{\bm{u}, a,(\hat{\VV}, \hat{\EE}), i, \tilde{r}_{\theta}, \tilde{r}_{\beta},\beta_{\mathcal{R}} \}$ \label{alg:recuri}
		\STATE TopDown$\{\bm{u}, a,(\hat{\VV}, \hat{\EE}), j, \tilde{r}_{\theta}, \tilde{r}_{\beta}, \beta_{\mathcal{R}} \}$ \label{alg:recurj}
		\STATE $\hat{\Psi}[k] \leftarrow \psi_{(a, \hat{\theta}, \hat{\beta})}$ \label{alg:psik}
	\end{algorithmic}
\end{algorithm} 

Several aspects of Algorithm~\ref{alg:topdown} are worth being addressed:
\begin{itemize}
	\item	The function TopDown recursively goes through all the forks of the tree $(\hat{\VV}, \hat{\EE})$ in the depth-first search manner, which can be seen from steps~\ref{alg:return_if_leaf}, \ref{alg:ij_children}, \ref{alg:recuri} and \ref{alg:recurj}. 
	\item The assumption for $i < j$ in step~\ref{alg:ij_children} is without loss of generality as in all remaining steps of the algorithm the exchange of $i$ and $j$ does not produce any change in their results.	
	\item According to Remark 2 in \cite{gorecki2017kendalls}, all pairs from the
	Cartesian product of $l_i$ and $l_j$ defined in steps \ref{alg:li} and \ref{alg:lj} have the same youngest common ancestor $k$. It follows from Proposition 3 in \cite{goreckihofertholena2016approachjiis} that bivariate margins corresponding to these pairs share the same copula, which motivates the mean aggregated ML estimator used in step~\ref{alg:meanaggmle}. 
	% TO OSTAP: step~\ref{alg:meanaggmle} is quite essential show we SHOULD explain where the motivation for it come from
	Note that such an aggregation approach in a one-parameter version has already been successfully used in Algorithm~3 in \cite{goreckihofertholena2016approachjiis}, see step 2 therein. 
	%Also note that using full ML estimation would be at least challenging, mainly due to substantial influence of the restrictions $\mathcal{R}_1$ and $\mathcal{R}_2$, which fundamentally influence constraints on the parameters, as already illustrated in Section~\ref{sec:bottom-up}.
	Also note that the viability of such an aggregation approach is studied in Section~\ref{sec:sim_study}.
	\item As the result of the argmax in step~\ref{alg:meanaggmle} is the vector of two components $(\theta_{\tilde{i}\tilde{j}}, \beta_{\tilde{i}\tilde{j}})$, all the operators to its left (the two sums and the division) are considered component-wise.
	\item The sets $l_i$ and $l_j$ are always disjoint, which follows from the fact that that node $i$ and node $j$ do not lie at the same branch of $(\hat{\VV}, \hat{\EE})$. This fact avoids that the same pair appears twice (first as $(i, j)$ and then as $(j, i)$) in the first two sums in step~\ref{alg:meanaggmle}.
	\item The \textbf{if} statement in step~\ref{alg:ifbeta1} decides which one of the restrictions $\mathcal{R}_1$ and $\mathcal{R}_2$ applies for the children $i$ and $j$ of node $k$ at the recursive steps~\ref{alg:recuri} and~\ref{alg:recurj}. As the parameters $\hat{\theta}$ and $\hat{\beta}$ are estimated by ML, we have asymptotic normality and the variances of these estimates. As $\theta_{child}$ appearing in  $\mathcal{R}_2$ is not yet available (it is estimated in further steps that depend on the decision in step~\ref{alg:ifbeta1}), it is thus convenient to test for $\mathcal{R}_1$ as it requires only the value of $\beta_{parent}$. In practice, however, testing the hypothesis $\beta_{parent} = 1$ would slow down the computations, therefore we decided to only check whether $\hat{\beta}$ lies in the prescribed interval $[1, \beta_{\mathcal{R}}]$. The involved parameter $\beta_{\mathcal{R}}$ also allows us to emphasize one of the restrictions -- we emphasize $\mathcal{R}_1$ with larger values of $\beta_{\mathcal{R}}$, whereas $\mathcal{R}_2$ with smaller ones. In the following, we use $\beta_{\mathcal{R}} = 1.05$ as it turned out to provide a convenient balance between $\mathcal{R}_1$ and $\mathcal{R}_2$.
	\item The output-triplet $(\hat{\VV}, \hat{\EE}, \hat{\Psi})$ contains the structure in $(\hat{\VV}, \hat{\EE})$, the family and the parameters of the generators in $\hat{\Psi}$ of the HOPAC estimate.
\end{itemize}

To illustrate the HOPAC estimation with the Top-Down approach, let $\bm{u}$ be the pseudo-observations of the sample from Figure~\ref{fig:BUdata}, with $d = 4$ and $n = 1000$. To estimate the structure, put $\bm{u}$ in Algorithm~\ref{alg:structure_estim}, resulting in the tree $(\hat{\VV}, \hat{\EE}) = (\{1, ..., 7\}, \{\{1, 5\}, \{2, 5\},$ $\{3, 6\}, \{4, 6\},$ $\{5, 7\}, \{6, 7\} \})$, which corresponds to the tree depicted in Figure~\ref{fig:BUmodel}. Let $a$ to be again the Clayton family, and recall that $\Theta_{\textrm{C}} = (0, \infty)$, see Table~\ref{tab:geners}. Finally, to obtain the parameter estimates, call TopDown$(\bm{u}, \textrm{C},(\hat{\VV}, \hat{\EE}), 7, \Theta_{\textrm{C}}, [1, \infty), 1.05)$.

In step~\ref{alg:ij_children}, $i \leftarrow 5$ and $j \leftarrow 6$. In further two steps, $l_5 \leftarrow  \{1, 2\}$ and $l_6 \leftarrow  \{3, 4\}$. Step~\ref{alg:meanaggmle} computes the argmax for $(\tilde{i}, \tilde{j}) \in ((1,3),$ $(2,3),$ $(1,4), (2,4))$. The four pairs of $(\theta_{\tilde{i}\tilde{j}}, \beta_{\tilde{i}\tilde{j}})$ are 
(0.99, 3.106),  (0.94, 3.124), (1.07, 3.006) and (0.976, 3.095). Using the component-wise mean results in $(\hat{\theta}, \hat{\beta}) \leftarrow$  (0.994,  3.083). In the next step (as $\hat{\beta} > \beta_{\mathcal{R}}$) the restriction $\mathcal{R}_2$ is applied, resulting in $r_\theta \leftarrow [0.994, 0.994]$ and $r_\beta \leftarrow [3.083, \infty)$.

\begin{figure}[t!]
	\centering
	\begin{subfigure}[t]{0.49\textwidth}
		\includegraphics[width=1\textwidth]{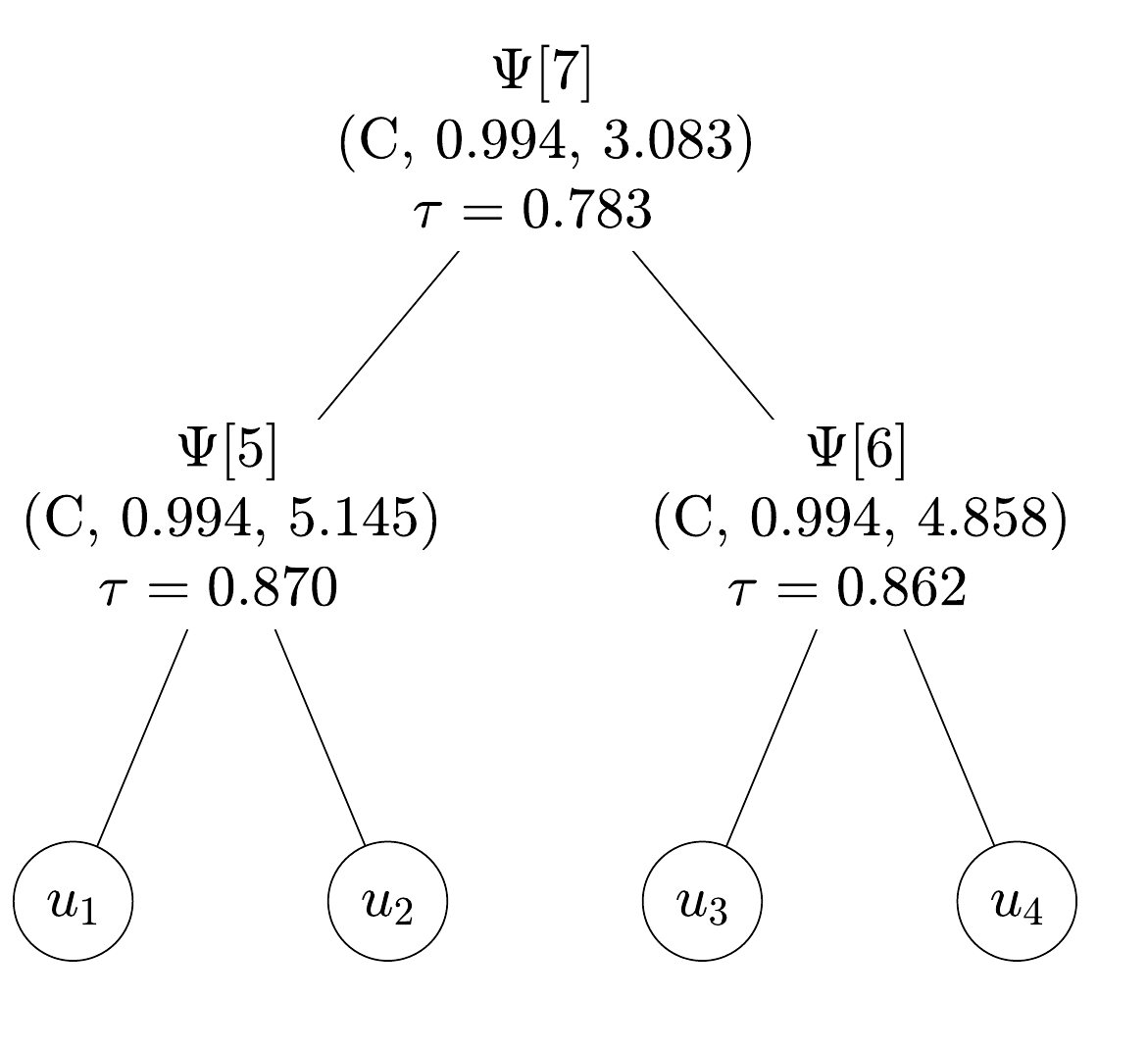}
		\caption{A HOPAC estimate from sample in Figure~\ref{fig:BUdata} under assumption of Clayton family.}
		\label{fig:TDest}
	\end{subfigure}
	\begin{subfigure}[t]{0.49\textwidth}
		\includegraphics[width=1\textwidth]{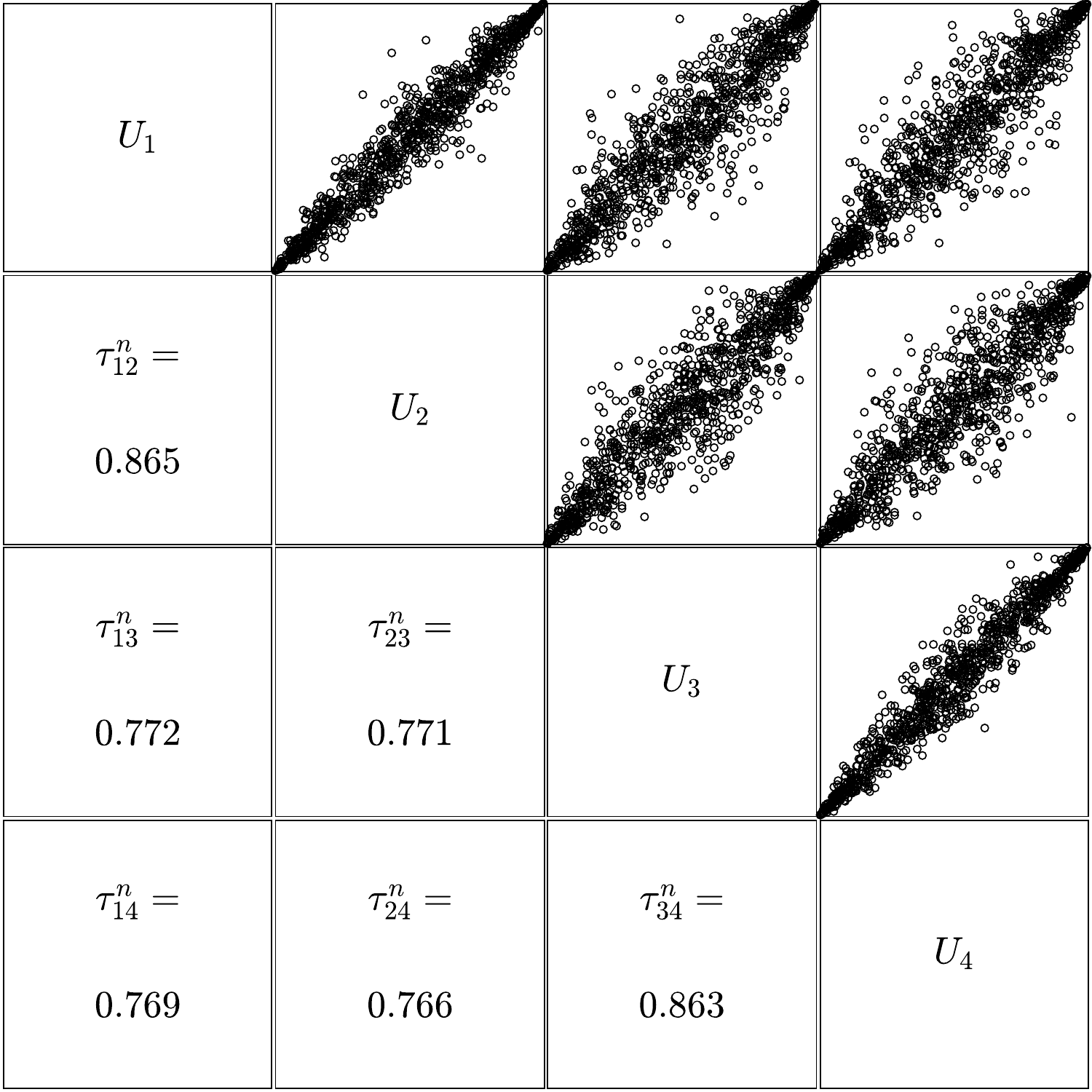}
		\caption{A sample of $n =1000$ observations from the HOPAC in Figure~\ref{fig:TDest}.}
		\label{fig:TDsample}
	\end{subfigure}
	\begin{subfigure}[t]{0.49\textwidth}
	\includegraphics[width=1\textwidth]{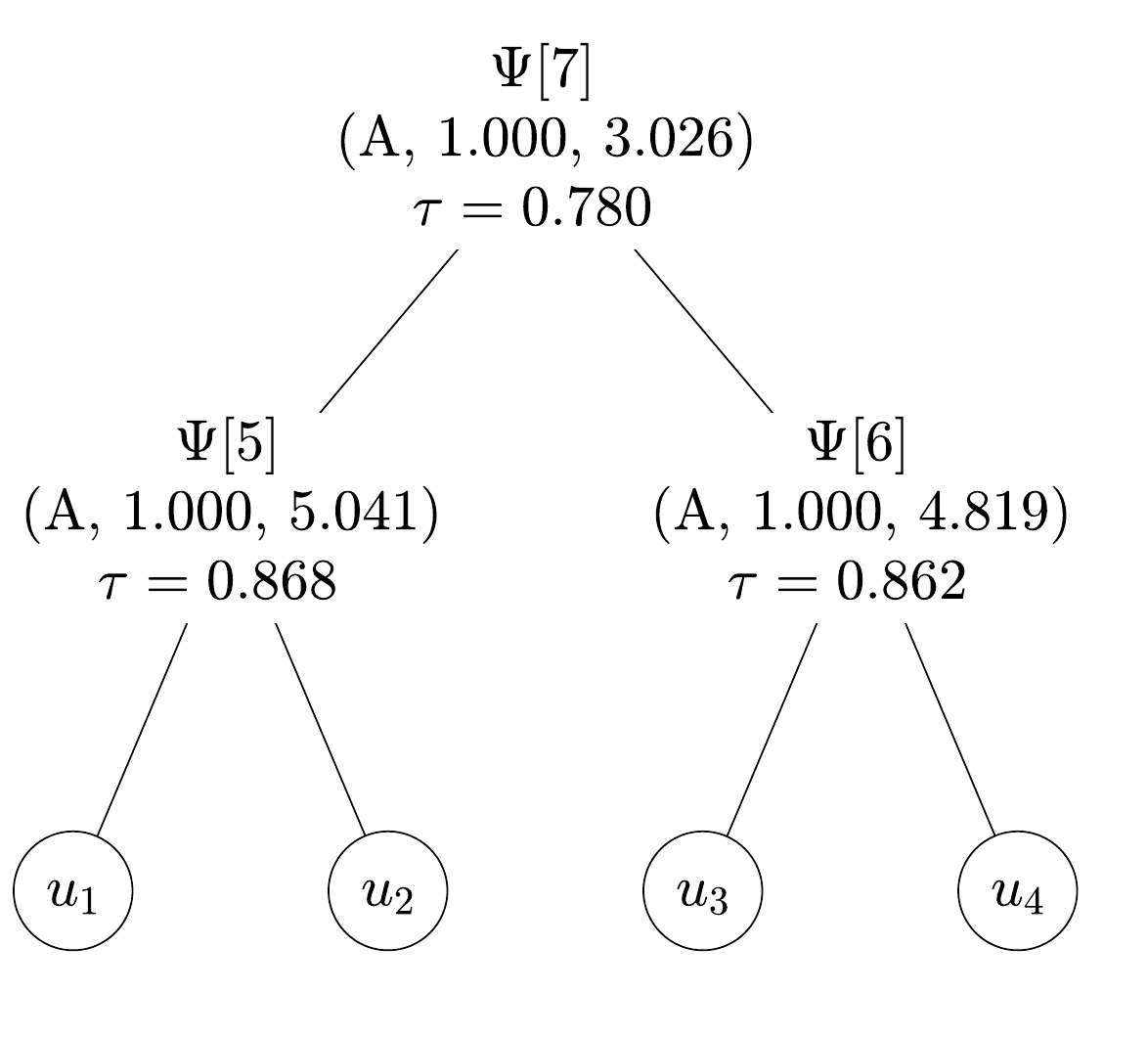}
	\caption{A HOPAC estimate from sample in Figure~\ref{fig:BUdata} under assumption of Ali-Mikhail-Haq family. Note that the values shown are rounded to 3 digits, hence, e.g., even if all $\theta$s are $< 1$, the rounded values are 1.000.}
	\label{fig:TDestA}
\end{subfigure}
	\caption{Top-Down estimates.}
	\label{fig:TDestsample}
	% MATLAB: outerpowerSTCO (with U = rnd(myHAC, 1000) for Clayton)
\end{figure}

As the recursive steps~\ref{alg:recuri} and \ref{alg:recurj} involve the estimation of bivariate OPACs for nodes 5 and 6 ($l_i \leftarrow \{i\}$ and $l_j \leftarrow\{j\}$ in both of the nested calls of TopDown), we just show the results of step~\ref{alg:psik}, which are $\hat{\Psi}[5] \leftarrow \psi_{(C, 0.994, 5.145)}$ and $\hat{\Psi}[6] \leftarrow \psi_{(C, 0.994, 4.858)}$. 
Finally, step~\ref{alg:psik} results in $\hat{\Psi}[7] \leftarrow \psi_{(C, 0.994,  3.083)}$. The resulting estimated HOPAC  $C_{(\hat{\VV}, \hat{\EE}, \hat{\Psi})}$ is depicted in Figure~\ref{fig:TDest}. We observe that the parameters are relatively close to the true parameters (shown in Figure~\ref{fig:BUmodel}), particularly, compared to the Bottom-Up analogue, for $\beta$ of node 7. This is further reflected via distributions of samples from these HOPACs -- compare the distributions and particularly the strength of the correlation in the pairs $(U_1, U_3)$, $(U_2, U_3)$, $(U_2, U_4)$ and $(U_1, U_4)$ shown in Figures~\ref{fig:BUdata}, \ref{fig:BUest2} and \ref{fig:TDsample} corresponding to the true copula, the Bottom-Up and Top-Down estimate, respectively. In contrast to the Bottom-Up estimate, the Top-Down one closely follows the true distribution.

It is clear that the arbitrary assumption of the Clayton family needs an extra attention. As suggested above, an extra criterion should be used to evaluate feasibility of such an assumption. For example, the goodness-of-fit test statistic used in the estimator defined in~\eqref{eq:snopac} can be used to this end; see \cite{Gen09}. 
Evaluating this criterion for the sample from Figure~\ref{fig:BUdata} and the Top-Down HOPAC estimate shown in Figure~\ref{fig:TDest}, we obtain $S_n = 0.0148$. For $a =$ A we get $S_n = 0.0148$, for $a =$ F we observe $S_n = 0.0694$ and for $a =$ J we receive $S_n = 0.1516$. It is not surprising that $S_n$ for the true family is minimal. What might be surprising is that the minimum is also obtained for the Ali-Mikhail-Haq family. However, looking at page 117 in \cite{Nel06}, Table 4.1 shows that for $\theta = 1$ the copulas from Clayton and Ali-Mikhail-Haq (there denoted 4.2.1 and 4.2.3, respectively) are both equal to $C(u, v) = uv/(u+v-uv)$. Looking further at the resulting Top-Down estimate for Ali-Mikhail-Haq shown in Figure~\ref{fig:TDestA}, and 
considering that the parameters $\theta$ for both families are relatively close to 1, this result rather confirms that the presented framework works correctly.

\subsection{Simulation study}
\label{sec:sim_study}
To evaluate the HOPAC estimator presented in the previous section, $N=100$ repetitions of the following routine for each of the families Ali-Mikhail-Haq, Clayton, Frank and Joe are performed. This routine first randomly generates a HOPAC model, then samples from it, computes several estimates based on the sample, and finally measures certain types of discrepancy between the model and the estimate, and eventually between the sample and the estimate. More precisely, the setup is as follows:
\begin{enumerate}
	\item Given a dimension $d \in \{5, 10, 20\}$, a correlation matrix of size $d \times d$ is randomly generated according to the sampling algorithm proposed in \cite{makalic2018efficient}.
	\item This matrix is then passed to Algorithm~\ref{alg:structure_estim}, which returns a binary tree with $d$ leaves, which serves as the \emph{structure} of the randomly generated HOPAC model. The algorithm also returns the estimates $\hat{\tau}_{d+1}, ..., \hat{\tau}_{2d-1}$ corresponding to the forks in that tree, which are used, in the next step, to generate the \emph{parameters} of the HOPAC model. 
	\item The forks in the structure are traversed depth-first starting from the root ($k = 2d-1$) and for each given fork $k \in \{d+1, ..., 2d-1\}$, the parameter $\beta$ is set equal to 1, i.e., to the case corresponding to  $\mathcal{R}_1$, with probability of 50\%. Hence, if $\beta$ is 1, the parameter $\theta$ is just adjusted in a way that Kendall's tau of this fork remains equal to $\hat{\tau}_{k}$. For the case corresponding to $\mathcal{R}_2$ (the remaining 50\%), the parameter $\theta$ is first generated randomly and then $\beta$ is adjusted to keep Kendall's tau equal to $\hat{\tau}_{k}$; see Figure~\ref{fig:hopacrnd} for examples.
	\begin{figure}[t!]
		\centering
		\includegraphics[width=0.45\textwidth]{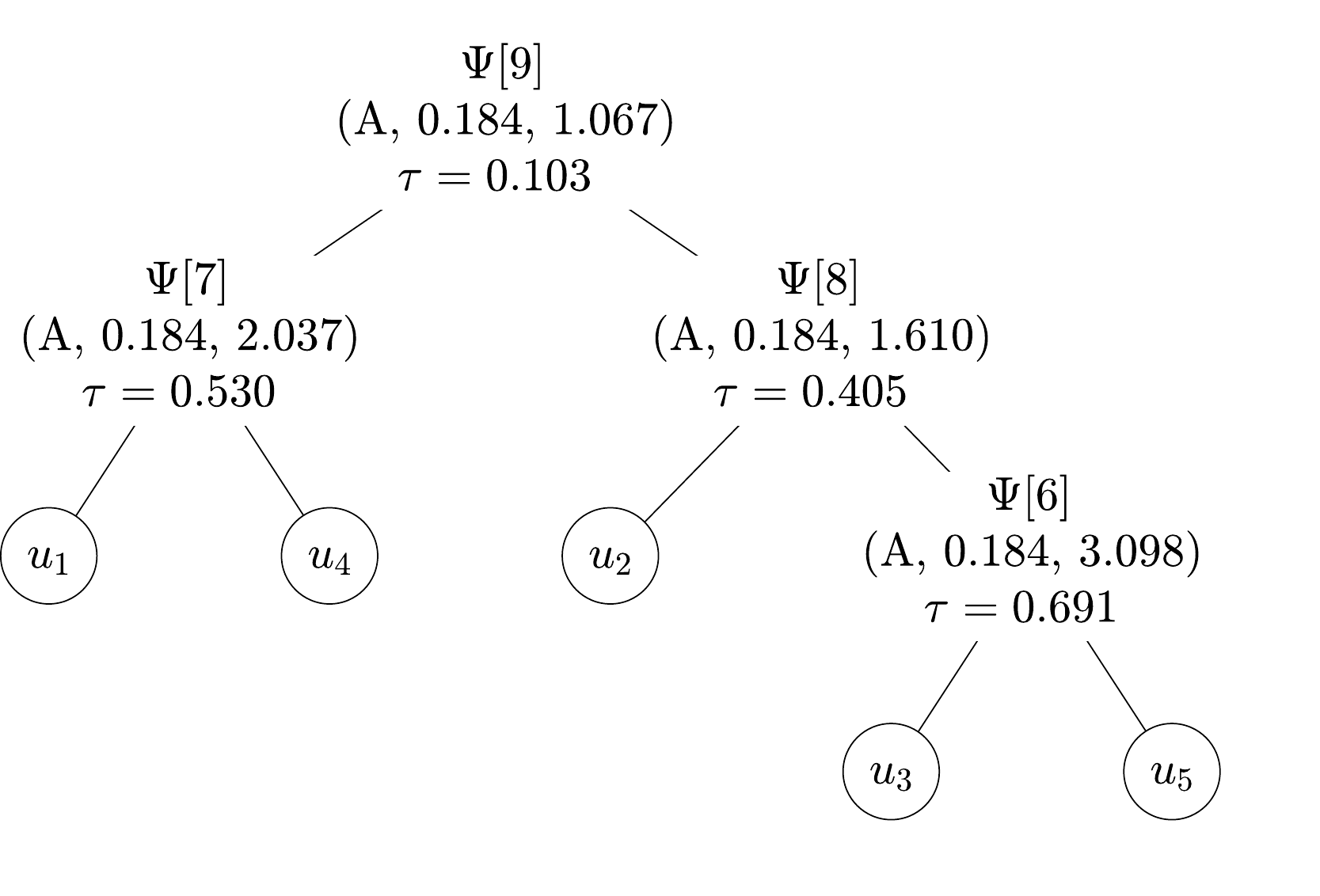}
		\includegraphics[width=0.53\textwidth]{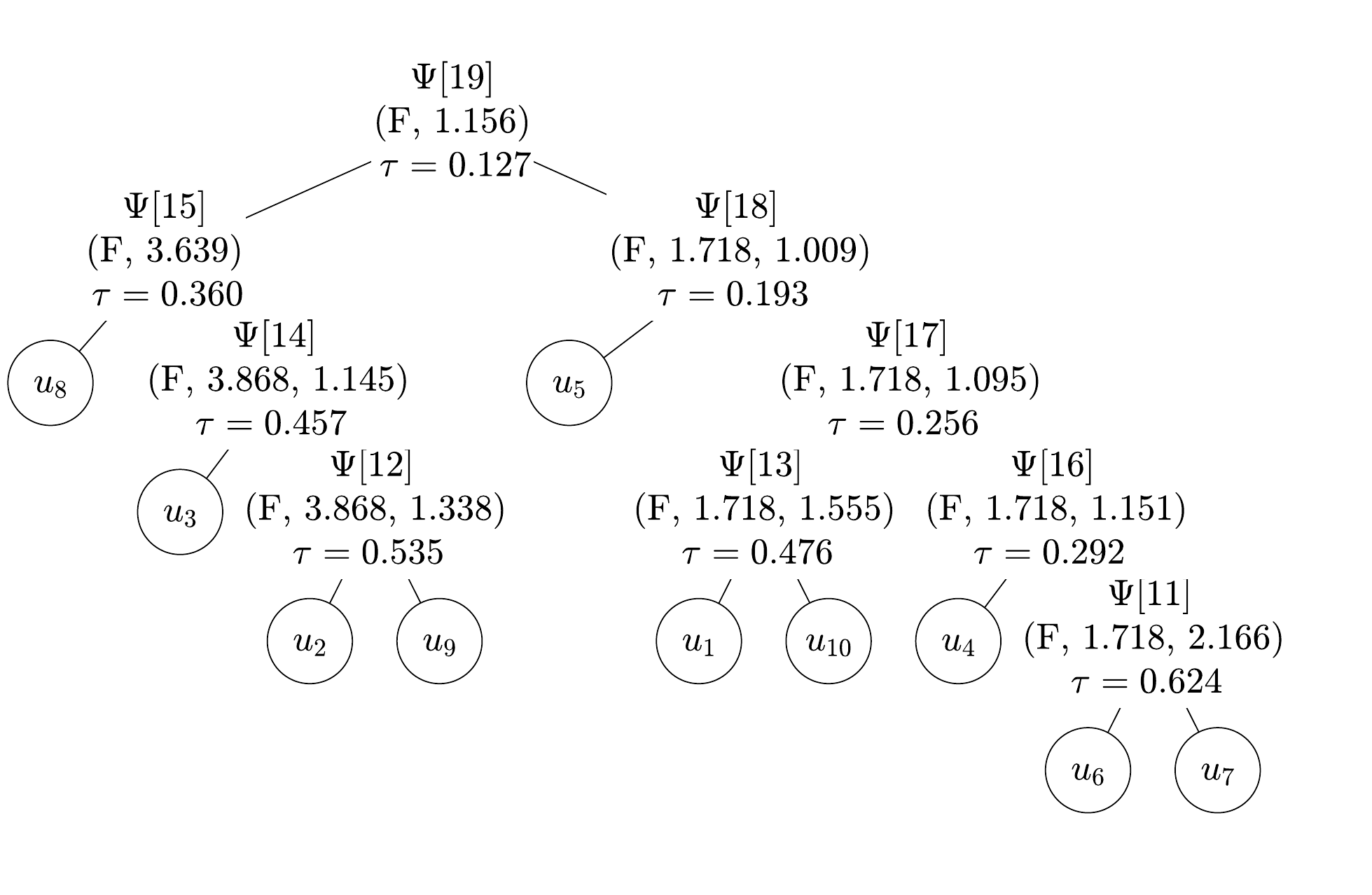}
		\hspace*{-5mm}
		\includegraphics[width=1.1\textwidth]{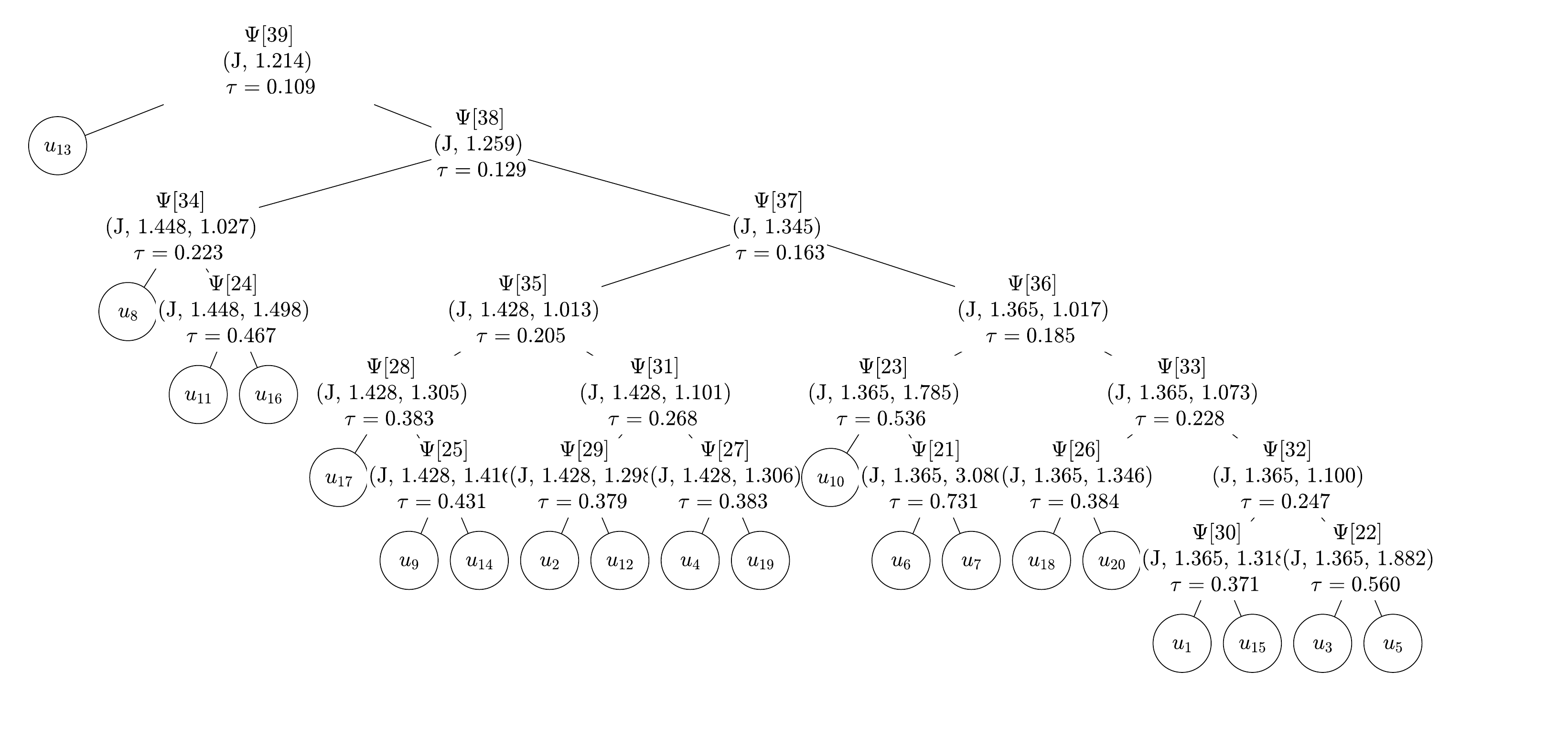}
		\caption{Three randomly generated HOPAC models.}
		\label{fig:hopacrnd}
		% MATLAB CSDAthreeRndHOPACs
	\end{figure}	
	\item Assume the same family $a \in $ \{A, C, F, J\} for each OPAC nested in the HOPAC model and sample from it with sizes $n \in \{200, 400, ..., 1000\}$.
	\item Based on these samples, compute realizations of the following estimators: 
		\begin{enumerate}%[a)]
			\item The OPAC estimator $(\hat{\theta}_{\textrm{ML}}, \hat{\beta}_{\textrm{ML}})$ defined by
			\begin{equation}
				%\hspace*{-5mm}
				\frac{1}{{d \choose 2}}\sum_{i=1}^{d} \sum_{j=i+1}^{d} \argmax_{(\theta_{ij}, \beta_{ij}) \in \Theta_a \times [1, \infty)}\sum_{m=1}^{n} \log c_{\psi_{(a,\theta_{ij},\beta_{ij})}}(u_{mi}, u_{mj}). 
				\label{eq:mledopac}
			\end{equation}						
			We include this estimator to the study in order to show to which level OPACs are (un)able to fit HOPACs, in other words, how important is the presence of hierarchy/structure in the copula model.
			Also note that accessing the density $c_{\psi}$ can be challenging
			due to need of differentiating the cumulative distribution function $d$-times. The estimator given by \eqref{eq:mledopac} is thus used instead of the standard (non-agregated) generalization of \eqref{eq:mleopac}. It is, however, worth noting here that this simple (OPAC) estimator shows an excellent improvement/complexity trade-off in the tail-dependence modeling application reported in  Section~\ref{sec:applic}, which hints at feasibility of such an aggregation approach in general.

			\item The HAC estimator (denoted HAC) based on one-parameter generators  given by Algorithm~\ref{alg:topdown}, where the OP transformation is avoided simply by setting the input argument $r_\beta$ equal to $[1, 1]$. Note that this estimator is included in our study in order to stress out the importance of having the OP transformation in the copula model.
			%NOTE: this is a GOOD idea, as what we want is to see what happens if there is NO OP transformation, rather than to compare with other HAC estimators!
			\item The HOPAC Top-Down $S_n$-based estimator (TD-Sn) given in Algorithm~\ref{alg:topdown}, where the OPAC ML estimator in step~\ref{alg:meanaggmle} is replaced by the distance-based estimator $S_n$ given by \eqref{eq:snopac}. 				
			\item The HOPAC Top-Down ML estimator (TD-ML) given exactly according to Algorithm~\ref{alg:topdown}.
		\end{enumerate}				
	\item For each sample (eventually replaced by the model) and estimator, evaluate the following 6 measures concerning their discrepancy in the distribution, Kendall's tau, upper-tail dependence coefficient and parameters. These measures can be divided into the following two groups:
		\begin{enumerate}%[i)]	
			\item \textbf{Sample versus estimate}. This group includes the three measures given at the top of Figures~\ref{fig:hopacest_opAC_567}~and~\ref{fig:hopacest_opFJ_567}, where $\hat{C}$ and $C_n$ denote the estimated and empirical copulas, respectively, $\hat{\tau}_{ij}$ and $\hat{\lambda}^u_{ij}$ denote the Kendall's tau and upper-tail dependence coefficient corresponding to the youngest common ancestor of leaves $i$ and $j$ in the estimated structure, respectively, $\tau^n_{ij}$ denotes the sample version of Kendall's tau corresponding to variables $U_i$ and $U_j$, and $\lambda_{ij}^{u,n,5\%}$ denotes the non-parametric estimate of the upper-tail dependence coefficient for variables $U_i$ and $U_j$
			at the level $k/n = 0.05$ according to (13) in \cite{schmidt2006non}, where $k \in \{1, ..., n\}$.
			\item \textbf{True versus estimate}. This group includes the three measures given at the top of Figures~\ref{fig:hopacest_opAC_123}~and~\ref{fig:hopacest_opFJ_123},where the pair $(\theta_i, \beta_i)$ and $\tau_i$ and $\lambda_i^u$ correspond to the fork $i$ in the copula model whereas the pair $(\hat{\theta}_i, \hat{\beta}_i)$ and $\hat{\tau}_i$ and $\hat{\lambda}_i^u$ correspond to the fork $i$ in the copula estimate. Note that the lower-tail dependence coefficient is not considered as it equals 0 for all families considered except Clayton, see Table~\ref{tab:geners}.
			Also note that these measures require that the structure of the model and the estimate match and are thus evaluated only for TD-Sn and TD-ML.
			To generate $N=100$ estimates matching the true structure, 	
		  	a new sample according to the model is generated in each out of  $N=100$ repetition until the structure returned by Algorithm~\ref{alg:structure_estim} equals the true structure. The ratio of true structures returned out of $N=100$ tries is depicted in Figure~\ref{fig:hopacest_structures1} on page~\pageref{fig:hopacest_structures1}.				
			\begin{figure}[h!]
				\centering
				Ali-Mikhail-Haq\\
				\vspace*{1mm}
				\includegraphics[width=1\textwidth]{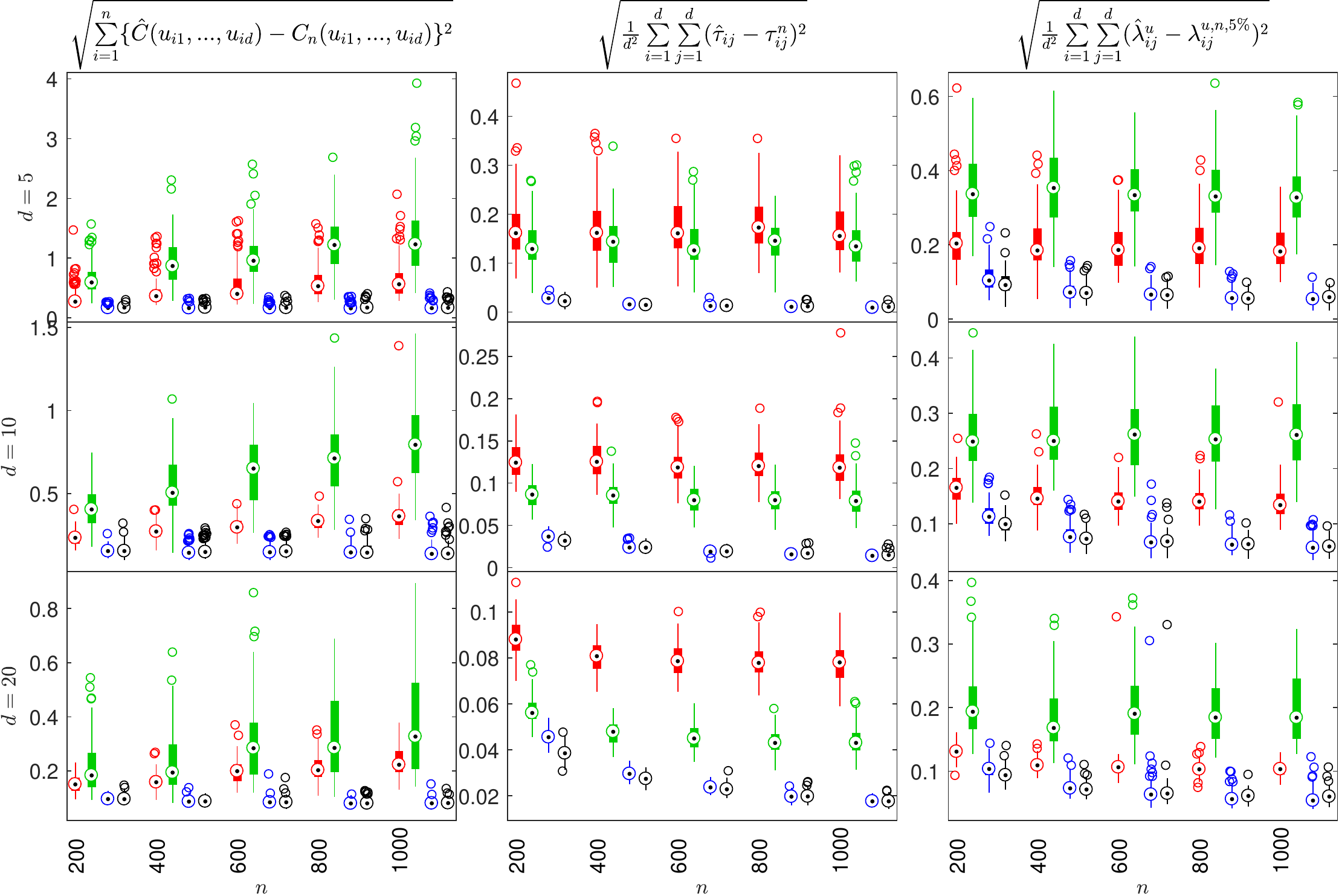}
				Clayton\\
				\vspace*{1mm}
				\includegraphics[width=1\textwidth]{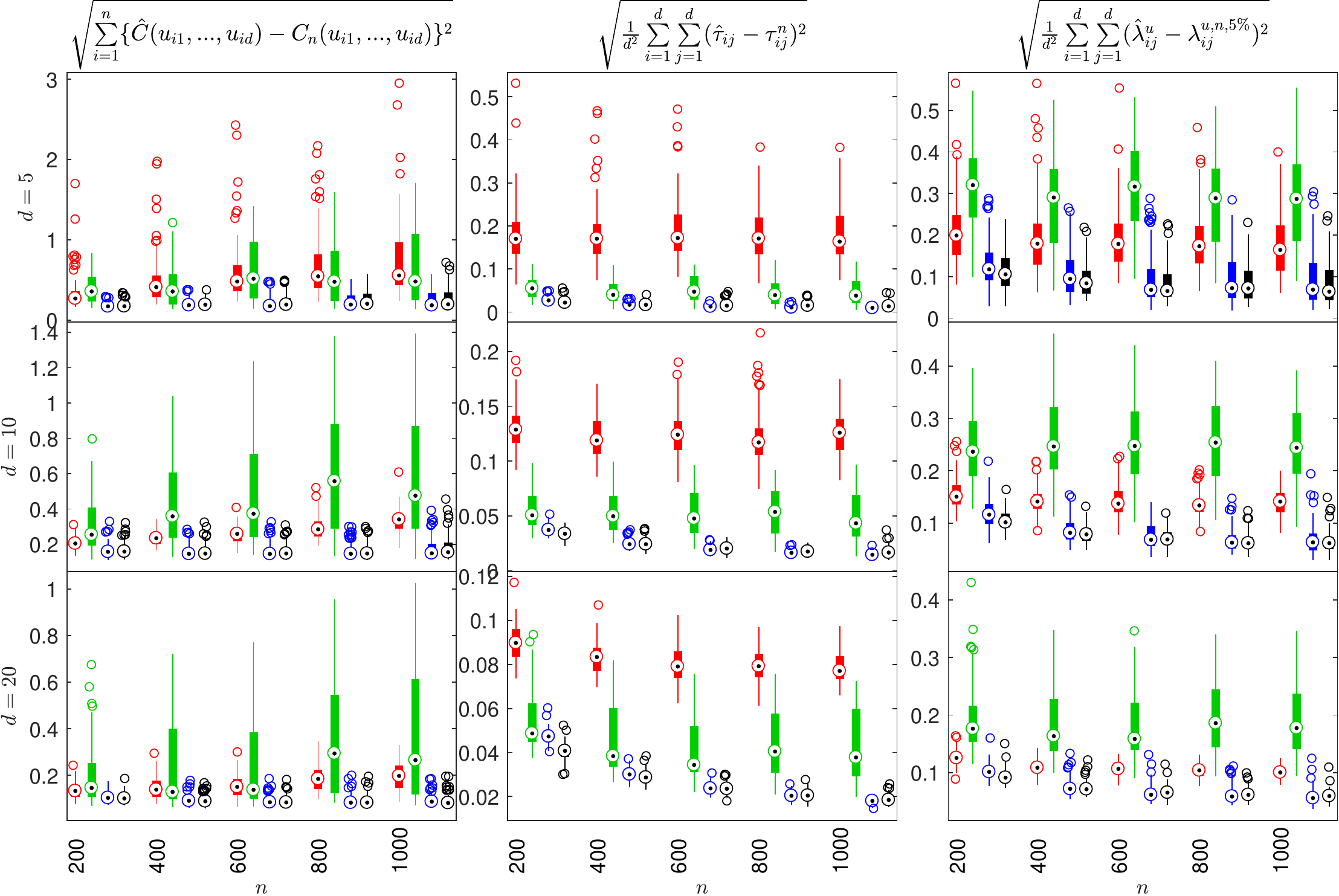}
				\caption{Realizations of the \textbf{sample versus estimate} measures for Ali-Mikhail-Haq and Clayton copulas estimated by  {\color{myRed}OPAC}, {\color{myGreen}HAC}, {\color{myBlue}TD-Sn} and TD-ML.}
				\label{fig:hopacest_opAC_567}
			\end{figure}
			\begin{figure}[h!]
				\centering
				Frank\\
				\vspace*{1mm}
				\includegraphics[width=1\textwidth]{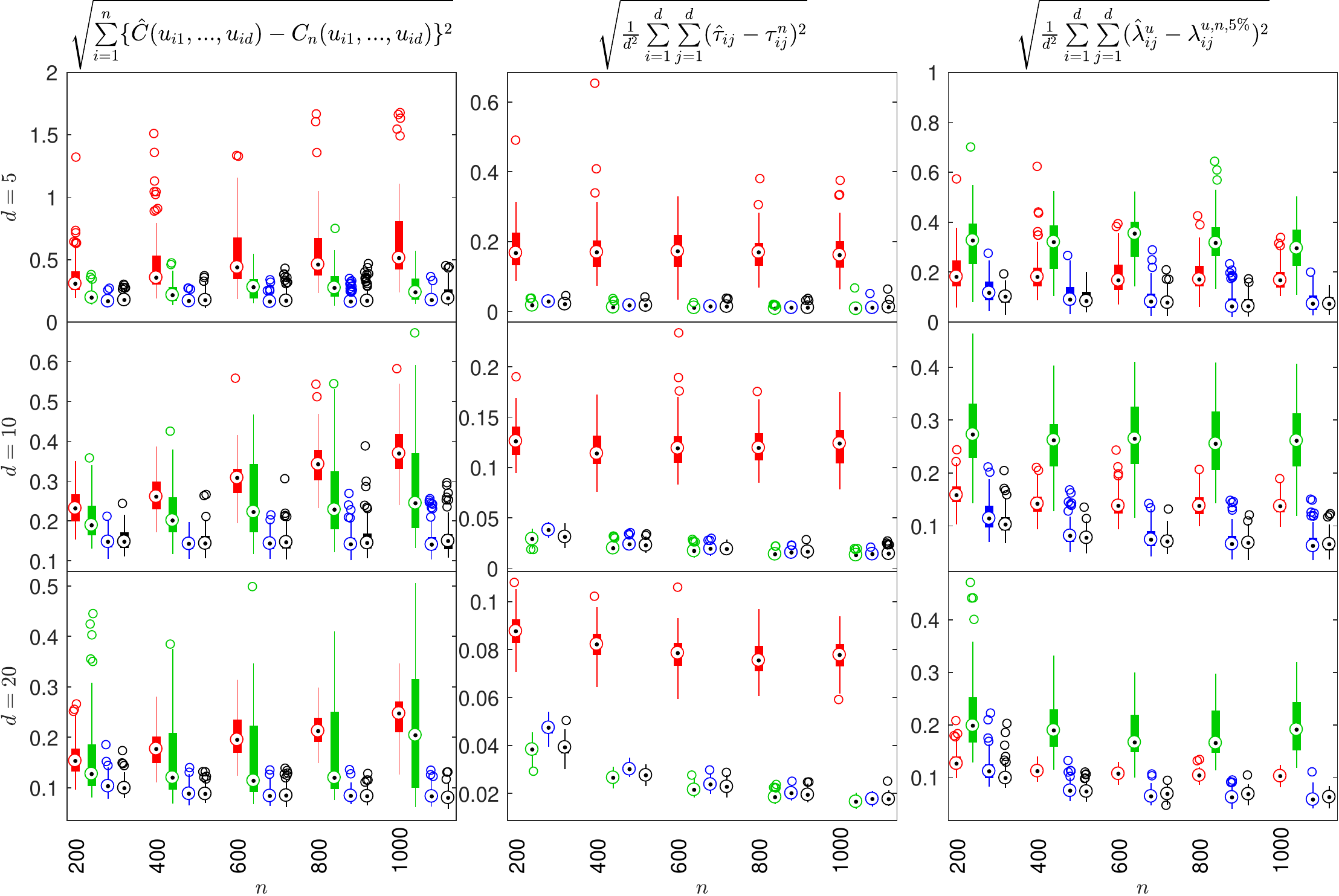}
				Joe\\
				\vspace*{1mm}
				\includegraphics[width=1\textwidth]{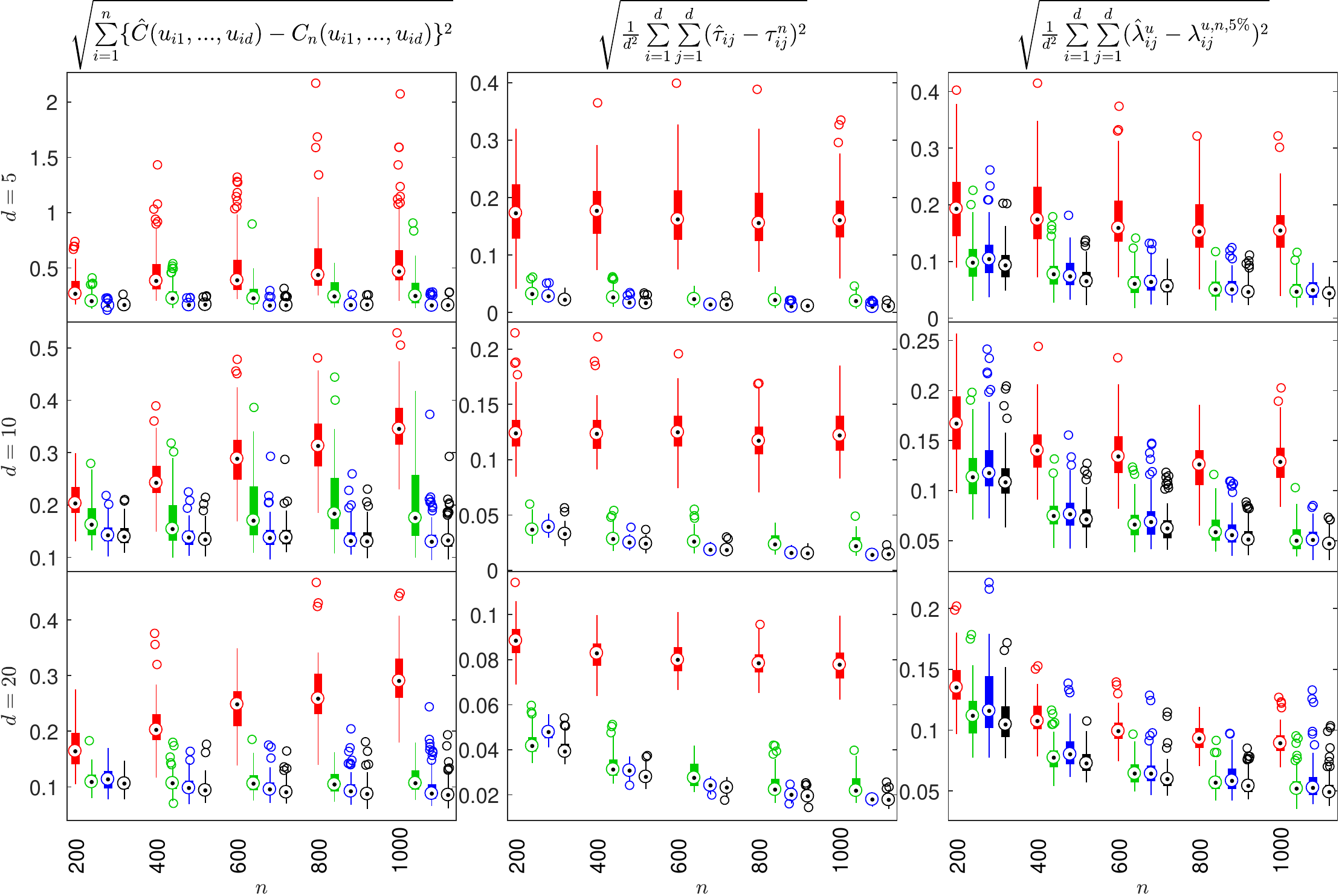}
				\caption{Realizations of the \textbf{sample versus estimate} measures for Frank and Joe copulas estimated by  {\color{myRed}OPAC}, {\color{myGreen}HAC}, {\color{myBlue}TD-Sn} and TD-ML.} 
				\label{fig:hopacest_opFJ_567}
			\end{figure}
			
			\begin{figure}[h!]
				\centering
				Ali-Mikhail-Haq\\
				\vspace*{1mm}
				\includegraphics[width=1\textwidth]{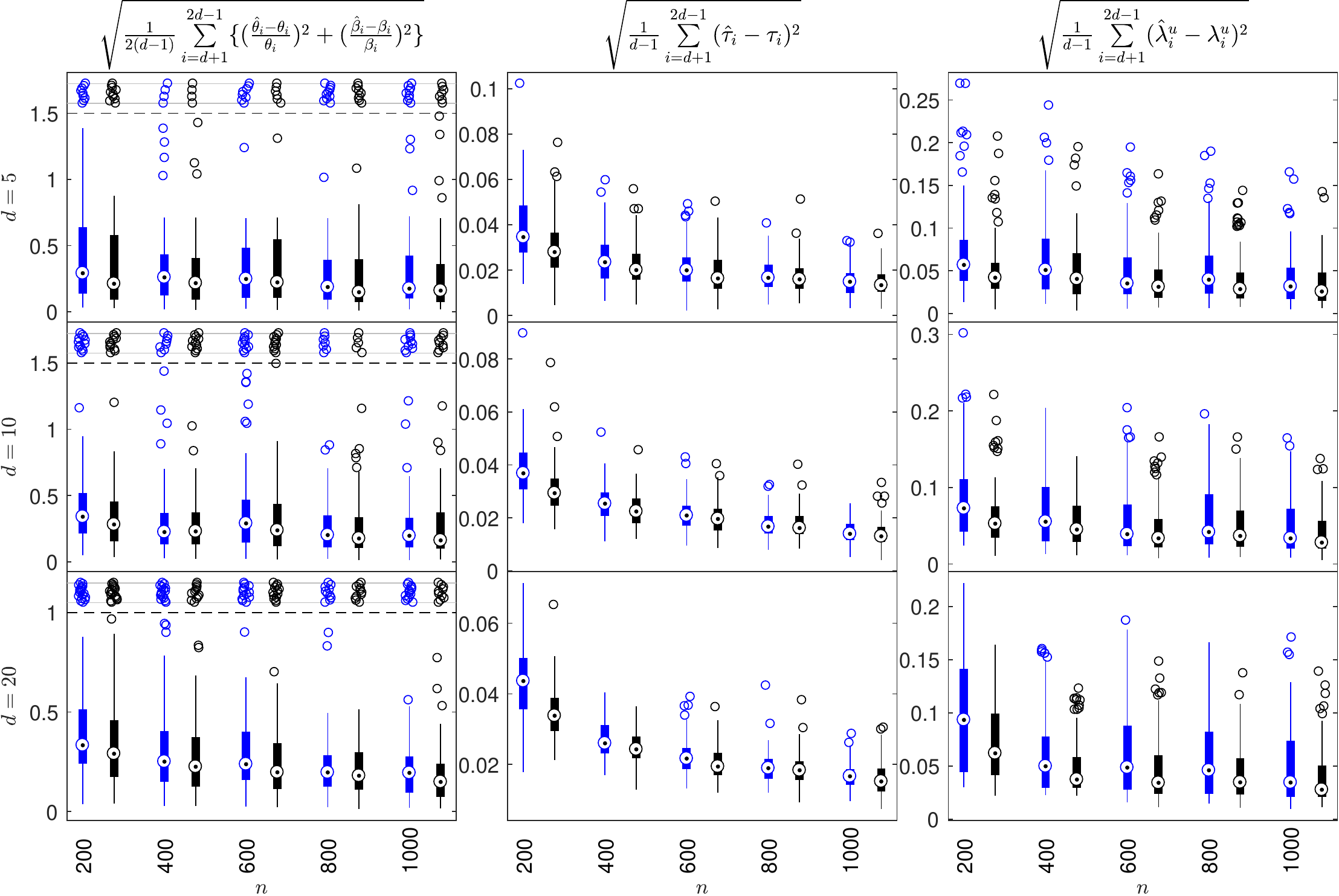}
				Clayton\\
				\vspace*{1mm}
				\includegraphics[width=1\textwidth]{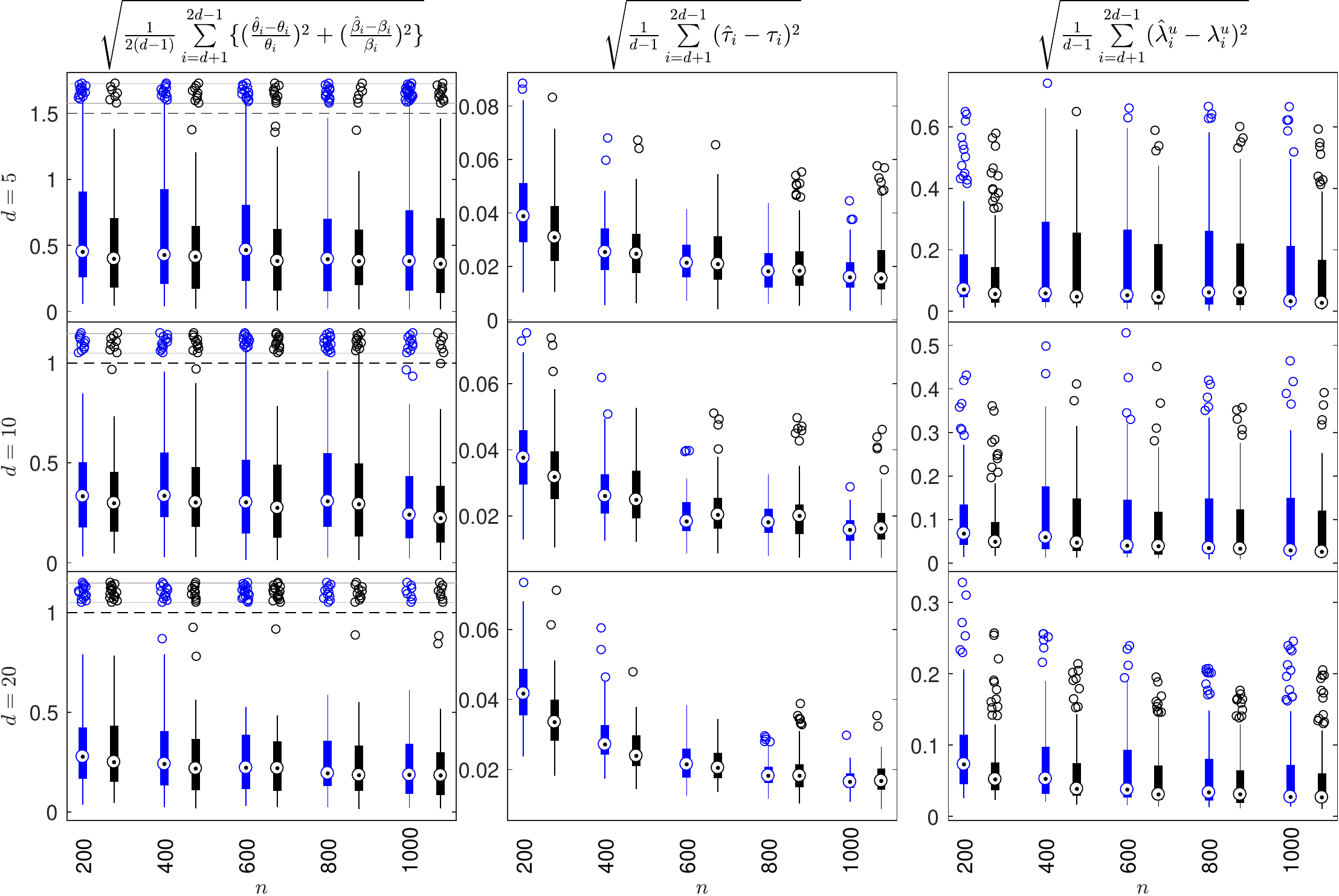}
				\caption{Realizations of the \textbf{true versus estimate} measures for Ali-Mikhail-Haq and Clayton copulas estimated by {\color{myBlue}TD-Sn} and TD-ML.}
				\label{fig:hopacest_opAC_123}
				% MATLAB: hopacestim_plot.m
			\end{figure}
			\begin{figure}[h!]
				\centering
				Frank\\
				\vspace*{1mm}
				\includegraphics[width=1\textwidth]{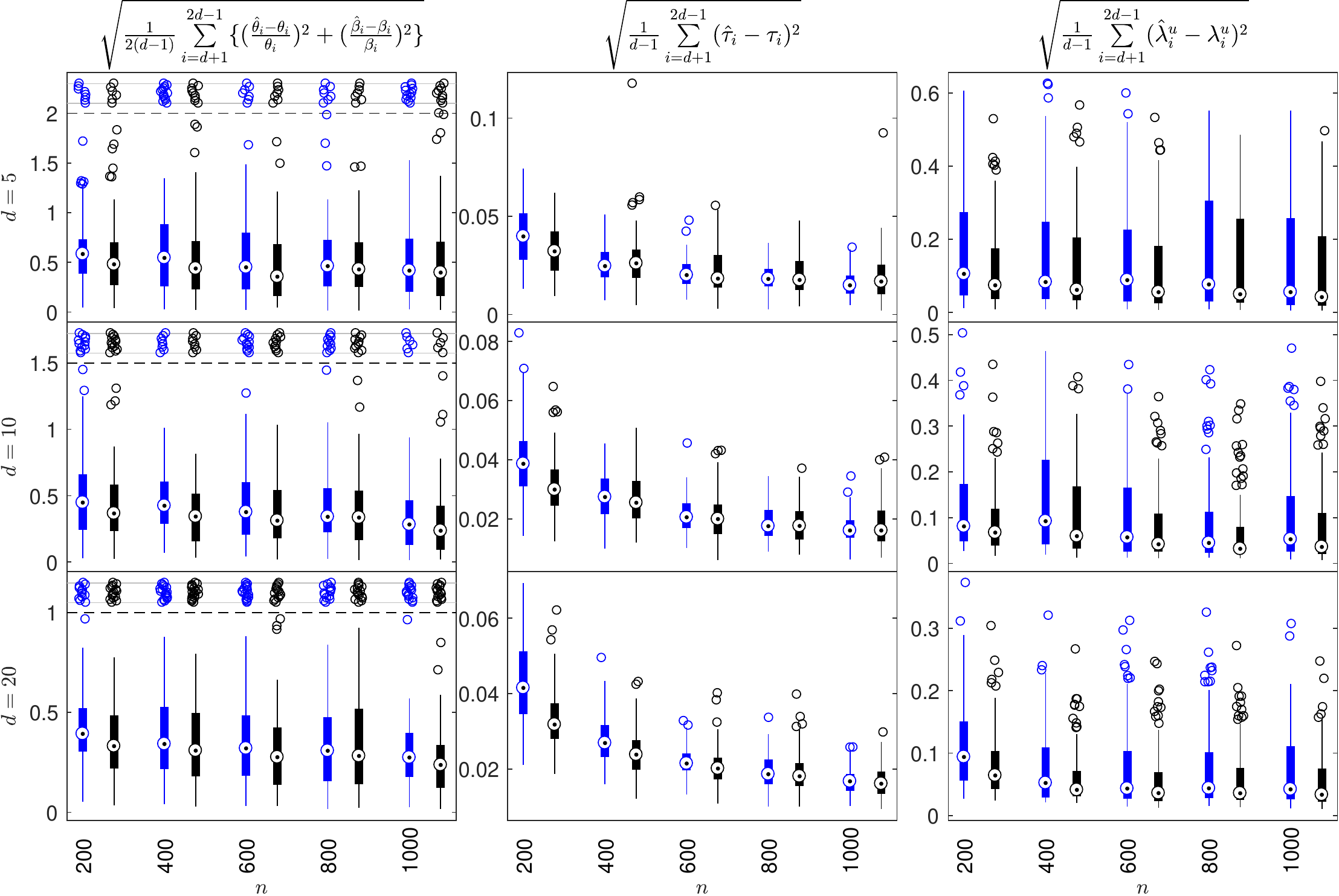}
				Joe\\
				\vspace*{1mm}
				\includegraphics[width=1\textwidth]{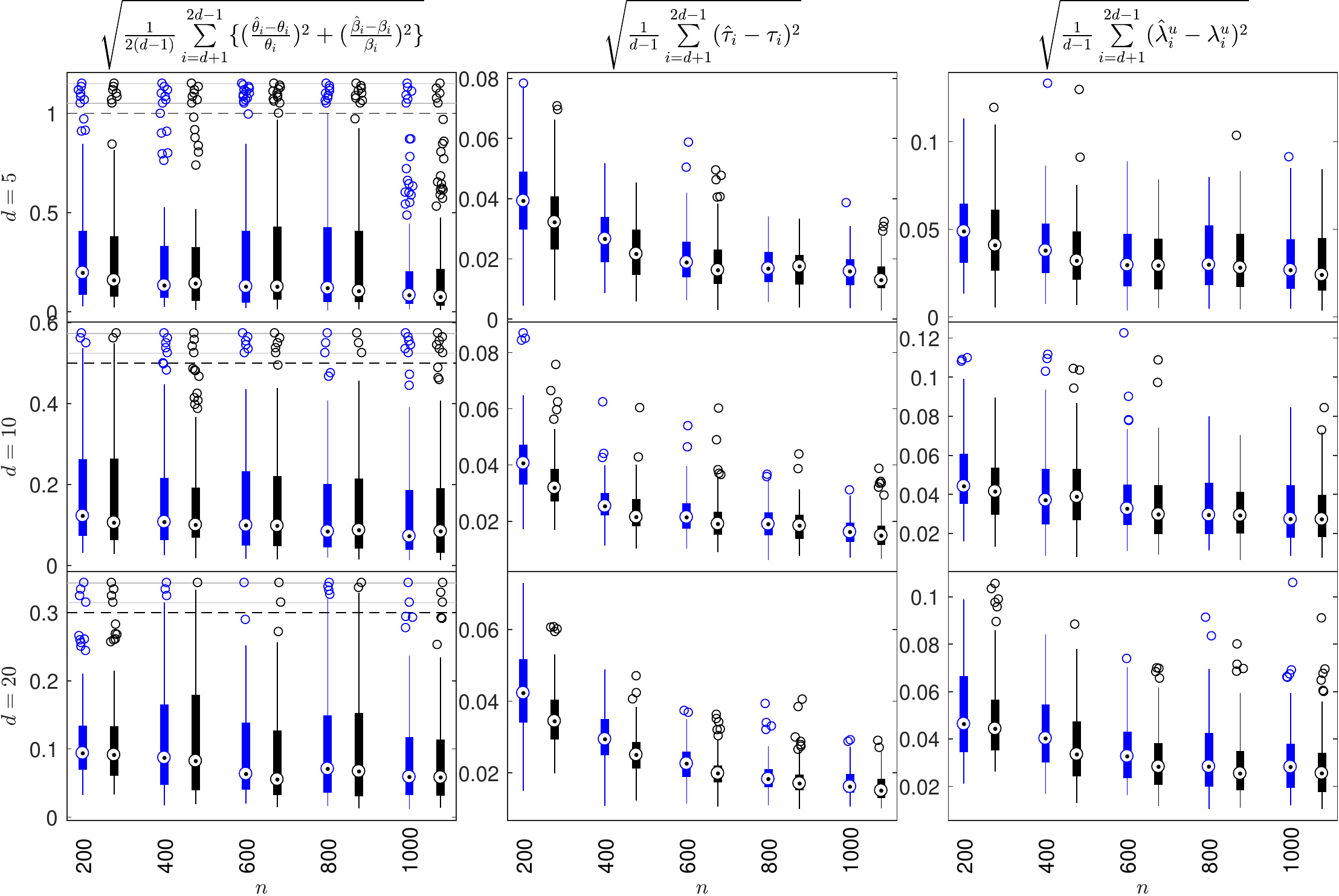}
				\caption{Realizations of the \textbf{true versus estimate} measures for Frank and Joe copulas estimated by {\color{myBlue}TD-Sn} and TD-ML.}
				\label{fig:hopacest_opFJ_123}
				% MATLAB: hopacestim_plot.m				
			\end{figure}					 		
		\end{enumerate}
\end{enumerate}
It can be observed that:
\begin{itemize}
	\item As $n$ increases, all measures decrease (converge to 0) for both HOPAC estimators TD-Sn and TD-ML.
	\item TD-ML shows lowest means and standard errors in most of the cases.
	\item The estimators either without hierarchy (OPAC) or with no OP transformation available (HAC) are unable to model HOPAC data, as is clear from Figures~\ref{fig:hopacest_opAC_567}~and~\ref{fig:hopacest_opFJ_567}.
	\item All the previous observations are independent of the underlying family.
\end{itemize}
	
The quality of the structure estimator (Algorithm~\ref{alg:structure_estim}) is also evaluated, see Figure~\ref{fig:hopacest_structures}. Note that each bar in Figure~\ref{fig:hopacest_structures1} shows the value $N/m \times 100$, where $m$ is the number of sampling repetitions until $N = 100$ true structures have been recovered. 
As such an equal-or-not criterion is too strict as it does not take into account \emph{how much} the estimated structure differs from the true structure. An extra proportional criterion based on a trivariate decomposition of the full structure according to \cite{Segers2014nonparametric} is also evaluated, see Figure~\ref{fig:hopacest_structures2}. There, each bar shows the value $r/m \times 100$, where $r = \sum_{i=1}^{m}r_j$ with $r_j$ being the ratio of the trivariate structures from the decomposition of the estimated structure matching the trivariate structures from the decomposition of the true structure to ${d \choose 3}$. Note that such a criterion has already been used, e.g., in \cite{uyttendaele2016estimation}.
As can be observed, the ratio of estimated true structures is:
\begin{itemize}
	\item Independent of the family underlying the sample.
	\item Increasing with $n$. 
	\item Converging to 100 (in $n$); this convergence is slower as $d$ increases. The impact of increasing $d$ is substantially lower for the proportional equal-or-not criterion.
\end{itemize}
\begin{figure}
	\begin{subfigure}[t]{0.45\linewidth}
		\centering
		\includegraphics[width=1\textwidth]{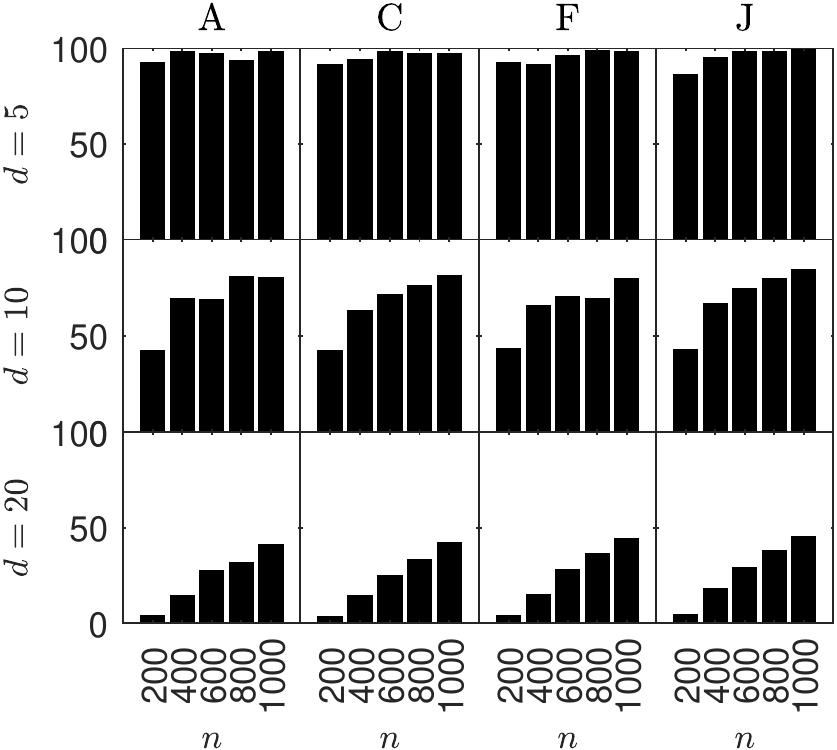}
		\caption{Equal-or-not criterion for the full structure.}
		\label{fig:hopacest_structures1}			
	\end{subfigure}
	\begin{subfigure}[t]{0.45\linewidth}
		\centering
		\includegraphics[width=1\textwidth]{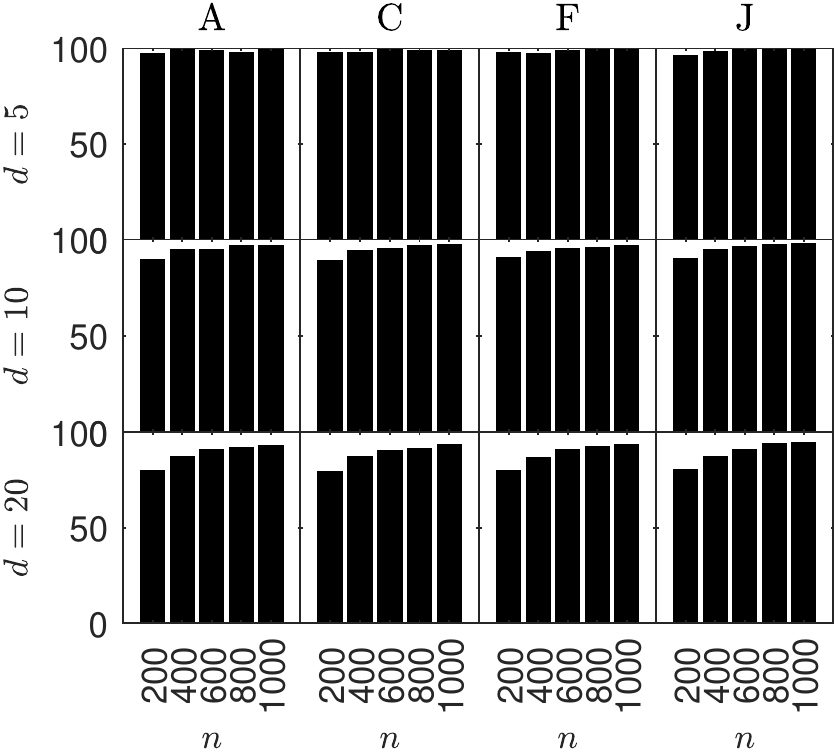}
		\caption{Proportional equal-or-not criterion based on a trivariate decomposition of the full structure \citep{Segers2014nonparametric}.}
		\label{fig:hopacest_structures2}			
	\end{subfigure}
	\caption{The y-axis shows the ratio of estimated true structures according to a selected criterion.}
	\label{fig:hopacest_structures}			
	% MATLAB hopacestim_plot_struc.m
\end{figure}

Finally note that other classes of copulas, e.g., elliptical or vine \citep{Cza10,joe2011dependence}, could be included in this simulation study. These were however not included as:
\begin{enumerate}
	\item It follows directly from their theoretical construction that radially symmetric elliptical copulas cannot fit asymmetric HOPACs;
	\item For vine copulas, realizations of the sample versus estimate measures require either the cumulative distribution function ($\hat{C}$), which is computationally demanding already for $d = 5$, or to access bivariate margins (to get $\hat{\tau}_{ij}$ or $\hat{\lambda}^u_{ij}$), which is, in general, not possible. %The latter could be done by sampling what would require huge samples for each comparison, therefore becomes again computationally demanding, and for $d=20$ even not possible by the implementation \citep{colblenz2019matvines}.
\end{enumerate}
These two important classes of copulas are included as benchmarks in the application reported in Section~\ref{sec:applic}, where they are compared to HOPACs in their ability of tail dependence modeling.

% \subsection{Summary}
% \label{sec:summaryHOPAC}
% 
% The highlights mentioned in Section~\ref{sec:summaryOPAC} that are brought to ACs by introducing the OP transformation \emph{are kept} also for HOPACs: 
% 	\begin{itemize}
% 		\item Theoretical background.
% 		\item Clear relation to dependence measures expressible analytically.
% 		\item Ali-Mikhail-Haq's $\tau$ covers \emph{whole} [0, 1).
% 	\end{itemize}
% The main contributions of Section~\ref{sec:nestedcase}:
% 	\begin{itemize}
% 		\item Efficient sampling strategy for HOPACs is provided.
% 		\item Two viable HOPAC estimators (TD-ML and TD-Sn) are provided and studied by simulation.
% 	\end{itemize}

%%%%%%%%%%%%%%%%
\section{Empirical Study}\label{sec:applic}
Value-at-Risk (VaR) has been established as an important risk measure in Quantitative Risk Management. In our study, we consider two different datasets of daily stock prices downloaded from Alpha Vantage\footnote{www.alphavantage.co}. The first one contains the five time series of stock prices of ADI (Analog Devices, Inc.), AVB (Avalonbay Communities Inc.), EQR (Equity Residential), LLY (Eli Lilly and Company) and TXN (Texas Instruments Inc.). It is important here that the clustering of the companies is given by their industry sector: ADI and TXN belong to the high-tech industry, AVB and EQR to the real estate industry and LLY to the pharmaceutical industry. Therefore, we would expect that the structure of the HOPAC or HAC used in the study will resemble these groupings and thus the copula structure will play an important role. The other dataset contains the first 10 time series of daily stock prices from the S\&P500 according to their market capitalization. For both datasets we use the time span 2002-02-01--2019-02-01. 

The negative profit-and-loss function of the portfolio is defined as $L_{t+1}=\sum_{j=1}^d b_j P_{j,t}\allowbreak(1-e^{R_{j,t+1}})$, where $P_{j,t}$ and $R_{j,t}$ are the price and log-return respectively of the asset $j$ at time point $t$, $d$ is the dimension of the portfolio and is equal to $5$ for the first dataset and to $10$ for the second one. Weights of the assets in the portfolio are denoted by $b_j$, for $j = 1, \ldots, d$ with $\sum_{j=1}^db_j = 1$. As the study aims at proving the general power of the HOPAC model and not the comparison between different portfolio allocation schemes we consider only the equally weighted portfolio $b_j = \frac{1}{d},~ j = 1,\ldots, d,$ advocated by \cite{demiguel2009optimal}. Let $F_L$ denote the distribution function of $L_{t+1}$. This leads to the VaR of the portfolio at level $\alpha$, $\VaR(\alpha) = F^{-1}_L(\alpha)$. We focus on $\alpha = \{0.95, 0.99\}$. The distribution function $F_L$ is estimated by simulating the path of the asset return from the underlying multivariate process estimated in the rolling window fashion on the windows of widths $w = \{126,252,504\}$. This corresponds to half a year, one year and two years of data. 
We fit all copulas (except for the historical simulation method \citep[Chapter~2]{mcneil2015quantitative}) to pseudo-observations constructed from 
the i.i.d.~standardized residuals and the underlying temporal dependency is modeled by marginal the GARCH(1, 1) method with $t$-distributed innovations:
\begin{eqnarray*}
  R_{j, t} = \mu_{j, t} + \sigma_{j, t}Z_{j, t}\quad\mbox{ with }\quad \sigma_{j, t}^2 = \omega_j + \alpha_j\sigma_{j, t-1}^2 + \beta_j(R_{j, t-1}-\mu_{j, t-1})^2, 
\end{eqnarray*}
and $\omega>0, \alpha_j\geq 0, \beta_j\geq 0, \alpha_j+\beta_j<1$. Afterwards, various copula models are estimated on the standardized residuals $Z_{j, t}$. Thus, the estimated $\widehat{\VaR}_{t, w}(\alpha)$ at a given time point $t$ window width $w$ and level $\alpha$ is computed as follows: a) we estimate the GARCH(1, 1) for all univariate time series of log-returns on the time interval $(t-w-1, t-1]$; b) extract standardised residuals and estimate the copula; c) simulate a sample of size $n = 1000$ from the estimated copula and plug them into the estimated GARCH obtaining 1000 predictions of log-returns for the time point $t$; d) compute empirical quantiles at level $\alpha$ of the 1000 predicted negative profit-and-loss functions obtained from predicted log-returns. Further, the evaluation of each model is made on the basis of the \emph{VaR violation ratio}:
\begin{equation*}
  \hat\alpha = \frac{1}{t_e - t_s+ 1}\sum_{t = t_s}^{t_e}\mathbbm{1}_{\{L_t>\widehat{\VaR}_{t, w}(\alpha)\}},
\end{equation*}
where $t_s$ is the first day (always set to 505; maximal window width plus 1) and $t_e$ is the last day (4279; last day in the data) of the back-testing period. The closer $\hat\alpha$ is to the theoretical level $\alpha$, the better the model. We thus compare the absolute deviations $|\hat\alpha-\alpha|$ for the different models. We are aware of the various tests in the spirit of \cite{Kupiec1995} but as they did not give any new insights visible from pure deviations we decided not to present them in the paper. 

All in all, our study considers 20 models: a) AC, OPAC, HAC and HOPAC for Ali-Mikhail-Haq, Clayton, Frank and Joe copulas; b) Gaussian  and $t$-copulas; c) R-Vine copulas, see \cite{colblenz2019matvines} and d) the quantile-based historical estimator (denoted ``Historical'') which is computed directly on the true profit-and-loss function without any underlying time-series model. 

\begin{figure}
	\begin{minipage}[t]{0.49\linewidth}
		\centering
		\footnotesize
		$d=5$\\
		~\\
		\vspace*{-2mm}
		\includegraphics[width=1\textwidth]{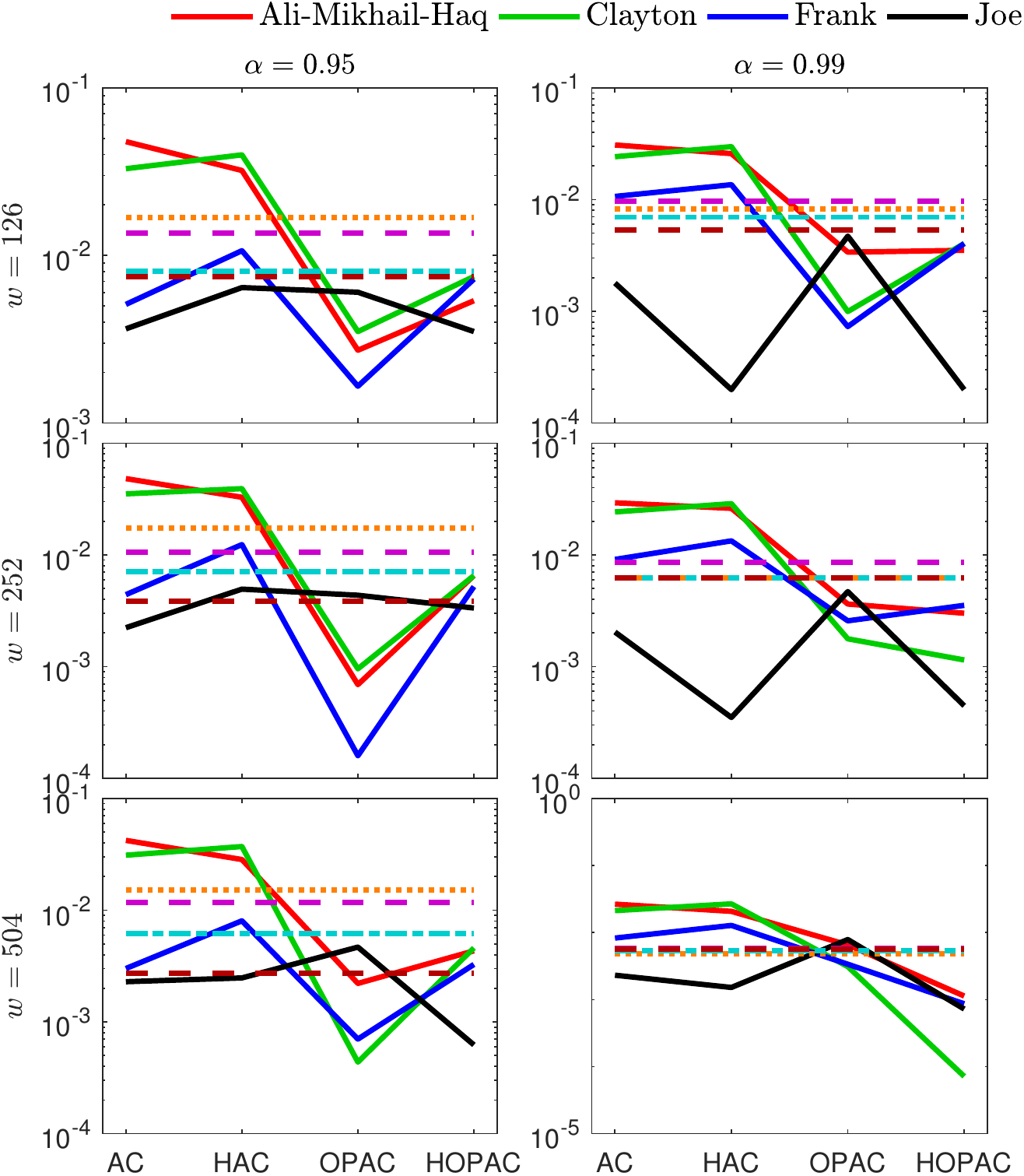}
	\end{minipage}	
	\begin{minipage}[t]{0.49\linewidth}
		\centering
		\footnotesize
		$d=10$\\
		~\\
		\vspace*{-2mm}
		\includegraphics[width=1\textwidth]{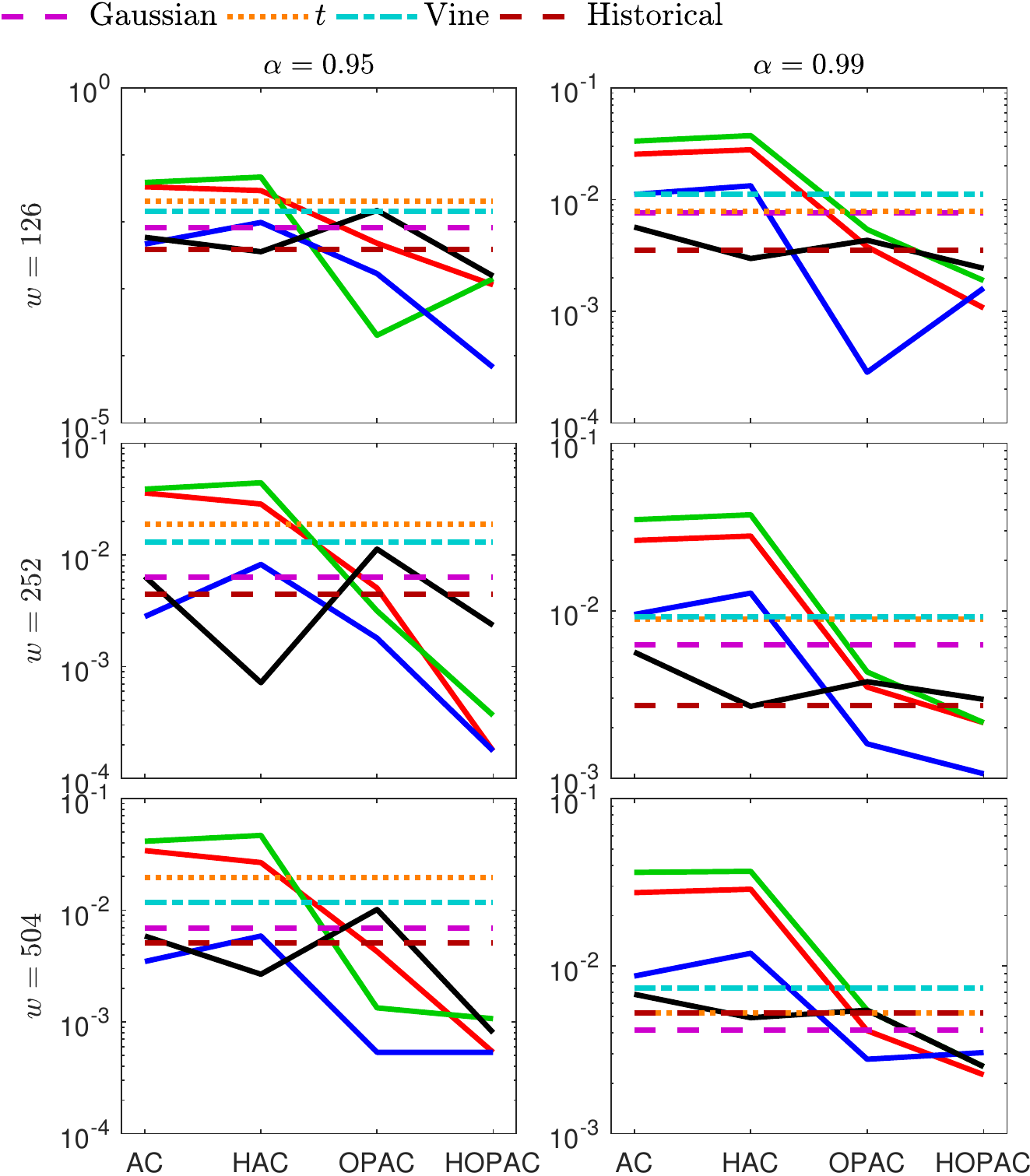}
	\end{minipage}	
	\caption{Comparison of the models on the basis of $|\hat\alpha - \alpha|$ for $d = \{5, 10\}$, $\alpha = \{0.95, 0.99\}$, $w = \{126, 252, 504\}$.}\label{fig:VaR}
\end{figure}

The results are summed up in Figure \ref{fig:VaR}, where the first two columns of panels correspond to the 5d portfolio and the second two columns to the 10d portfolio. The first column of panels for each portfolio depicts the results for $\alpha = 0.95$ and the second for $\alpha = 0.99$. The rows of panels show the results for the different widths $w = \{126, 252, 504\}$ of windows. Each panel represents deviations $|\hat\alpha-\alpha|$ for all the models on the log-scale. Solid lines represent all the (H)(OP)AC models based on Ali-Mikhail-Haq (red), Clayton (green), Frank (blue) and Joe (black) families and the particular model (AC, HAC, OPAC or HOPAC) is shown on the horizontal axes. All the remaining models are represented with non-solid lines (these models are neither hierarchical nor OP transformed). 

We clearly see from Figure \ref{fig:VaR} that the OPAC and HOPAC estimators outperform all the remaining estimators in almost all cases, almost independently of the type of the copula generator. Moreover, this implies that the OP transformation consistently improves the non-OP (H)AC estimators. In particular, the OP transformation is crucial for families that are unable to model upper-tail dependence, such as Ali-Mikhail-Haq, Clayton or Frank, or stronger correlations (e.g., Ali-Mikhail-Haq). For the Joe family, we observe good results also for the non-OP estimators. The OP-based estimators more frequently outperform the benchmark estimators for bigger $\alpha$s, particularly when compared to the historical estimator. An improvement given by considering hierarchies (i.e., from OPAC to HOPAC) is observed mainly for $d = 10$ what may be explained by the fact that it becomes more important to model hierarchies in higher dimensions. Surprisingly, for $d = 5$ where the structure has such a clear role in the selection of the stocks and $\alpha = 0.95$, exchangeable OPACs provide better results than HOPACs. For the AC-based estimators, no substantial influence of the size of the time window ($w$) is observed and can be explained by the relative robustness of these models over time.	

%%%%%%%%%%%%%%%%

\section{Conclusion}
\label{sec:conclu}
We demonstrated the improvements the OP transformations can bring to exchangeable ACs and hierarchical ACs. For the exchangeable case, a simplified way to compute the tail dependence coefficients was proposed. Also, two feasible OPAC estimators were considered via simulations.
Then, a new way of construction, an efficient sampling strategy and an estimator were provided for HOPACs, including a simulation study confirming the feasibility of the proposed estimator. Excellent abilities of the (H)OPAC models were finally demonstrated on an application from risk management.

Finally note that there also exist other, more general transformations for ACs, e.g.,
\begin{itemize}
	\item the \emph{tilted} OP transformation given by $\widetilde{\psi}(t) = \psi\{(c^{\beta} + t)^{1/\beta} - c\}$, where $c \in [0, \infty)$, see \cite{Hof11}, or
	\item the \emph{regularly varying transformed generator}, where the transform is given by a whole function, see \cite{DIBERNARDINO201689}. 
\end{itemize}
Their interpretation is, however, less clear, and the same applies to the hierarchical case.